\documentclass{emulateapj}
\usepackage{natbib}
\usepackage{graphicx}
\usepackage{appendix}
\usepackage{epsfig}
\usepackage{epsf}
\usepackage{amsmath}
\usepackage{amssymb}
\usepackage{url}
\usepackage{aas_macros}

\newcommand{\Hal}{\rm H\alpha}
\newcommand{\timp}{t_\mathrm{imp}}
\newcommand{\rhocsm}{\rho_\mathrm{csm}}
\newcommand{\rhoej}{\rho_\mathrm{ej}}
\newcommand{\csm}{\mathrm{csm}}

\slugcomment{To appear in ApJ. XX.}
\shorttitle{Delayed CSM Interaction in SNe\,Ia}
\shortauthors{Graham et al.}

\begin{document}
\title{Delayed Circumstellar Interaction for Type Ia SN~2015cp \\ Revealed by an {\it HST} Ultraviolet Imaging Survey}
\author{
M.~L.~Graham$^{1}$,
C.~E.~Harris$^{2,3}$,
P.~E.~Nugent$^{2,3}$, 
K.~Maguire$^{4}$,
M.~Sullivan$^{5}$,
M.~Smith$^{5}$,
S.~Valenti$^{6}$,
A.~Goobar$^{7}$,
O.~D. Fox$^{8}$,
K.~J.~Shen$^{2}$,
P.~L.~Kelly$^{9}$,
C.~McCully$^{10,11}$,
T.~G.~Brink$^{2}$,
and A.~V.~Filippenko$^{2,12}$
}

\altaffiltext{1}{Department of Astronomy, University of Washington, Box 351580, U.W., Seattle, WA 98195-1580, USA}
\altaffiltext{2}{Department of Astronomy, University of California, Berkeley, CA 94720-3411, USA}
\altaffiltext{3}{Lawrence Berkeley National Laboratory, 1 Cyclotron Road, MS 90R4000, Berkeley, CA 94720, USA}
\altaffiltext{4}{Astrophysics Research Centre, School of Mathematics and Physics, Queens University Belfast, Belfast BT7 1NN, UK}
\altaffiltext{5}{Department of Physics and Astronomy, University of Southampton, Southampton, SO17 1BJ, UK}
\altaffiltext{6}{Department of Physics, University of California, Davis, CA 95616, USA}
\altaffiltext{7}{Department of Physics, Oskar Klein Centre, Stockholm University, SE-106 91 Stockholm, Sweden}
\altaffiltext{8}{Space Telescope Science Institute, 3700 San Martin Drive, Baltimore, MD 21218, USA}
\altaffiltext{9}{School of Physics and Astronomy, University of Minnesota, 116 Church Street SE, Minneapolis, MN 55455, USA}
\altaffiltext{10}{Las Cumbres Observatory, 6740 Cortona Dr, Suite 102, Goleta, CA 93117-5575, USA}
\altaffiltext{11}{Department of Physics, University of California, Santa Barbara, CA 93106-9530, USA}
\altaffiltext{12}{Miller Senior Fellow, Miller Institute for Basic Research in Science, University of California, Berkeley, CA 94720, USA}

\begin{abstract}

The nature and role of the binary companion of carbon-oxygen white dwarf stars that explode as Type Ia supernovae (SNe\,Ia) are not yet fully understood. Past detections of circumstellar material (CSM) that contain hydrogen for a small number of SN\,Ia progenitor systems suggest that at least some have a nondegenerate companion. In order to constrain the prevalence, location, and quantity of CSM in SN\,Ia systems, we performed a near-ultraviolet (NUV) survey with the {\it Hubble Space Telescope (HST)} to look for the high-energy signature of SN\,Ia ejecta interacting with CSM. Our survey revealed that SN~2015cp, a SN 1991T-like overluminous SN\,Ia, was experiencing late-onset interaction between its ejecta and surrounding CSM at $664$ days after its light-curve peak. We present ground- and space-based follow-up observations of SN~2015cp that reveal optical emission lines of H and Ca, typical signatures of ejecta-CSM interaction. We show how SN~2015cp was likely similar to the well-studied SN\,Ia-CSM event PTF11kx, making it the second case in which an unambiguously classified SN\,Ia was observed to interact with a distant shell of CSM that contains hydrogen ($R_{\rm CSM} \gtrsim 10^{16}$~cm). The remainder of our {\it HST} NUV images of SNe\,Ia were nondetections that we use to constrain the occurrence rate of observable late-onset CSM interaction. We apply theoretical models for the emission from ejecta-CSM interaction to our NUV nondetections, and place upper limits on the mass and radial extent of CSM in SN\,Ia progenitor systems.

\end{abstract}


\section{Introduction} \label{sec:intro}

Supernovae of Type Ia (SNe\,Ia; see, e.g., \citealt{1997ARA&A..35..309F} for a review of SN classification) are thermonuclear explosions of carbon-oxygen white dwarfs (WDs), and are valuable as standardizable candles for dark-energy cosmology studies \citep[e.g.,][]{1998AJ....116.1009R,1999ApJ...517..565P}, but their progenitor scenario and their explosion mechanism are not yet well understood (e.g., \citealt{2011NatCo...2E.350H}). Most SN\,Ia progenitor hypotheses involve binary systems with either another WD, or a red giant, helium, or main sequence star, and an explosion that occurs after a stellar merger with, or mass accretion from, the companion. As the mass-transfer process is unlikely to be perfectly efficient, some circumstellar material (CSM) is expected to remain in the system, and its amount and location depend on the progenitor scenario. For example, the merger of two carbon-oxygen WDs may have none, and only a small amount of CSM is produced by a He WD companion ($<0.02$ $\rm M_{\odot}$ at $>5\times10^{17}$ cm; \citealt{2013ApJ...770L..35S}), but a nondegenerate companion appears to be able to produce a larger mass of H-rich CSM (up to several $\rm M_{\odot}$; \citealt{2003Natur.424..651H,2012Sci...337..942D}).

The signatures of ejecta-CSM interaction in SNe\,Ia have so far been observed in only a few cases. \cite{2003Natur.424..651H} discovered the first example, SN~2002ic, in which H$\alpha$ emission indicated several solar masses of CSM and a nondegenerate companion star (but see also the alternative interpretation offered by \citealt{2006ApJ...653L.129B}). The second clear detection of a SN~2002ic-like event was SN~2005gj, which was initially classified as a SN\,IIn but later determined to be a 1991T-like SN\,Ia with hydrogen emission from CSM interaction \citep{2005CBET..302....1P}. \cite{2006ApJ...648L.119I} observed CSM interaction to begin at 35~d after explosion for one of eight SNe\,Ia surveyed with $Swift$ X-ray and UV imaging (SN~2005ke): in this case, the associated emission was bright in the near-ultraviolet (NUV) but not seen in optical filters $UBV$, and lasted at least 3 months (but see also the X-ray reanalysis by \citealt{2007ApJ...670.1260H}). \cite{2013ApJS..207....3S} first established SNe\,Ia-CSM as a subclass of objects that exhibit a clear SN\,Ia-like optical spectrum with superimposed narrow- to medium-width hydrogen emission lines; this work also summarized the eight SNe\,Ia-CSM discovered prior to and/or externally to the Palomar Transient Factory (PTF), which began in 2010, and added eight more PTF-discovered SNe\,Ia-CSM, including the remarkable PTF11kx. \cite{2012Sci...337..942D} found that the ejecta of SN\,Ia PTF11kx began interacting with H-rich CSM $\sim50$~d after explosion, and presented follow-up observations that show CSM in multiple shells starting at a distance of $\sim10^{16}$ $\rm cm$ from the WD (assuming the fastest SN\,Ia ejecta have a velocity of $2\times10^4$ $\rm km\ s^{-1}$). Over the past decade, wide-field low-redshift surveys have since accumulated $\sim20$ SNe\,Ia which exhibit similar signatures of CSM interaction \citep{2013ApJS..207....3S,2015MNRAS.447..772F}, although for some the classification of the SN as Type Ia is contested \citep[e.g.,][]{2016MNRAS.459.2721I}. This is a small fraction of the hundreds of nearby SNe\,Ia discovered annually, but demonstrates that at least some SN\,Ia progenitor systems contain CSM gas from a nondegenerate companion. For all identified SN\,Ia-CSM events to date the interaction was observed to begin within 2 months of explosion, which sets a physical constraint that the CSM be within a distance of $<10^{16}$ $\rm cm$. The concern here is that this might be an observational bias, since most normal SNe\,Ia are monitored only for a few months; if CSM is more prevalent or typically at a larger distance from the WD, we might be systematically missing the signatures of this interaction. 

There are additional good reasons to suspect that we might be systematically missing the interaction of CSM with SN\,Ia ejecta. For example, SN\,Ia 2006X exhibited evolution in its \ion{Na}{1}~D absorption line that was interpreted as the photoionization of more distant CSM \citep{2007Sci...317..924P}. In a compilation of 17 SNe\,Ia with multi-epoch spectra, \cite{2014MNRAS.443.1849S} find that $\sim18$\% exhibit time-variable \ion{Na}{1}~D. Additionally, studies of large samples of SNe\,Ia ($20$--$35$ events) found \ion{Na}{1}~D to be blueshifted $\sim20\%$ more often than redshifted, and argued that this was best explained by outflowing CSM from the progenitor system \citep{2011Sci...333..856S,2013MNRAS.436..222M}, but see \cite{2018MNRAS.479.3663B} for an alternative explanation. In these SNe\,Ia the CSM is estimated to be $5\times10^{16}$ to $10^{18}$ $\rm cm$ away from the explosion, which means that the fastest SN ejecta would not interact until $>1$ yr. Additional support for late-time CSM interaction is presented by \cite{2013ApJ...772L...6F}, who interpret an increase in $Spitzer$ mid-infrared photometry for SN\,Ia 2005gj at 500--1000~d past peak brightness as dust heated by hot blackbody emission from new or renewed CSM interaction. Additional evidence for nondegenerate companion stars comes from detections of, e.g., ``blue bumps'' in the light curve in the week after explosion \citep{2017ApJ...845L..11H}, and/or late-time H$\alpha$ emission from material swept off the companion star and embedded in the ejecta \citep{2016MNRAS.457.3254M,2018arXiv180403666S}. It is important to note that there is plenty of evidence that many -- or even most -- SN\,Ia progenitor systems host a degenerate binary companion and/or do not contain CSM \citep[e.g.,][]{2011ApJ...741...20B,2014ARA&A..52..107M,2016ApJ...821..119C}. In this work we aim to put upper limits on the fraction of SNe\,Ia with CSM, and we are not suggesting that the systematic oversight of CSM interaction extends to {\it all} SNe\,Ia. 

There is further motivational support in searching for late-onset CSM interaction in SNe\,Ia from recent observations of delayed CSM interaction in some core-collapse SNe. For example, \cite{2015ApJ...815..120M} present observations of SN~2014C, a SN\,Ib that appeared to explode into an empty cavity surrounded by CSM, which transitioned to a SN\,IIn as the ejecta caught up with the previously released material. 
Similarly, \cite{2018MNRAS.478.5050M} find H$\alpha$ emission from the onset of CSM interaction in SN\,Ib 2004dk 13\,yr after the original explosion. \cite{2017ApJ...837...62V} present the results of a systematic narrow-band imaging survey of Type Ibc, IIb, and Ia SNe to look for H$\alpha$ emission due to CSM interaction years to decades after explosion. Most of their potential detections are associated with core-collapse events, but of particular relevance to this work is their potential detection of variable late-time H$\alpha$ for SN\,Ia 1981B.

In order to search for emission from late-onset CSM interaction in SNe\,Ia, we performed a Snapshot survey with the {\it Hubble Space Telescope} ({\it HST}) during Cycle 24, using WFC3 and the near-ultraviolet (NUV) F275W filter (GO-14779; PI Graham). We targeted mainly 1--3~yr old nearby SNe\,Ia, and chose to survey in the NUV for its sensitivity to high-energy emission from CSM interaction as well as the low surface brightness of the host galaxies in the UV --- removing the need to perform image subtraction to detect a new transient. The UV signature of late-onset CSM interaction would last months or more (e.g., SN~2005ke; \citealt{2006ApJ...648L.119I}), and easily outshine the low flux from the nebular SN\,Ia. Our goals with this survey were to assay the prevalence of late-onset CSM interaction for SNe\,Ia and constrain the amount of CSM mass in the progenitor systems of SNe\,Ia. 

Our {\it HST} Snapshot program obtained images for $72$ of our $80$ proposed targets, which is a very high completeness rate. Two of our observations yielded NUV detections at the locations of the SNe: one SN\,Ia (SN~2015cp) and one SN\,IIn (ASASSN-15og; but we note that \citealt{2017MNRAS.467.1098H} reanalyzed early-time data and reclassified ASASSN-15og as a SN\,Ia-CSM). Our observations of SN\,Ia 2015cp mark only the second time that the ejecta from an unambiguous Type Ia SN were observed to interact with CSM containing H, after the inaugural PTF11kx \citep{2012Sci...337..942D}.

In Section \ref{sec:obs} we present our targets and describe our {\it HST} observations. We analyze our images in Section \ref{sec:ana}, presenting and characterizing our two detections and evaluating the limiting magnitudes for the rest. In Section \ref{sec:15cp} we present SN\,Ia-CSM 2015cp, providing an in-depth analysis of our NUV detection and NUV-optical follow-up observations along with a physical interpretation of the CSM in its progenitor system. Additionally, we refer the reader to the X-ray and radio follow-up observations and full physical model for SN~2015cp presented by Harris et al. (2018, submitted; hereafter H18). In Section \ref{sec:disc} we convert our observations into limits on the intrinsic spectral luminosity and compare with other SNe with CSM interaction. We also put an upper limit on the fraction of SNe\,Ia with CSM, and the amount of CSM mass that may have experienced interaction with the SN ejecta but yielded NUV emission below our detection thresholds. Our conclusions are presented in Section \ref{sec:conc}.

\section{Observations}\label{sec:obs}

To search for the NUV emission from late-onset interaction between SN ejecta and distant CSM, we conducted an {\it HST} Snapshot program in Cycle 24 (GO-14779; PI Graham). In Section \ref{ssec:targs} we describe how we built our initial target list of 80 SNe, we describe our {\it HST} {\it WFC3} instrument configuration and exposure times in Section \ref{ssec:exps}, and in Section \ref{ssec:survey} we summarize our survey results.

\subsection{Target Selection}\label{ssec:targs}

To generate an initial list of potential targets we used discoveries from both professional and amateur surveys, such as the original and intermediate Palomar Transient Factory (PTF and iPTF; \citealt{2009PASP..121.1334R}), the La Silla Quest survey (LSQ; \citealt{2013PASP..125..683B}), the All-Sky Automated Survey for Supernovae (ASAS-SN\footnote{\url{http://www.astronomy.ohio-state.edu/$\sim$assassin/index.shtml}}), the Panoramic Survey Telescope and Rapid Response System (PanSTARRS; \citealt{2016arXiv161205560C}), the Optical Gravitational Lensing Experiment (OGLE; \citealt{2015AcA....65....1U}), the Lick Observatory Supernova Search (LOSS) with the Katzman Automatic Imaging Telescope (KAIT; \citealt{2001ASPC..246..121F}), the Master Global Robotic Network \citep{2010AdAst2010E..30L}, and SNHunt which is part of the Catalina Real Time Survey (CRTS; \citealt{2009ApJ...696..870D}). We also used the private PTF/iPTF Marshal, the private Las Cumbres Observatory's Supernova Exchange database, the public Astronomer's Telegrams, the public Transient Name Server, the public supernova database WISeREP \citep{2012PASP..124..668Y}, and the public Open Supernova Catalog \citep{2017ApJ...835...64G} to form an initial list of possible targets. 

\begin{table*} 
\begin{center} 
\caption{All Targets Requested for SNAP GO-14779} 
\label{tab:targets} 
\begin{tiny} 
\begin{tabular}{lcccll} 
\hline 
\hline 
Target & Coordinates & Redshift & Distance & Initial Type  & Reference(s) \\ 
Name  & ($\alpha$, $\delta$; J2000) &  &  [Mpc]  &  &  \\ 
\hline 
  ASASSN-14co  &  15:57:29.91 +01:06:33.74 &  0.033 &  146        &  Ia norm.                & \protect{\cite{2014ATel.6245....1B,2014ATel.6255....1L}} \\ 
  ASASSN-14dc  &  02:18:38.06 +33:36:58.30 &  0.044 &  195        &  IIn, Ia-CSM           & \protect{\cite{2014ATel.6267....1H,2014ATel.6660....1O,2014ATel.6284....1C}}  \\ 
  ASASSN-14eu  &  15:00:36.84 -03:50:48.90 &  0.02272 &  99       &  Ia pec.                  & \protect{\cite{2014ATel.6360....1C,2014ATel.6363....1M}}  \\ 
  ASASSN-14ew &  20:22:00.84 -51:47:46.30 &  0.0196 &  85         &  Ia norm.                 & \protect{\cite{2014ATel.6367....1C,2014ATel.6384....1T}}  \\ 
  ASASSN-14lo   &  11:51:53.19 +18:32:31.00 &  0.01993 &  87      &  Ia norm.                 & \protect{\cite{2014ATel.6794....1K,2014ATel.6812....1S}}  \\ 
  ASASSN-14lq   &  22:57:19.41 -20:58:00.80 &  0.026 &  114         &  Ia 91T                    & \protect{\cite{2014ATel.6798....1N,2014ATel.6814....1Z}}  \\ 
  ASASSN-14lw   &  01:06:49.17 -46:58:59.10 &  0.016 &  69           &  Ia 91T, HV-feat.     & \protect{\cite{2014ATel.6809....1K,2014ATel.6832....1C}}  \\ 
  ASASSN-15de  &  11:28:49.65 +29:54:52.50 &  0.05172 &  230     &  Ia 91T                    & \protect{\cite{2015ATel.7094....1B,2015ATel.7103....1C}}  \\ 
  ASASSN-15hy  &  20:10:02.35 -00:44:21.30 &  0.025 &  109          &  Ia 06gz                   & \protect{\cite{2015ATel.7450....1H,2015ATel.7452....1F}}  \\ 
  ASASSN-15jo   &  14:06:44.67 -34:27:17.40 &  0.015 &  65            &  Ia, intracluster?      & \protect{\cite{2015ATel.7540....1H,2015ATel.7553....1H}}  \\ 
  ASASSN-15nr  &  17:26:41.84 +13:54:35.50 &  0.023 &  100          &  Ia 91T                    & \protect{\cite{2015ATel.7881....1M,2015ATel.7882....1B}}  \\ 
  ASASSN-15og  &  03:21:07.44 -31:18:45.60 &  0.06831 &  308      &  IIn, Ia-CSM, 05gj   & \protect{\cite{2015ATel.7912....1B,2015ATel.7932....1M,2017MNRAS.467.1098H}}  \\ 
  ASASSN-15sh  &  19:32:07.01 -62:26:29.10 &  0.033 &  145          &  Ia 91T                     & \protect{\cite{2015ATel.8237....1B,2015ATel.8268....1D}}  \\ 
  ASASSN-15ut   &  00:21:21.09 -48:38:30.30 &  0.011 &  51            & Ia 91T, HV               & \protect{\cite{2015ATel.8479....1K,2016ATel.8495....1F}}  \\ 
  LSQ14fmg        &  22:16:46.10 +15:21:14.10 &  0.06 &  269             &  Ia 91T                   & \protect{\cite{2014ATel.6495....1T}}  \\ 
  LSQ15aae        &  16:30:15.70 +05:55:58.70 &  0.05 &  222             &  Ia 91T                    & \protect{\cite{2015ATel.7325....1T}}  \\ 
  LSQ15adm        &  20:14:21.82 -58:08:20.20 &  0.073 &  330           &  Ia-CSM/pec.          & \protect{\cite{2015ATel.7363....1C}}; WISeREP  \\ 
  LSQ15bxe        &  12:27:52.38 -17:06:25.00 &  0.044 &  195            &  Ia 91T                    & \protect{\cite{2016ATel.8518....1F}}  \\ 
  MasterOT0442 &  04:42:12.20 +23:06:16.70 &  0.016 &  69             &  IIn, Ia-CSM, 05gj   & \protect{\cite{2014ATel.6484....1T,2014ATel.6487....1S,2014ATel.6488....1O}}  \\ 
  OGLE-2014-SN-107 &  00:42:28.76 -64:45:51.00 &  0.067 &  301   &  Ia 91T                    & \protect{\cite{2014ATel.6596....1W,2014ATel.6612....1T}}; WISeREP  \\ 
  OGLE-2014-SN-141 &  05:37:18.64 -75:43:17.00 &  0.063 &  283   &  Ia 91T                    & \protect{\cite{2014ATel.6722....1W,2014ATel.6706....1D}} \\ 
  PS1-13dsg        &  01:18:05.59 +27:11:25.70 &  0.05 &  222             &  Ia 91T, 99a           & \protect{\cite{2013ATel.5437....1D,2013ATel.5456....1D}} \\ 
  PS1-14oo         &  13:07:45.12 +16:14:05.00 &  0.05 &  222             &  Ia 91T                   & \protect{\cite{2014ATel.5937....1C}} \\ 
  PS15cwx          &  05:14:47.80 +07:03:01.30 &  0.046 &  204            &  Ia 91T/pec., Ic?    & \protect{\cite{2015ATel.8299....1W,2015ATel.8300....1K,2015ATel.8301....1T}} \\ 
  PS15sv            &  16:13:11.74 +01:35:31.10 &  0.038 &  167             &  Ia 91T                   & \protect{\cite{2015ATel.7280....1S,2015ATel.7308....1W}} \\ 
  PSNJ02+42     &  02:45:17.11 +42:13:50.30 &  0.03 &  131               &  Ia 91T                   & \protect{\cite{2014ATel.6378....1S}} \\ 
  PSNJ08+48     &  08:35:16.68 +48:19:01.10 &  0.043 &  190             &  Ia, HV?                 & \protect{\cite{2014ATel.6852....1Z}} \\ 
  PSNJ23-15     &  23:53:25.61 -15:39:17.60 &  0.058 &  259              &  Ia 91T, 98es         & \protect{\cite{2015ATel.7905....1P,2015ATel.7934....1Z}} \\ 
  PTF11kx          &  08:09:12.87 +46:18:48.80 &  0.046 &  205             &  Ia-CSM               & \protect{\cite{2012Sci...337..942D}} \\ 
  iPTF13asv      &  16:22:43.19 +18:57:35.00 &  0.036 &  158             &  Ia, OL, 91T, SC    & \protect{\cite{2013ATel.5061....1C,2013CBET.3543....1Z,2016ApJ...823..147C}} \\ 
  iPTF13daw     &  02:43:31.29 +01:59:03.90 &  0.07 &  318               &  Ia OL?                   & PTF  \\ 
  iPTF13dud     &  23:43:55.75 +27:20:33.00 &  0.06 &  269               &  Ia, HV?                   & \protect{\cite{2013ATel.5579....1C}}; PTF \\ 
  iPTF13ebh     &  02:21:59.98 +33:16:13.70 &  0.01327 &  57           &  Ia, HV?, 91bg-like   & \protect{\cite{2013ATel.5580....1C,2013ATel.5584....1M,2015A&A...578A...9H}}  \\ 
  iPTF13s         &  13:32:53.30 +35:57:33.70 &  0.06 &  269               &  Ia 91T                    & \protect{\cite{2013ATel.4805....1S,2013ATel.4810....1G}}; PTF  \\ 
  iPTF14abk     &  12:28:44.91 +64:37:14.60 &  0.035 &  155             &  Ia norm.                & PTF \\ 
  iPTF14aqs     &  13:18:58.67 +42:32:23.50 &  0.08 &  366               &  Ia norm., OL         & PTF \\ 
  iPTF14atg     &  12:52:44.84 +26:28:13.00 &  0.0051 &  93              &  Ia 91bg, non-degen?     & \protect{\cite{2014ATel.6168....1H,2014ATel.6203....1W,2015Natur.521..328C}}  \\ 
  iPTF14bdn    &  13:30:44.88 +32:45:42.40 &  0.01558 &  68            &  Ia 91T, HV              & \protect{\cite{2014ATel.6175....1C}} \\ 
  iPTF14fpg     &  00:28:12.00 +07:09:43.50 &  0.02 &  86                  &  Ia norm.                  & \protect{\cite{2014ATel.6503....1T,2014ATel.6534....1J}}  \\ 
  iPTF14fyq     &  00:03:21.72 +43:41:10.60 &  0.065 &  280              &  Ia norm.                  & \protect{\cite{2014ATel.6534....1J}}  \\ 
  iPTF14gmo    &  02:21:18.35 +40:51:15.90 &  0.066 &  300             &  Ia, OL?                    & \protect{\cite{2014ATel.6638....1B,2014ATel.6651....1P}}; PTF  \\ 
  iPTF14gnl     &  00:23:48.33 -03:51:27.90 &  0.0537 &  241             &  Ia, OL?                    & \protect{\cite{2014ATel.6598....1P}}; PTF \\ 
  iPTF14ibo     &  23:04:59.31 +17:29:09.50 &  0.08 &  366                &  Ia, OL?                     & \protect{\cite{2014ATel.6764....1F}}; PTF  \\ 
  iPTF14sz      &  03:05:17.75 +00:13:03.10 &  0.03 &  132                &  Ia norm.                   & PTF  \\ 
  iPTF15agv    &  10:58:59.07 +46:40:25.20 &  0.035 &  155              &  Ia norm.                   & \protect{\cite{2015ATel.7478....1P}}  \\ 
  iPTF15akf     &  11:55:18.14 +50:48:02.00 &  0.0527 &  235            &  Ia 91T                     & PTF \\ 
  iPTF15clp     &  22:39:10.88 +34:18:25.70 &  0.02 &  86                  &  Ia norm.                   & \protect{\cite{2015ATel.7410....1G}}; PTF \\ 
  iPTF15eod    &  09:10:08.76 +50:03:39.40 &  0.026 &  114              &  Ia pec., SC, 09dc    & \protect{\cite{2015ATel.8263....1T}}  \\ 
  iPTF15go      &  14:26:36.78 +24:03:10.20 &  0.03 &  132                &  Ia norm.                  & \protect{\cite{2015ATel.7027....1C}}  \\ 
  iPTF15wd      &  11:38:53.74 +55:03:17.10 &  0.05735 &  256         &  Ia 91T                      & PTF \\ 
  iPTF16abc     &  13:34:45.49 +13:51:14.30 &  0.023 &  100             &  Ia 91T, tidal tail  & \protect{\cite{2016ATel.8907....1M,2016ATel.8909....1C}}  \\ 
  SN2012cg     &  12:27:12.83 +09:25:13.20 &  0.00147 &  6.3          &  Ia, non-degen?       & \protect{\cite{2012CBET.3111....1K,2016ApJ...820...92M}}  \\ 
  SN2012gl      &  10:12:50.32 +12:40:56.70 &  0.00937 &  40           &  Ia norm.                   & \protect{\cite{2012CBET.3302....2S}}; PTF  \\ 
  SN2013I        &  02:49:42.17 +00:45:35.70 &  0.036 &  160             &  Ia-CSM, IIn              & \protect{\cite{2013ATel.4741....1B,2013CBET.3386....1T}}  \\ 
  SN2013bh      &  15:02:13.09 +10:38:45.30 &  0.07436 &  320        & Ia pec., 00cx           & \protect{\cite{2013ATel.4955....1M,2013MNRAS.436.1225S}}  \\ 
  SN2013dn      &  23:37:45.74 +14:42:37.10 &  0.052 &  233            &  Ia-CSM                   & \protect{\cite{2013CBET.3570....1D,2015MNRAS.447..772F}}  \\ 
  SN2013dy      &  22:18:17.60 +40:34:09.60 &  0.00388 &  17          &  Ia, slow decline       & \protect{\cite{2013CBET.3588....1C,2015MNRAS.452.4307P}}  \\ 
  SN2013gh      &  22:02:21.84 -18:55:00.40 &  0.0088 &  38             &  Ia, NaID                  & \protect{\cite{2013CBET.3706....1H,2016PASA...33...55C}}; WISeREP  \\ 
  SN2013gv      &  03:09:57.27 +19:12:48.50 &  0.033 &  145            &  Ia, HV?                  & \protect{\cite{2013CBET.3739....1K,2013ATel.5636....1T}} \\ 
  SN2013hh      &  11:29:04.37 +17:14:09.50 &  0.012 &  52              &  Ia 91T                    & \protect{\cite{2013CBET.3754....1K,2013ATel.5656....1C}}  \\ 
  SN2014E      &  12:03:31.29 +02:02:34.00 &  0.01896 &  82           &  Ia norm.                  & \protect{\cite{2014CBET.3784....1H,2014ATel.5741....1C,2014ATel.5742....1T}}  \\ 
  SN2014I        &  05:42:19.80 -25:32:39.90 &  0.03 &  131               &  Ia, HV-feat.              & \protect{\cite{2014ATel.5777....1C}}  \\ 
  SN2014J       &  09:55:42.14 +69:40:26.00 &  0.00068 &  3            &  Ia, HV, CSM, NaID  & \protect{\cite{2014CBET.3792....2I,2014ATel.5791....1T}}  \\ 
  SN2014R      &  09:30:12.30 +55:51:13.50 &  0.02515 &  110        &  Ia 91T                      & \protect{\cite{2014CBET.3808....1G,2014CBET.3809....3Z}}  \\ 
  SN2014ab      &  13:48:05.99 +07:23:16.40 &  0.02323 &  101      &  IIn, 12ca                   & \protect{\cite{2014CBET.3826....1H,2014ATel.5964....1H,2014ATel.5968....1F}}  \\ 
  SN2014ai      &  09:19:44.20 +33:45:49.60 &  0.023 &  100           &  Ia norm.                   & \protect{\cite{2014CBET.3838....1F,2014ATel.6008....1F}}  \\ 
  SN2014aj      &  05:11:43.96 +67:29:29.40 &  0.02 &  87               &  Ia, HV                      & \protect{\cite{2014CBET.3844....2E,2014ATel.6007....1T}}  \\ 
  SN2014ap      &  11:30:13.37 +24:10:07.20 &  0.0235 &  102        &  Ia                             & \protect{\cite{2014CBET.3856....1C,2014ATel.6001....1C}}  \\ 
  SN2014aw      &  16:10:23.62 +47:04:40.40 &  0.037 &  164         &  Ia, OL                      & \protect{\cite{2014CBET.3866....1D,2014ATel.6113....1K}}; PTF  \\ 
  SN2014bn      &  21:15:13.26 +02:11:22.20 &  0.05 &  200            &  Ia, OL                      & \protect{\cite{2014CBET.3897....1C,2014ATel.6214....1T}}; PTF  \\ 
  SN2014ch      &  15:58:31.10 +12:51:59.60 &  0.044 &  195          &  Ia, CII, HV?, OL?    & \protect{\cite{2014CBET.3946....1D,2014ATel.6205....1W}}  \\ 
  SN2014dl      &  16:29:46.09 +08:38:30.60 &  0.033 &  145           &  Ia 91T                     & \protect{\cite{2014CBET.3995....1D,2014ATel.6507....1N,2014ATel.6508....1M}}  \\ 
  SN2014dt      &  12:21:57.57 +04:28:18.50 &  0.00525 &  23         &  Iax                          & \protect{\cite{2014CBET.4011....1N,2014ATel.6648....1O}}  \\ 
  SN2014eg      &  02:45:09.27 -55:44:16.90 &  0.018 &  78             &  Ia 91T                    & \protect{\cite{2014ATel.6711....1K,2014ATel.6739....1S}}  \\ 
  SN2015F       &  07:36:15.76 -69:30:23.00 &  0.004 &  24              &  Ia norm.                 & \protect{\cite{2015CBET.4081....1M,2015ATel.7209....1F}}  \\ 
  SN2015aw      &  02:06:22.53 -52:01:26.70 &  0.01962 &  85         &  Ia 91T/norm          & \protect{\cite{2015ATel.7815....1M}}  \\ 
  SN2015bd      &  11:23:45.88 -01:06:21.20 &  0.0186 &  81            &  Ia 91T/99aa          & \protect{\cite{2015ATel.8393....1S}}  \\ 
  SN2015bp      &  15:05:30.07 +01:38:02.40 &  0.004 &  18             &  Ia 91bg                  & \protect{\cite{2015ATel.7251....1J}}  \\ 
  SN2015bq      &  12:35:06.26 +31:14:35.60 &  0.02818 &  123       &  Ia 91T/99aa           & \protect{\cite{2015ATel.7109....1Z,2015ATel.7119....1F}}  \\ 
  SN2015cp      &  03:09:12.75 +27:31:16.90 &  0.038 &  167           &  Ia 91T                   & \protect{\cite{2016ATel.8498....1F,2016ApJ...827L..40S}}  \\ 
 \hline 
\end{tabular} 
\end{tiny} 
\end{center} 
\end{table*}

Our goal in choosing targets was to maximize our chances of a detection, so we prioritized SNe\,Ia that were reported to exhibit characteristics known or suspected to correlate with the single-degenerate scenario and/or the presence of CSM. One of the characteristics of SNe\,Ia that is known to correlate with such qualities is an overluminous light curve and/or a spectrum similar to those of SN\,Ia 1991T \citep{2015A&A...574A..61L,2015ApJ...805..150F,2015ApJ...808...49K}, a subtype that occurs more often in relatively young stellar populations (e.g., \citealt{2009ApJ...691..661H}). SNe\,Ia with a total-to-selective extinction ($\rm R_V$) lower than Galactic values exhibit higher photospheric velocities than normal, and these SNe\,Ia are more often associated with a relatively young, metal-rich population \citep{2013Sci...340..170W}, suggesting (indirectly) that they may have a CSM-producing progenitor (i.e., assuming CSM would have a different value of $R_V$ than Milky Way dust). For these reasons, we have prioritized SNe\,Ia that exhibit a ``91T-like" light curve or spectrum, high photospheric velocities (HV), blueshifted \ion{Na}{1}~D absorption (see Section \ref{sec:intro}), and/or hosts with younger stellar populations. As a control set we also included a small fraction of SNe that exhibited strong signatures of CSM interaction around their time of discovery (SNe\,Ia-CSM and SNe\,IIn). 

Our final sample, listed in Table \ref{tab:targets}, includes $39$ SNe\,Ia with a (tentative, at least) classification of SN\,1991T-like (i.e., a spectroscopic match or an overluminous peak brightness), $16$ SNe\,Ia classified as normal, and $9$ SNe\,IIn and/or SNe\,Ia-CSM; the remaining $16$ are SNe\,Ia of other subtypes. In column 5 of Table \ref{tab:targets} we provide an abbreviated classification for each of our targets, and Table \ref{tab:targtypes} defines those abbreviations. Column 6 of Table \ref{tab:targets} provides the discovery and/or classification references for each of our targets, when they're publicly available; in a few cases the full classification relies on private survey data (e.g., PTF). The redshifts and distances listed in columns 3 and 4 represent the best measurements or estimates available: host galaxy spectra lead to redshifts with $>3$ significant figures, whereas SN spectral fits are less precise, and most of the distance estimates are based on the redshift (assuming a flat cosmology with $\Omega_M=0.3$ and $\Omega_{\Lambda}=0.7$) except for a few nearby SNe with redshift-independent distances (eg., 2014J). In pursuit of our goal to maximize our chances of detection, we also limited our sample to SNe within a distance of $\lesssim350$ Mpc and with phases of $\sim 1$--3 yr post-explosion during Cycle 24, i.e., after the supernova’s emission had faded and later than most SNe\,Ia are routinely monitored. 

\begin{table}
\begin{center}
\caption{Descriptions of Initial Type Details for Table \ref{tab:targets}}
\label{tab:targtypes}
\begin{tabular}{ll}
\hline
Descriptor & Meaning \\
\hline
Ia                 & e.g., \protect{\cite{1997ARA&A..35..309F}} \\
Iax                & \cite{2013ApJ...767...57F} \\
Ia-CSM        &  \cite{2013ApJS..207....3S} \\
Ib/c               & e.g., \protect{\cite{2016ApJ...827...90L,1997ARA&A..35..309F}}  \\
IIn                & e.g., \protect{\cite{1997ARA&A..35..309F}} \\
SLSN-II         & \cite{2012Sci...337..927G} \\
norm.             & SN\,Ia with no special attributes \\
91T               & \cite{1992ApJ...384L..15F} \\
91bg              & \cite{1992AJ....104.1543F} \\
98es             & 91T-like; \cite{1998IAUC.7054....2J} \\
99aa              & overluminous; \cite{2000ApJ...539..658K} \\
00cx              & \cite{2001PASP..113.1178L} \\
02ic               & SN\,Ia-CSM; \cite{2003Natur.424..651H} \\
05gj              & SN\,Ia-CSM/IIn: \cite{2006ApJ...650..510A} \\
06gz             & SC-like; \cite{2007ApJ...669L..17H} \\
09dc              & SC-like; \cite{2009ApJ...707L.118Y} \\
12ca              & SN\,Ia-CSM or SN\,IIn; see Section \ref{sssec:disc_compNUV} \\
SC                & super-Chandra candidate; \cite{2006Natur.443..308H} \\
HV                  & high photospheric silicon velocity  \\
HV-feat.          & high-velocity features present (\ion{Ca}{2} or \ion{Si}{2}) \\
CII                   & carbon features present \\
NaID              & narrow sodium features present \\
CSM               & showed a signature of CSM \\
OL                 & overluminous; peak brightness $<-19.3$ mag\tablenote{i.e., photometrically, potentially 91T-like} \\
slow decline     & smaller $\Delta m_{15}(B)$ than typical  \\
non-degen.     & evidence of a nondegenerate companion \\
pec.               & spectroscopically peculiar \\
tidal tail         & in a special environment: galaxy tidal tail \\
intracluster     & in a cluster; no obvious host galaxy \\
?                     & denotes tentative subclassification \\
\hline
\end{tabular}
\end{center}
\end{table}

In choosing targets for this program we also considered several conditions for technical feasibility. To maximize the probability of a detection we rejected any SNe that appeared to be in regions of high host-galaxy UV background. To ensure that in the case of any detection we could make as thorough a physical interpretation as possible, we also prioritized objects with denser spectroscopic and photometric coverage around peak brightness. 

\subsection{Instrument Configuration and Exposure Times} \label{ssec:exps}

For our survey we chose to use {\it HST} WFC3+F275W because it covers the expected emission region and is wide enough to offer a decent signal-to-noise ratio. Images with F275W will also be free from the light-leak contamination suffered by F225W. Three other advantages of an NUV search are as follows. (1) We are unlikely to mistake a SN\,Ia light-echo for CSM interaction because near-peak SNe\,Ia do not exhibit much NUV flux. While SNe\,Ia with CSM might be more likely to produce a light-echo, it would be very faint (e.g., as seen for SN~2006X; \citealt{2008ApJ...677.1060W}). (2) Late-time SNe\,Ia without CSM interaction are very faint in the NUV and will not contaminate our results. (3) Although a shocked companion star might cause excess UV flux, it would be below the detection threshold for most of our targets \citep{2012ApJ...760...21P,2015MNRAS.454.1948G}.

As this was the first late-time NUV imaging program attempted for SNe\,Ia, and any data at all would constrain the amount of CSM, we designed our program to be able to observe a large number of SNe\,Ia. {\it HST} Snapshot observations with durations in the $20$--$30$~min range have the highest completion rate, and a $20$~min exposure time would be sufficient for our science goals because it would yield limiting magnitudes of $m_{\rm F275W} \gtrsim 25$ mag (AB). In order to minimize overheads and maximize integration time, we kept individual exposures $>348$~s to avoid the 5.8~min buffer dump, and to minimize the effects of cosmic rays (CR) we obtained two frames instead of just one. After including overheads, our 20~min Snapshot visits were all comprised of two individual exposures that each have an integration time of $429$~s. 

\subsection{Survey Results}\label{ssec:survey}

This {\it HST} Snapshot survey ran through Cycle 24, with observations occurring randomly in late 2016 and throughout 2017. Since Snapshot programs stay live for two cycles, there is also one observation in May of 2018 (SN~2014ch). The observation dates for successful targets are listed in Table \ref{tab:obs}. Our program obtained 72 out of our 80 requested targets, for an overall completion rate of 90\%. This is a very high completion rate for a Snapshot program, which typically achieve closer to 30\% completeness. 

\begin{table*} 
\begin{center} 
\caption{$HST$ WFC3 NUV Observations} 
\label{tab:obs} 
\begin{tiny} 
\begin{tabular}{lcccclc} 
\hline 
\hline 
Target & Observation & \multicolumn{2}{c}{Astrometric Uncertainty} & \multicolumn{2}{l}{Potential Sources} & NUV Limiting \\ 
Name & Date & Guide Star & Total & \# & Comments & Magnitude \\ 
 & [UT] & [\arcsec] & [\arcsec] & & & [AB mag] \\ 
\hline 
      ASASSN-14co &   2017-09-01 &  0.0424 & 1.13 &          0 & confirmed no sources & $25.94\pm0.08$ \\ 
       ASASSN-14dc &   2017-09-29 &  0.0361 & 1.11 &          1 & all sources are cosmic rays & $25.97\pm0.08$ \\ 
       ASASSN-14eu &   2017-04-26 &  0.0849 & 1.25 &          1 & all sources are cosmic rays & $25.8\pm0.07$ \\ 
       ASASSN-14ew &   2017-04-03 &  0.5381 & 2.61 &          2 & all sources are cosmic rays & $25.81\pm0.05$ \\ 
       ASASSN-14lo &   2017-02-25 &  0.4187 & 2.26 &          3 & all sources are cosmic rays & $25.82\pm0.07$ \\ 
       ASASSN-14lq &   2017-05-01 &    0.05 & 1.15 &          0 & confirmed no sources & $25.81\pm0.08$ \\ 
       ASASSN-14lw &   2017-01-27 &   0.488 & 2.46 &          1 & all sources are cosmic rays & $25.82\pm0.05$ \\ 
       ASASSN-15de &   2016-12-05 &  0.3754 & 2.13 &          4 & image needs further review & $25.79\pm0.07$ \\ 
       ASASSN-15hy &   2017-04-05 &  0.4742 & 2.42 &          2 & all sources are cosmic rays & $25.8\pm0.07$ \\ 
       ASASSN-15jo &   2017-03-13 &  0.0224 & 1.07 &          1 & all sources are cosmic rays & $25.77\pm0.08$ \\ 
       ASASSN-15nr &   2017-01-31 &  0.4384 & 2.32 &          0 & confirmed no sources & $25.77\pm0.05$ \\ 
       ASASSN-15og &   2016-11-20 &   0.396 & 2.19 &          5 & image needs further review & $25.91\pm0.08$ \\ 
       ASASSN-15sh &   2017-03-24 &  0.1131 & 1.34 &          0 & confirmed no sources & $25.75\pm0.07$ \\ 
       ASASSN-15ut &   2016-11-29 &   0.446 & 2.34 &          3 & all sources are cosmic rays & $25.81\pm0.05$ \\ 
          LSQ14fmg &         none &         &      &            &            &            \\ 
          LSQ15aae &   2017-04-22 &  0.6512 & 2.95 &          6 & all sources are cosmic rays & $25.8\pm0.07$ \\ 
          LSQ15adm &   2017-03-22 &  0.5532 & 2.66 &          1 & all sources are cosmic rays & $25.85\pm0.06$ \\ 
          LSQ15bxe &   2017-03-04 &  0.0283 & 1.08 &          0 & confirmed no sources & $25.79\pm0.06$ \\ 
      MasterOT0442 &   2017-03-04 &  0.1487 & 1.45 &          3 & all sources are cosmic rays & $25.8\pm0.07$ \\ 
  OGLE-2014-SN-107 &   2016-11-20 &  0.0424 & 1.13 &          0 & confirmed no sources & $25.84\pm0.07$ \\ 
  OGLE-2014-SN-141 &   2016-11-16 &   0.389 & 2.17 &          2 & all sources are cosmic rays & $25.82\pm0.07$ \\ 
         PS1-13dsg &   2017-09-20 &  0.3895 & 2.17 &          2 & all sources are cosmic rays & $25.94\pm0.07$ \\ 
          PS1-14oo &   2017-03-01 &  0.1131 & 1.34 &          0 & confirmed no sources & $25.86\pm0.08$ \\ 
           PS15cwx &   2017-08-01 &  0.4486 & 2.35 &          1 & all sources are cosmic rays & $25.94\pm0.06$ \\ 
            PS15sv &   2017-09-28 &  0.0424 & 1.13 &          2 & image needs further review & $\sim25.8$ \\ 
         PSNJ02+42 &   2017-01-09 &  0.0424 & 1.13 &          0 & confirmed no sources & $25.77\pm0.08$ \\ 
         PSNJ08+48 &   2017-03-05 &  0.0283 & 1.08 &          0 & confirmed no sources & $25.79\pm0.07$ \\ 
         PSNJ23-15 &         none &         &      &            &            &            \\ 
           PTF11kx &   2016-12-01 &  0.5025 & 2.51 &          3 & all sources are cosmic rays & $25.84\pm0.07$ \\ 
         iPTF13asv &   2017-07-05 &   0.389 & 2.17 &          4 & all sources are cosmic rays & $25.98\pm0.06$ \\ 
         iPTF13daw &         none &         &      &            &            &            \\ 
         iPTF13dud &         none &         &      &            &            &            \\ 
         iPTF13ebh &   2017-09-08 &  0.4528 & 2.36 &          2 & all sources are cosmic rays & $25.98\pm0.07$ \\ 
           iPTF13s &   2017-05-27 &  0.5587 & 2.68 &          2 & all sources are cosmic rays & $25.98\pm0.06$ \\ 
         iPTF14abk &         none &         &      &            &            &            \\ 
         iPTF14aqs &   2017-05-22 &  0.3394 & 2.02 &          3 & all sources are cosmic rays & $25.94\pm0.06$ \\ 
         iPTF14atg &   2017-06-07 &  0.1345 &  1.4 &          0 & confirmed no sources & $26.0\pm0.07$ \\ 
         iPTF14bdn &   2017-04-05 &  0.5587 & 2.68 &          2 & all sources are cosmic rays & $25.83\pm0.05$ \\ 
         iPTF14fpg &   2017-09-09 &  0.4597 & 2.38 &          0 & confirmed no sources & $25.94\pm0.06$ \\ 
         iPTF14fyq &   2016-12-26 &  0.0424 & 1.13 &          0 & confirmed no sources & $25.87\pm0.08$ \\ 
         iPTF14gmo &   2017-03-04 &  0.1703 & 1.51 &          1 & all sources are cosmic rays & $25.8\pm0.07$ \\ 
         iPTF14gnl &   2017-09-10 &  0.4243 & 2.27 &          2 & all sources are cosmic rays & $25.95\pm0.07$ \\ 
         iPTF14ibo &   2017-07-04 &  0.4314 & 2.29 &          0 & confirmed no sources & $25.97\pm0.06$ \\ 
          iPTF14sz &   2017-02-10 &  0.4904 & 2.47 &          1 & all sources are cosmic rays & $25.82\pm0.06$ \\ 
         iPTF15agv &   2017-02-16 &  0.3748 & 2.12 &          1 & all sources are cosmic rays & $25.83\pm0.07$ \\ 
         iPTF15akf &   2016-11-25 &  0.1131 & 1.34 &          0 & confirmed no sources & $25.8\pm0.08$ \\ 
         iPTF15clp &   2017-07-17 &  0.1204 & 1.36 &          2 & all sources are cosmic rays & $25.91\pm0.07$ \\ 
         iPTF15eod &   2016-12-01 &  0.5657 &  2.7 &          2 & all sources are cosmic rays & $25.82\pm0.08$ \\ 
          iPTF15go &   2017-03-04 &  0.1273 & 1.38 &          3 & all sources are cosmic rays & $25.79\pm0.07$ \\ 
          iPTF15wd &   2017-05-03 &  0.3466 & 2.04 &          1 & all sources are cosmic rays & $25.78\pm0.07$ \\ 
         iPTF16abc &   2016-12-22 &  0.4104 & 2.23 &          5 & all sources are cosmic rays & $25.76\pm0.06$ \\ 
          SN2012cg &   2017-03-05 &  0.0283 & 1.08 &          1 & all sources are cosmic rays & $25.77\pm0.08$ \\ 
          SN2012gl &   2017-02-28 &  0.4742 & 2.42 &          1 & all sources are cosmic rays & $25.82\pm0.06$ \\ 
           SN2013I &   2017-08-22 &  0.4314 & 2.29 &          1 & all sources are cosmic rays & $25.98\pm0.06$ \\ 
          SN2013bh &   2017-07-18 &  0.3466 & 2.04 &          4 & all sources are cosmic rays & $25.94\pm0.07$ \\ 
          SN2013dn &   2017-09-06 &  0.3538 & 2.06 &          5 & image needs further review & $25.91\pm0.07$ \\ 
          SN2013dy &   2017-01-26 &  0.0922 & 1.28 &          1 & all sources are cosmic rays & $25.79\pm0.07$ \\ 
          SN2013gh &   2017-09-10 &  0.4669 &  2.4 &          6 & all sources are cosmic rays & $25.97\pm0.07$ \\ 
          SN2013gv &   2017-01-18 &  0.0781 & 1.23 &          0 & confirmed no sources & $25.79\pm0.07$ \\ 
          SN2013hh &   2017-06-11 &  0.0602 & 1.18 &          2 & all sources are cosmic rays & $25.93\pm0.07$ \\ 
           SN2014E &         none &         &      &            &            &            \\ 
           SN2014I &   2016-12-10 &   0.488 & 2.46 &          2 & all sources are cosmic rays & $25.84\pm0.06$ \\ 
           SN2014J &   2016-12-15 &  0.3821 & 2.15 &          1 & all sources are cosmic rays & $25.77\pm0.06$ \\ 
           SN2014R &   2017-01-18 &  0.4601 & 2.38 &          0 & confirmed no sources & $25.73\pm0.07$ \\ 
          SN2014ab &   2017-03-10 &  0.0922 & 1.28 &          0 & image needs further review & $\sim25.5$ \\ 
          SN2014ai &   2017-01-15 &  0.4031 & 2.21 &          0 & confirmed no sources & $25.82\pm0.06$ \\ 
          SN2014aj &   2016-12-10 &  0.3677 &  2.1 &          1 & all sources are cosmic rays & $25.8\pm0.07$ \\ 
          SN2014ap &         none &         &      &            &            &            \\ 
          SN2014aw &   2017-03-21 &  0.3818 & 2.15 &          3 & all sources are cosmic rays & $25.82\pm0.07$ \\ 
          SN2014bn &   2017-06-08 &  0.4245 & 2.27 &          0 & confirmed no sources & $25.96\pm0.05$ \\ 
          SN2014ch &   2018-05-02 &  0.4319 &  2.3 &          2 & image needs further review & $25.98\pm0.07$ \\ 
          SN2014dl &   2017-06-28 &  0.6224 & 2.87 &          5 & all sources are cosmic rays & $25.95\pm0.07$ \\ 
          SN2014dt &   2017-02-28 &  0.0943 & 1.28 &          0 & confirmed no sources & $25.71\pm0.07$ \\ 
          SN2014eg &         none &         &      &            &            &            \\ 
           SN2015F &   2016-11-16 &  0.1414 & 1.42 &          3 & all sources are cosmic rays & $25.81\pm0.07$ \\ 
          SN2015aw &   2016-12-16 &  0.0361 & 1.11 &          1 & all sources are cosmic rays & $25.81\pm0.09$ \\ 
          SN2015bd &   2017-01-22 &  0.1273 & 1.38 &          0 & image needs further review & $\sim25.5$ \\ 
          SN2015bp &   2017-04-23 &  0.1063 & 1.32 &          0 & confirmed no sources & $25.81\pm0.06$ \\ 
          SN2015bq &   2017-02-23 &   0.027 & 1.08 &          0 & confirmed no sources & $25.8\pm0.08$ \\ 
          SN2015cp &   2017-09-12 &  0.3748 & 2.12 &          3 & image needs further review & $25.97\pm0.06$ \\ 
 \hline 
\end{tabular} 
\end{tiny} 
\end{center} 
\end{table*}

\section{Analysis}\label{sec:ana}

Of the 72 images obtained by this survey, 64 were unambiguously devoid of a NUV point source at the site of the SN. In Section \ref{ssec:ana_method} we demonstrate our methodology for confirming that the majority of our observations were nondetections, and in Section \ref{ssec:limmags} we describe our technique for measuring the point-source limiting magnitudes of each of our observations. For the 8 images that were not unambiguously devoid of a NUV point source, we make case-by-case evaluations of all sources detected within the astrometric uncertainty radius. In Section \ref{ssec:ana_sndet} we present the two cases that led to confirmed detections of SN~2015cp and ASASSN-15og, and in Section \ref{ssec:ana_snnondet} we present the three cases that led to confirmed nondetections for ASASSN-15de, SN~2013dn, and SN~2014ch. In Section \ref{ssec:ana_snhostuv} we present the three targets (SNe~2014ab, 2015bd, and PS15sv) that turned out to have a higher amount of NUV host-galaxy emission than expected, and describe how we calculate a point-source limiting magnitude for those images in a way that accounts for the elevated host background. In all cases we use and report {\it HST} magnitudes in the AB system.

\subsection{Image-Processing Methodology}\label{ssec:ana_method}

We downloaded the processed data from the Mikulski Archive for Space Telescopes (MAST), and used the images that were combined with {\tt AstroDrizzle} and corrected for charge-transfer efficiency (the {\sc drc} files). Since only two exposures are obtained per visit the cosmic ray (CR) removal is not complete, but the CRs are still easily identified and are not a source of contamination for our survey. Our choice of relatively short exposure times and the F275W filter results in quite sparsely populated images. As described below, we find that the original SN coordinates from their discovery surveys, combined with the inherent astrometric solution of the {\tt AstroDrizzled} images, are sufficient to confirm that $64$ of our {\it HST} observations are nondetections. We find this to be the case even though these two components lead to astrometric uncertainty radii of $1$--$2$\arcsec, which is significantly larger than the astrometric uncertainties derived from (for example) coregistering the {\it HST} images to ground-based images that contain the SN. For the $8$ observations with potential detections, however, we {\it do} use images containing the SN to revise the astrometry of our {\it HST} images (described later, in Sections \ref{ssec:ana_sndet} to \ref{ssec:ana_snhostuv}). We do not need to refine the astrometry of our {\it HST} images unless there is a marginal detection.

To determine the absolute astrometric uncertainty radius for the location of the SN site in the {\it HST} images, we combined the uncertainties from the original ground-based imaging that discovered the SN with those for our WFC3 images. Ground-based imaging provides absolute astrometry of $\lesssim1\arcsec$ --- this is typical even for the ASAS-SN project which has large pixels \citep{2017MNRAS.467.1098H}. Usually the astrometric precision is much better, and so we use $1\arcsec$ as the $3\sigma$ upper limit on the absolute astrometric uncertainty from ground-based surveys' reported coordinates. 

The absolute astrometric uncertainty of the {\it HST} WFC3 images is dominated by the guide stars' astrometric errors, and the next largest contribution is from any failures during guiding or guiding difficulties that may cause the telescope to roll; other sources such as geometric distortion are a much smaller contribution \citep{WFC3_Data_Handbook}. We use the image headers to verify that fine-lock guiding was successful, and use the telescope jitter data to verify that the roll angle was stable during the observations (i.e., we use the {\sc jif} tables). We obtain information about the guide stars from the headers and query the Guide Star Catalog 2.3 through the Vizier\footnote{\url{http://vizier.u-strasbg.fr/}, \cite{2000A&AS..143...23O}} service to obtain their coordinate uncertainties. We find that for all images, the guide-star coordinates are the largest source of error in the {\it HST} absolute astrometry (i.e., hundreds instead of several to tens of milliarcsec), and so these errors represent the absolute astrometric precision at the 1$\sigma$ level\footnote{This interpretation of 1$\sigma$ was corroborated through conversations with the {\it HST} STScI help desk.}. For the final absolute astrometric uncertainty radius in which we search for NUV sources we add $1.0\arcsec$ to 3 times the guide-star error (i.e., 3$\sigma$), and use this as the combined ground- and space-based absolute astrometric uncertainty. This radius is listed in Table \ref{tab:obs} under ``Total Astrometric Uncertainty." 

We use Source Extractor \citep{1996A&AS..117..393B} to detect sources in our images, using the parameter values set as listed in Table \ref{tab:SEp}. We list the number of detected sources within the astrometric uncertainty radius for each of our SNe in Table \ref{tab:obs}. We then visually review all images, reduced and raw frames, to classify these sources. In almost all cases, detected sources can be easily identified as a cosmic ray residual (i.e., it appeared in only one of our two original frames, or had a flux profile that deviated obviously from the {\it HST} point-spread function [PSF]). The results of our visual review are reported in column 6 of Table \ref{tab:obs}. Cases which are even slightly ambiguous are treated individually in Sections \ref{ssec:ana_sndet} to \ref{ssec:ana_snhostuv}, and all other observations are considered confirmed nondetections.

\begin{table}
\begin{center}
\caption{Source Extractor Parameters}
\label{tab:SEp}
\begin{tabular}{lc}
\hline
\hline
Parameter & Value \\
\hline
DETECT\_MINAREA   & 5    \\  
DETECT\_THRESH    &  2    \\ 
ANALYSIS\_THRESH &  2    \\
BACK\_SIZE                & 64   \\
BACK\_FILTERSIZE   &  3     \\
\hline
\end{tabular}
\end{center}
\end{table}

\subsection{Limiting Magnitudes}\label{ssec:limmags}

In order to determine the limiting magnitude of our {\it HST} images, we inject simulated WFC3 point sources with a variety of fluxes, and measure the apparent magnitude at which we recover a point source $50\%$ of the time.

Our pre-selection of candidates in regions of low NUV surface brightness also means that for an estimate of the point-source limiting magnitudes, we can plant fake sources in the surrounding region and not only the {\it exact} pixel location of the SN, which is helpful to minimize the amount of computational processing. In order to plant enough fakes we do 1000 realizations, injecting 10 simulated point sources each time, with random locations within a $\sim4$\arcsec\ box around the SN location. We use fluxes that cover a range of values, from bright enough to always be detected, to faint enough to never be detected. We use the TinyTim PSF \citep{1993ASPC...52..536K} via the TinyTim web service\footnote{\url{http://tinytim.stsci.edu/cgi-bin/tinytimweb.cgi}}, and generate the correct PSF for the WFC3 UVIS detector at the chip location for the site of the SN to ensure that the PSF distortion is accounted for. We inject directly into the post-{\sc AstroDrizzled} {\sc drc} images. Although this will result in slightly less accurate limiting magnitudes than injecting into the {\sc flc} images, this effect is $\ll 0.1$ mag. Injecting fakes directly into the {\sc drc} images saves a significant amount of processing time and is completely sufficient for our science goals (which are achieved with a mean error in limiting magnitude of $0.07$ mag, as described below).

We then run Source Extractor on each image realization, using the same parameters as in Table \ref{tab:SEp}. We match the detected sources to the injected sources with a 2 pixel radius, and determine the fraction of injected sources detected as a function of magnitude in bins of $0.05$ mag (using the input, and not the recovered, magnitude in order to automatically incorporate the aperture correction). We then perform a spline fit to this histogram and calculate the apparent magnitude of a source with a 50\% probability of detection, and use that as our limiting magnitude. We convert Poisson errors for the number of objects per bin into an uncertainty on the limiting magnitude. The results are listed in column 7 of Table \ref{tab:obs}, in AB magnitudes. The limiting magnitudes of all of our images have a mean value of $25.85$ mag and a standard deviation of $0.08$ mag; the mean error in limiting magnitude is $0.07$ mag. These limits are used to constrain the frequency of late-onset CSM interaction and the physical characteristics of CSM in SN\,Ia progenitor systems in Section \ref{sec:disc}.

\begin{figure*}
\begin{center}
\includegraphics[width=8cm]{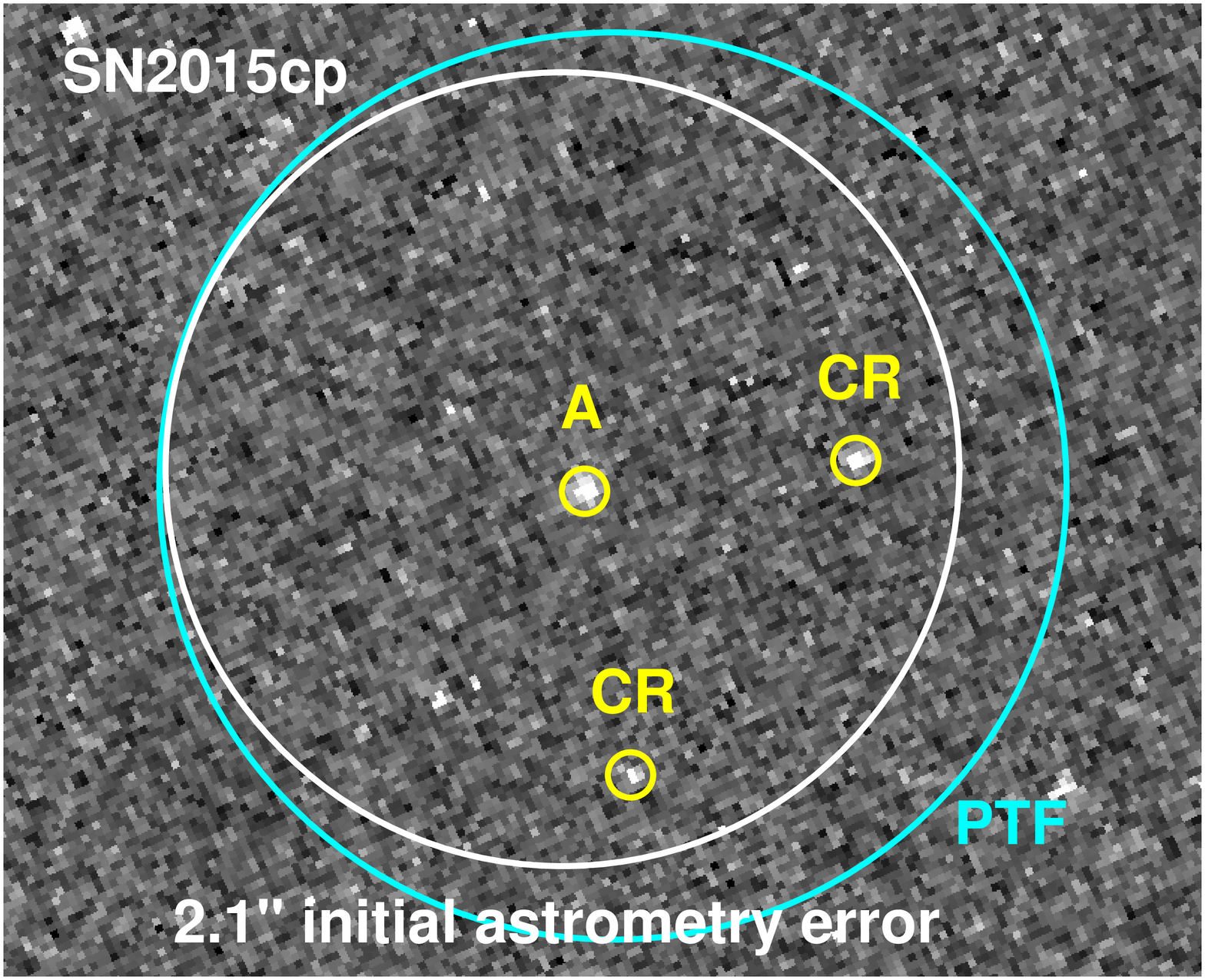}
\includegraphics[width=8cm]{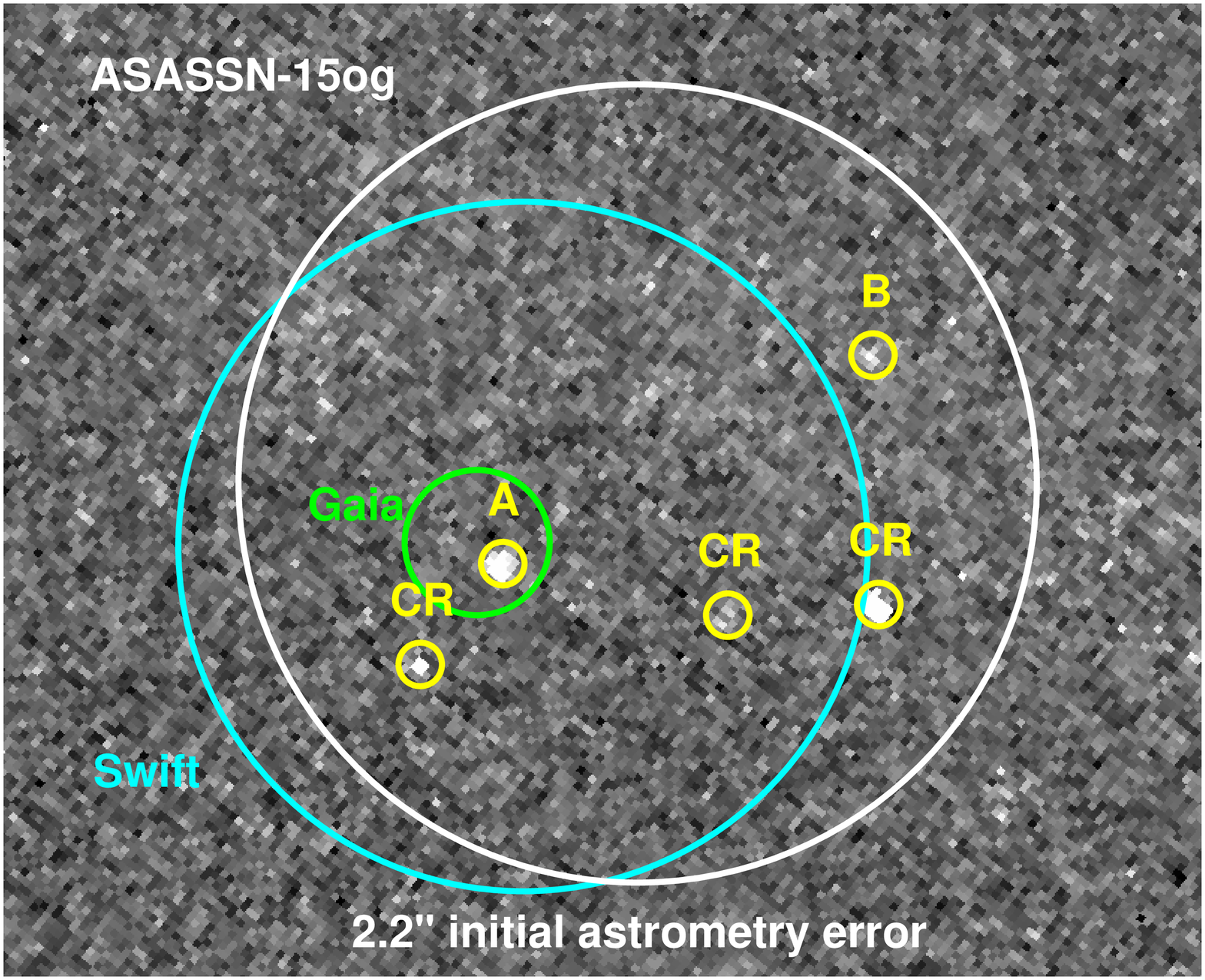}
\includegraphics[width=8cm]{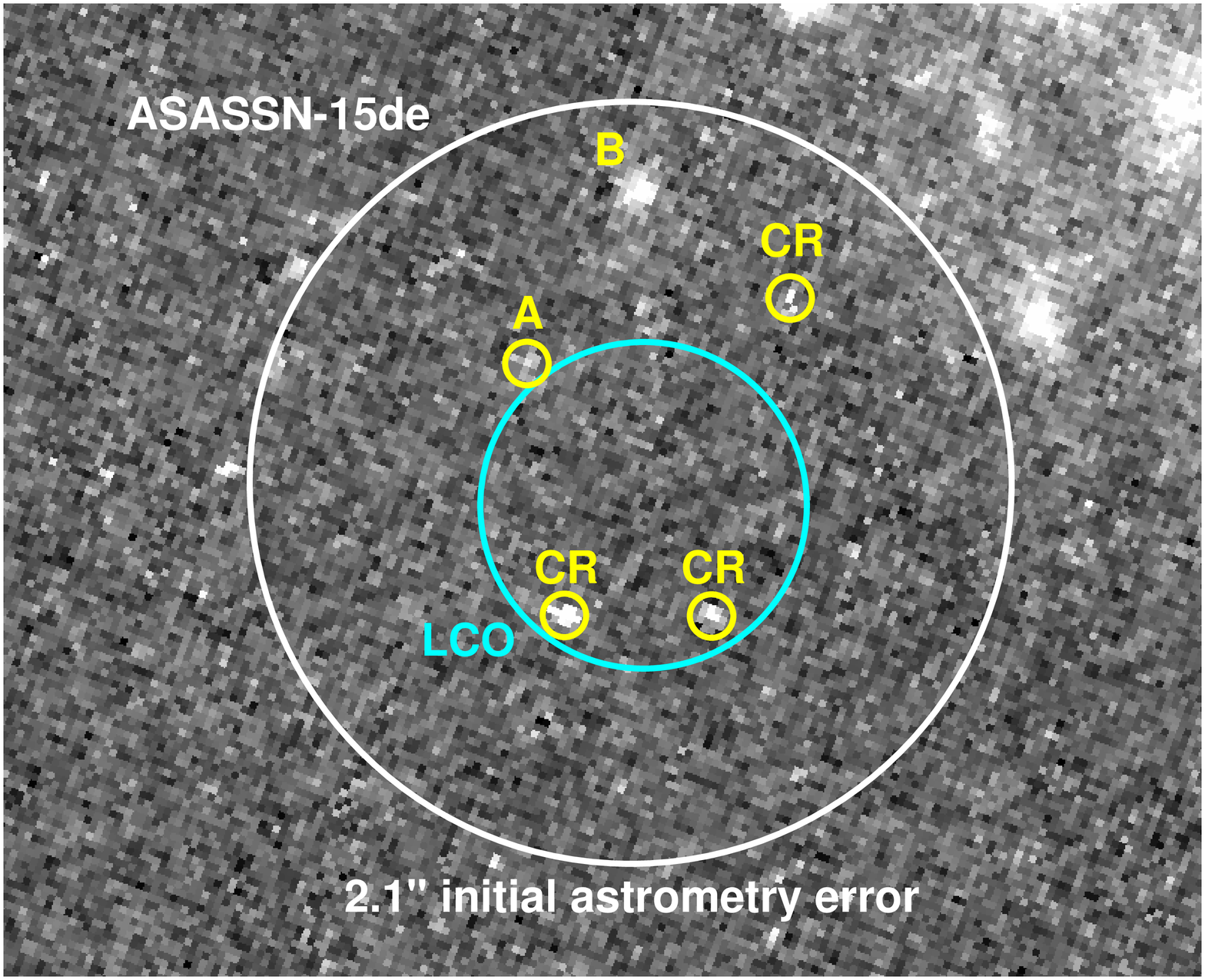}
\includegraphics[width=8cm]{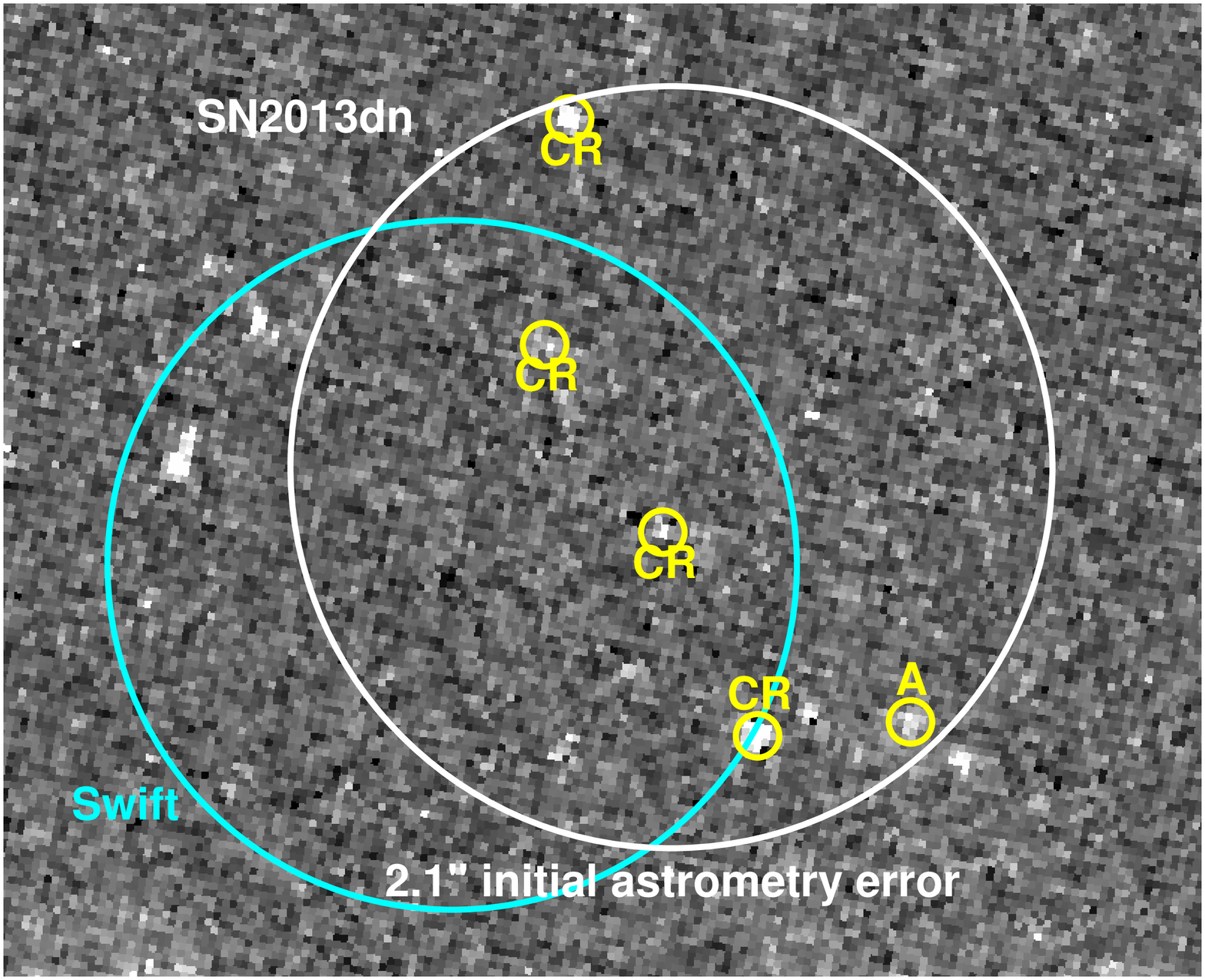}
\includegraphics[width=8cm]{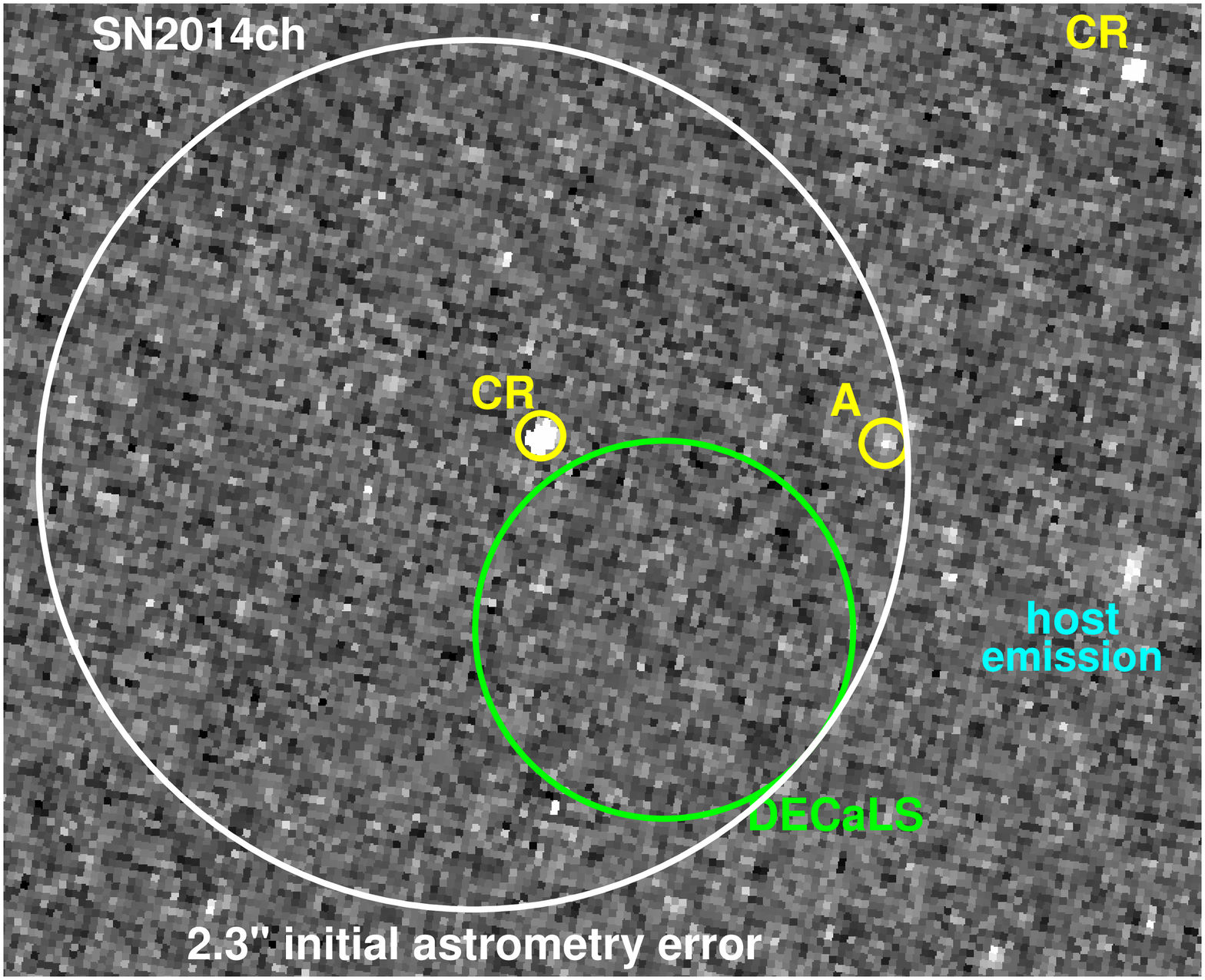}
\caption{Our {\it HST} NUV imaging for five SNe in which we detected one or more point sources within the initial astrometric uncertainty radius that was not associated with a cosmic ray residual: SN~2015cp, ASASSN-15og, ASASSN-15de, SN~2013dn, and SN~2014ch (top left to bottom). For these $5$ images, we refine our astrometry because there were detections within the initial astrometric error radius that could not be attributed to CR. In all panels, white circles mark the initial astrometric error radius, yellow circles mark the detected sources (labeled ``CR" if they have been associated with a cosmic ray, or A or B otherwise), and cyan or green circles represented revised absolute astrometry after comparing with images containing the SN (as labeled).}
\label{fig:nohostbg}
\end{center}
\end{figure*}

\subsection{Image Review: Confirmed Detections}\label{ssec:ana_sndet}

In this section we demonstrate our in-depth image review that confirmed the detections of SN\,Ia-CSM/SN\,IIn~ASASSN-15og and SN\,Ia~2015cp, and provide some analysis of their NUV point source. 

\subsubsection{ASASSN-15og}\label{sssec:review_asassn15og}

For ASASSN-15og we find that three of the five sources within the $2.2\arcsec$ astrometric uncertainty radius can be confirmed as cosmic ray residuals, as they appear in only the one of the two original {\sc flc} images. We label the other two sources as objects A and B in Figure \ref{fig:nohostbg}. As done in Section \ref{sssec:review_asassn15de}, the simplest first step is to try to improve the astrometric uncertainty, to see if an association with the SN's location can be ruled out for either or both of them.

Imaging of ASASSN-15og with the Neil Gehrels {\it Swift} Observatory is publicly available\footnote{\url{https://archive.stsci.edu/swiftuvot/file\_formats.html}}. 
We find that {\it Swift} images from the Ultraviolet and Optical Telescope (UVOT) in filters $UBV$ obtained on 2015 Aug. 16 (UT dates are used throughout) contain both the SN and a nearby star ($\sim55\arcsec$ W of the SN) which is also visible in our {\it HST} F275W image. 
The {\it Swift} $V$-band image has the smallest astrometric uncertainty.
We use Source Extractor to obtain the centroids of both the SN and the star in the {\it Swift} image, and the centroids of the star in our {\it HST} image, and use this information to revise the expected location of ASASSN-15og in our {\it HST} image\footnote{Note that the geometric distortion across the WFC3 UVIS chip is low, only 2\% at the very edges, and that this is corrected in the final drizzled {\sc drc} images \citep{MultiDrizzle_Handbook}.}. 
The revised absolute astrometry error radius is $\sqrt{ \sigma_{\alpha}^2 + \sigma_{\delta}^2 }$, where $\sigma$ is the standard deviation in the barycenter of ASASSN-15og as measured by Source Extractor in the {\it Swift} image (i.e., the square root of the variance or the second-order moment, {\tt X2\_WORLD}).
The new expected location of ASASSN-15og is shown in Figure \ref{fig:nohostbg}.
Although the {\it Swift} image did not decrease the radius of the astrometric error circle, it is now centered on object A and rules out an association with object B. 

ASASSN-15og was also observed by {\it Gaia} (as Gaia16ajy) and reported as a {\it Gaia} Alert at J2000 coordinates $\alpha = 03^h21^m07.49^s$, $\delta = -31^\circ 18' 45.''94$\footnote{\url{http://gsaweb.ast.cam.ac.uk/alerts/alert/Gaia16ajy/}} on 2015 Oct. 29, about 2.5 months after the initial discovery \citep{2015ATel.7912....1B}. We obtained the {\it Gaia} coordinates for the nearby star from the {\it Gaia} Data Release 1 source catalog\footnote{\url{https://gea.esac.esa.int/archive/}}, and used the {\it Gaia} coordinates for the star and ASASSN-15og to revise the expected SN location in our {\it HST} image, as shown in Figure \ref{fig:nohostbg}. The absolute astrometric uncertainty of the {\it Gaia} DR1 objects is extremely low ($<0.001\arcsec$), and the astrometric uncertainty on {\it Gaia} Alerts has been quoted as similarly low ($\sim0.05\arcsec$)\footnote{Private communication with the {\it Gaia} Alerts team.}, but we checked the absolute astrometric offset between these two catalogs ourselves. We searched the {\it Gaia} Alerts archive for objects in the vicinity of ASASSN-15og (using the maximum search radius of $20^\circ$), and found seven that appeared to be point-like nontransient sources, such as variable stars and quasars, that were also listed in the {\it Gaia} DR1 source catalog. We find that there is always a small offset in the coordinates for the same object in the Alerts and DR1 catalogs, and that this offset appears to be random and not systematic (i.e., is not always in the same direction). We use the standard deviation of these offsets to calculate that the $1\sigma$ error radius in the expected position of Gaia16ajy is $\sigma_{\rm Gaia} = 0.14\arcsec$. In Figure \ref{fig:nohostbg} the radius of the {\it Gaia} circle is $3\sigma_{\rm Gaia}$, and we can see that object A is consistent with the expected location of ASASSN-15og. 

We now endeavor to confirm that object A is an astrophysical point source and not, for example, the chance alignment of two cosmic rays. With SN~2015cp we obtained follow-up observations that confirmed the NUV source as a true detection, and such an analysis as presented below for ASASSN-15og was not necessary. As discussed in the final paragraph of this section, since ASASSN-15og had already exhibited signatures of CSM interaction, it was not prioritized for follow-up observations.

\begin{figure}
\begin{center}
\includegraphics[width=8.2cm]{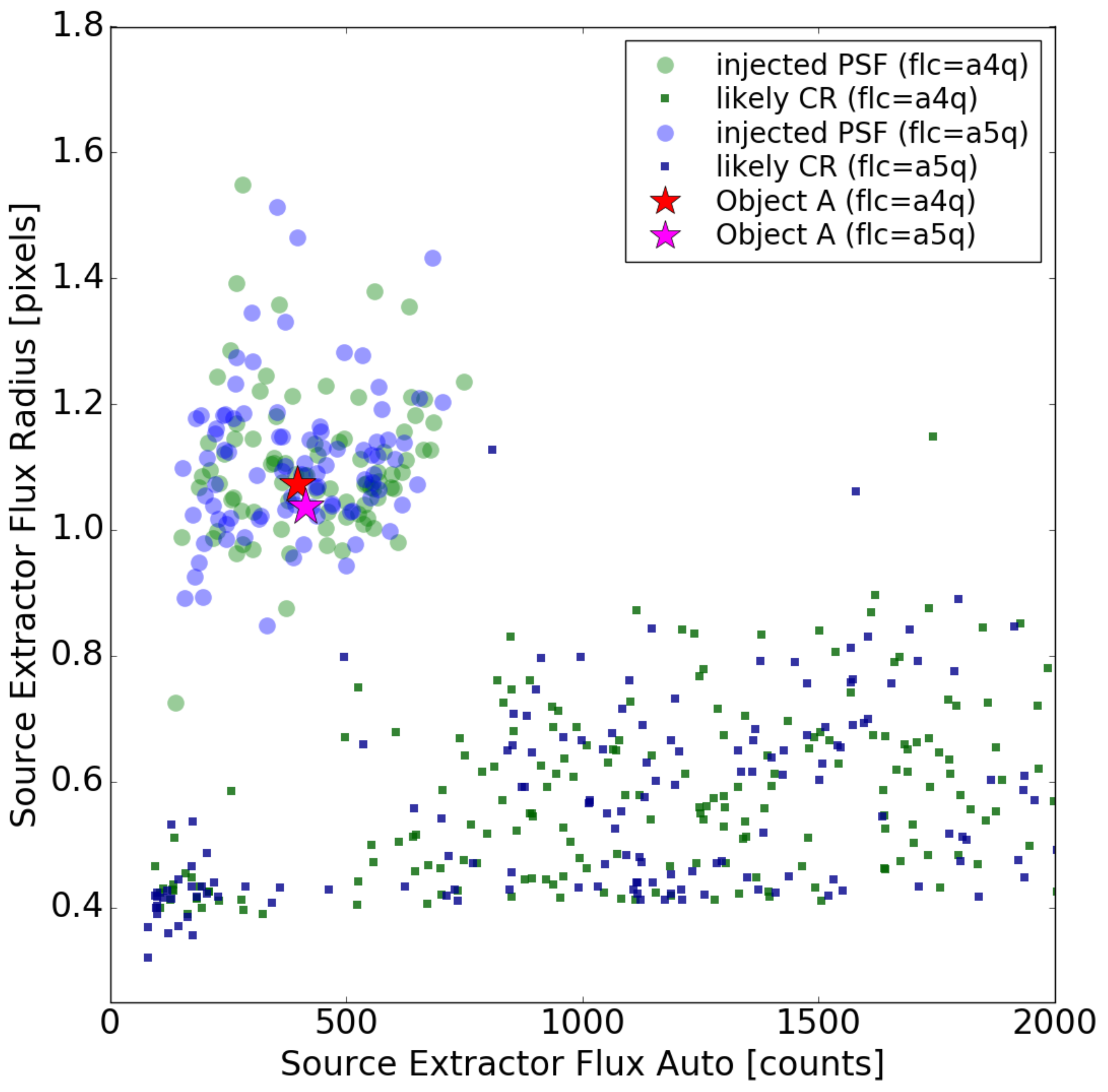}
\caption{An analysis to confirm that the source detected at the location of ASASSN-15og is point-like (Section \ref{sssec:review_asassn15og}; Figure \ref{fig:nohostbg}). The flux radius in pixels vs. the total source flux, both as measured by Source Extractor, for injected PSFs (circles), and for sources in the image (squares; i.e., mostly cosmic rays), with object A highlighted as a 5-point star. Blue and red points are from one {\sc flc} image and green and magenta points are for the other. Clearly, object A is much more like a PSF than a cosmic ray, and has equivalent flux in both {\sc flc} images.  \label{fig:asassn15og_objA}}
\end{center}
\end{figure}

First we checked the data-quality frames and determined that object A is not flagged as a possible cosmic ray. Then we injected a population of simulated point sources, again using the TinyTim PSF for WFC3 as in Section \ref{ssec:limmags}, with total fluxes similar to that of object A into the two {\sc flc} images. We used the mask frame to ensure that we only placed these simulated sources in clean pixels, and then we ran Source Extractor with the same settings as in Table \ref{tab:SEp}. In Figure \ref{fig:asassn15og_objA} we plot the measured flux radius versus the measured source flux for all objects detected in either {\sc flc} image, using circles to represent injected PSFs, five-point stars for object A, and squares for all the rest, which are most likely to be cosmic rays. Figure \ref{fig:asassn15og_objA} demonstrates that object A has a flux radius similar to a TinyTim PSF and unlike a cosmic ray. It also shows that the flux of object A is very similar in both {\sc flc} images, and lies in a flux range that is relatively unoccupied by noninjected sources in the image. We made similar comparisons with other Source Extractor output parameters such as the full width at half-maximum intensity (FWHM), aperture flux, and peak surface brightness (not shown), with similar results: it appears that object A is a real source for which we cannot rule out an association with the location of ASASSN-15og. Unfortunately, there is no previous {\it HST} imaging at these coordinates, and so we cannot say whether it is a pre-existing source unrelated to ASASSN-15og. 

To obtain accurate photometry for object A, we use the injected point sources to derive a correction between the true and measured flux for this image, similar to (but technically not) an aperture correction. For this we use Source Extractor's {\sc FLUX\_AUTO} because it demonstrates the tightest correlation with input flux (better than the aperture fluxes we tested, for 3, 5, and 10 pixel radii). We use the image zeropoint from the header keywords, and find that the observed apparent magnitude of object A is $m_{\rm F275W} = 23.94 \pm 0.08$ mag. The Galactic extinction for the coordinates of ASASSN-15og is $A_{\rm F275W} = 0.061 \pm 0.002$ mag \citep{2011ApJ...737..103S}, and so the extinction-corrected apparent magnitude is $m_{\rm F275W} = 23.88 \pm 0.08$ mag.

We also observed the location of ASASSN-15og with the Low Resolution Imaging Spectrometer at Keck Observatory on 2017 January 3, just under two months after the {\it HST} NUV image was obtained. Our $B$ and $R$ filter images do not show any point source at the site of the SN, but the location has a higher amount of host-galaxy surface brightness in the optical filters than in the NUV. We estimate that a point source of $g\approx20$ mag could have been detected at a signal-to-noise ratio of $\sim 1$, given its location in the host galaxy, but this does not put any further constraints on the nature or brightness evolution of our {\it HST} detection. 

This late-time detection of ASASSN-15og is not surprising: its peak brightness was overluminous \citep[Gaia Alerts;][]{2017ApJ...835...64G} and its classification optical spectrum exhibited a blue continuum and Balmer emission --- all indicators of CSM interaction. ASASSN-15og was initially classified as a SN\,IIn/SN\,Ia-CSM at $\sim 20$~d after peak brightness \citep{2015ATel.7932....1M}, and a reanalysis of its early-time data by \citealt{2017MNRAS.467.1098H} reclassified this event as a SN\,Ia-CSM (removing the SN\,IIn designation). We included a few such objects with early-onset CSM interaction in our target list as proof-of-concept because they have much higher probability of yielding a late-time NUV detection. Since the main science goal of this project was to find and study SNe\,Ia that appeared normal at the time of discovery and up until regular monitoring ended, but then exhibited interaction only at late times, we did not trigger any further follow-up observations of ASASSN-15og. 

\subsubsection{SN~2015cp}\label{sssec:review_sn2015cp}

For SN~2015cp, only two of the three sources detected within the error radius are confirmed as cosmic ray residuals; the third we label as object A in Figure \ref{fig:nohostbg}. As in the case of ASASSN-15og, we first attempt to refine the astrometry in order to rule out or confirm an association between object A and SN~2015cp. We find that optical images from the iPTF survey that contain the SN provide an astrometric error radius that is approximately equivalent to our initial estimate for the {\it HST} astrometry, as shown by the white ({\it HST}) and cyan ({\it PFT}) circles in Figure \ref{fig:nohostbg}. Instead, we applied the same techniques as described in Section \ref{sssec:review_asassn15og} and Figure \ref{fig:asassn15og_objA} to first verify that object A is a point source, and then to measure an apparent magnitude for object A of $m_{\rm F275W} = 24.28 \pm 0.09$ mag). The Galactic extinction for the coordinates of SN~2015cp is $A_{\rm F275W} = 1.28 \pm 0.01$ mag \citep{2011ApJ...737..103S}, so the extinction-corrected apparent magnitude is $m_{\rm F275W} = 23.0 \pm 0.10$ mag.

For SN~2015cp we obtained ground-based photometric and spectroscopic follow-up data that revealed an optical source dominated by H$\alpha$ emission had appeared at the location of SN~2015cp, confirming this NUV source as a true detection. These observations and a physical interpretation of the source of late-time emission in the SN~2015cp system are provided in Section \ref{sec:15cp}.

\subsection{Image Review: Confirmed Nondetections}\label{ssec:ana_snnondet}

In this section we demonstrate our in-depth image review that confirmed the nondetections of ASASSN-15de, SN~2013dn, and SN~2014ch. The point-source limiting magnitudes for these images were obtained with the method discussed in Section \ref{ssec:limmags}.

\subsubsection{ASASSN-15de}\label{sssec:review_asassn15de}

For ASASSN-15de we detect 4 sources at the 2$\sigma$ level within the astrometric error radius of 2.1\arcsec\ using Source Extractor parameters as listed in Table \ref{tab:obs}. In Figure \ref{fig:nohostbg} we show the {\it HST} NUV {\sc drc} image, in which we identify these detected sources within the astrometric error radius. By examining the individual {\sc flc} images we find that 3 of them can be confirmed as residuals of cosmic rays but that one faint source could potentially be real (labeled CR and object A in Figure \ref{fig:nohostbg}, respectively). As a side note, the brighter source at the northern end of the error circle (labeled object B) is spatially extended and not a candidate detection of CSM interaction. 

The simplest thing that we can do to resolve the issue of whether object A might be associated with the SN is to improve our absolute astrometric uncertainty. To do this we use a $200$~s $g$-band image of ASASSN-15de obtained on 2015 Mar. 2 with the $1.0$~m telescope at Las Cumbres Observatory (LCO; \citealt{2013PASP..125.1031B}). In this image, the seeing FWHM is $1.5\arcsec$ and the SN has $g\approx17.9$ mag. There are three bright stars that appear in both the LCO and and {\it HST} images. We run Source Extractor to obtain the barycenter in world coordinates and its first-order moments (i.e., the variance, of which we take the square root and use as the positional uncertainty) for these three stars and the SN in the LCO image, and for these three stars in the {\it HST} image. We use this information to revise the expected SN location and astrometric error radius to be $0.9\arcsec$ in the {\it HST} image (the square root of the sum of the squares of the uncertainty in the SN's and the star's centroids), as shown with the labeled circle in Figure \ref{fig:nohostbg}. This improvement in absolute astrometry allows us to confirm that object A is in fact outside of the error circle, and that we have a nondetection for ASASSN-15de.

\subsubsection{SN~2013dn}

For SN~2013dn we find that four of the five sources within the $2.1\arcsec$ initial astrometric uncertainty radius can be confirmed as cosmic ray residuals, as they appear in only one of the two original {\sc flc} images. The fifth source appears to be extended (labeled object A in Figure \ref{fig:nohostbg}), but to rule it out we refine the expected location of SN~2013dn in our {\it HST} image. We find that SN~2013dn and a nearby star are detected in both a {\it Swift} $U$-band image from 2013 July 16 and our NUV {\it HST} image. As with ASASSN-15og, we use the star's location and uncertainty in the {\it Swift} image to derive a revised error circle for the location of SN~2013dn in our {\it HST} image, as shown in Figure \ref{fig:nohostbg}. A very similar new location is derived also from {\it Swift} $B$-band images acquired at the same time. We can see that the {\it Swift} astrometry removes object A from consideration but includes two new sources in the error radius --- however, we confirm they are both cosmic ray residuals, and that our late-time NUV imaging of SN~2013dn is a nondetection.

\subsubsection{SN~2014ch}

For SN~2014ch we find that one of the two sources within the $2.3\arcsec$ initial astrometric uncertainty radius is a cosmic ray, appearing in only one of the two {\sc flc} images. The second, labeled object A in Figure \ref{fig:nohostbg}, appears to be a real detection. To ensure that object A is of interest, we searched for publicly available images of SN~2014ch and found that the Dark Energy Camera's Legacy Survey (DECaLS) covered this field in the $z$ band on 2014 Aug. 12 (data release 5, brick {\tt 2395p127}). This was about $2.5$ months after the SN was discovered \citep{2014CBET.3946....1D}, and SN~2014ch is clearly visible in the image. As in previous cases, we use stars in both images to derive an astrometrically improved prediction for the location of SN~2014ch in our {\it HST} image. We find that this new radius excludes object A, as shown in Figure \ref{fig:nohostbg}. Furthermore, it appears likely that object A is simply a NUV-bright knot in the host galaxy, which has extended emission in the region. We therefore conclude that our late-time imaging of SN~2014ch is a nondetection.

\begin{figure}
\begin{center}
\includegraphics[width=8.2cm]{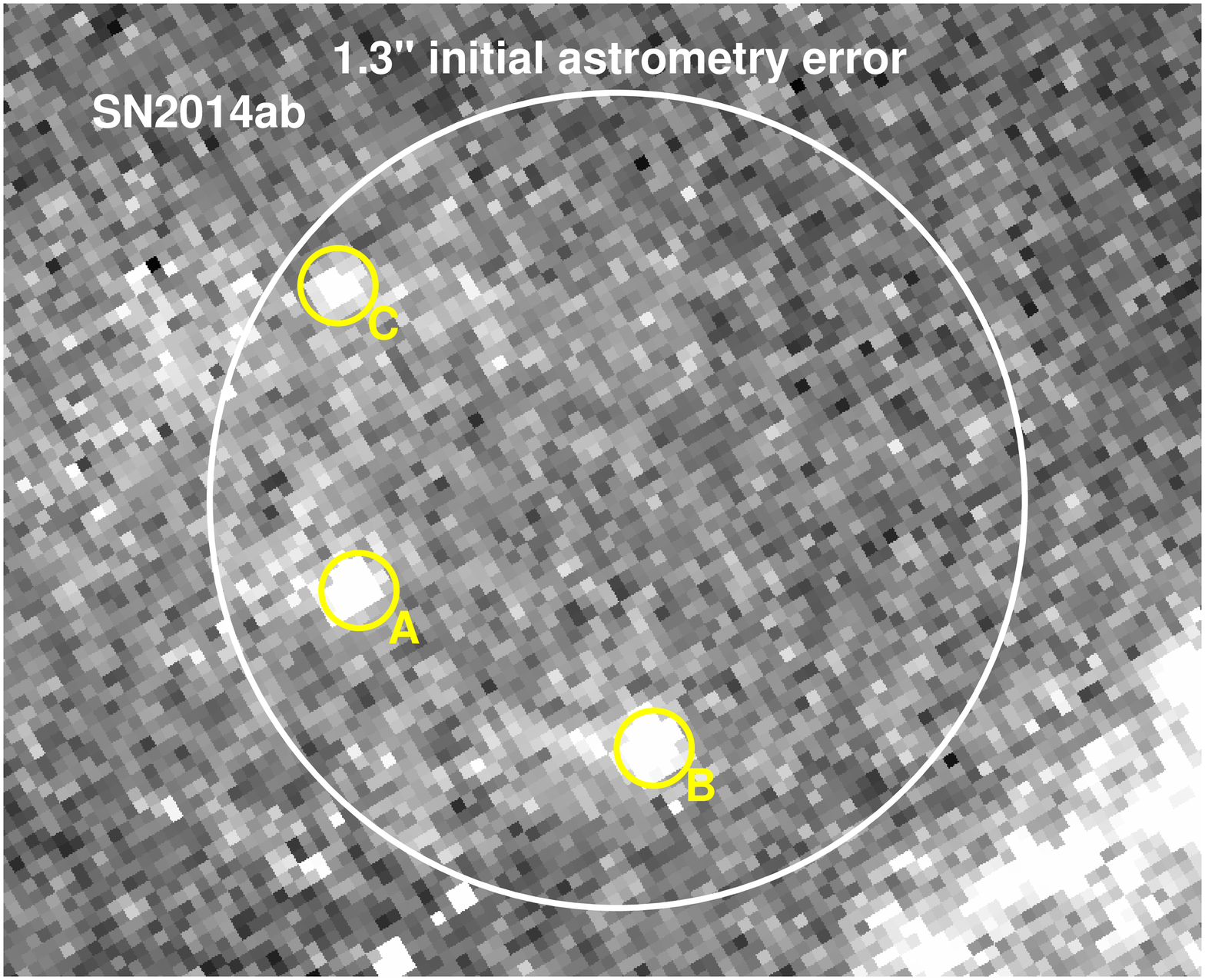}
\includegraphics[width=8.2cm]{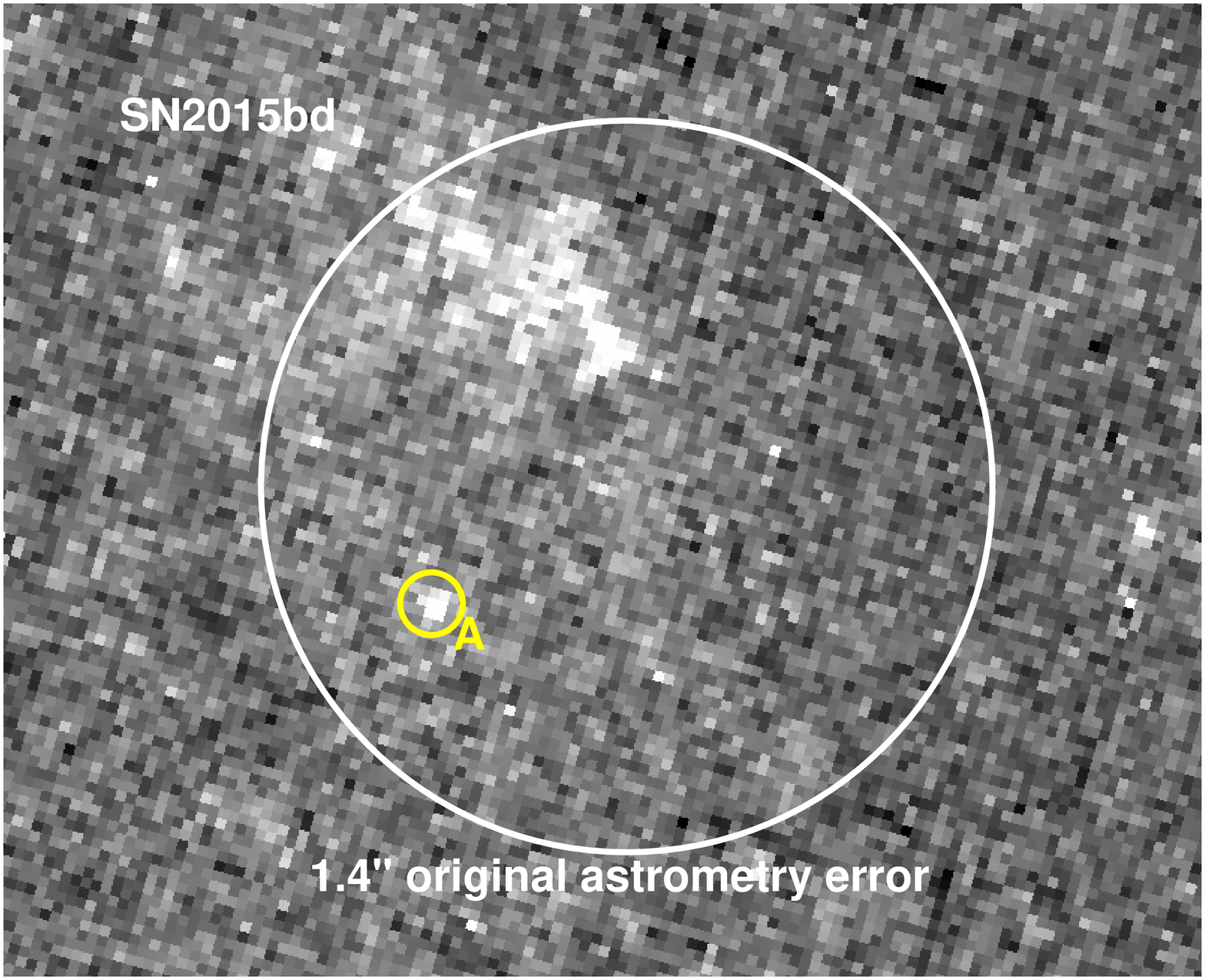}
\includegraphics[width=8.2cm]{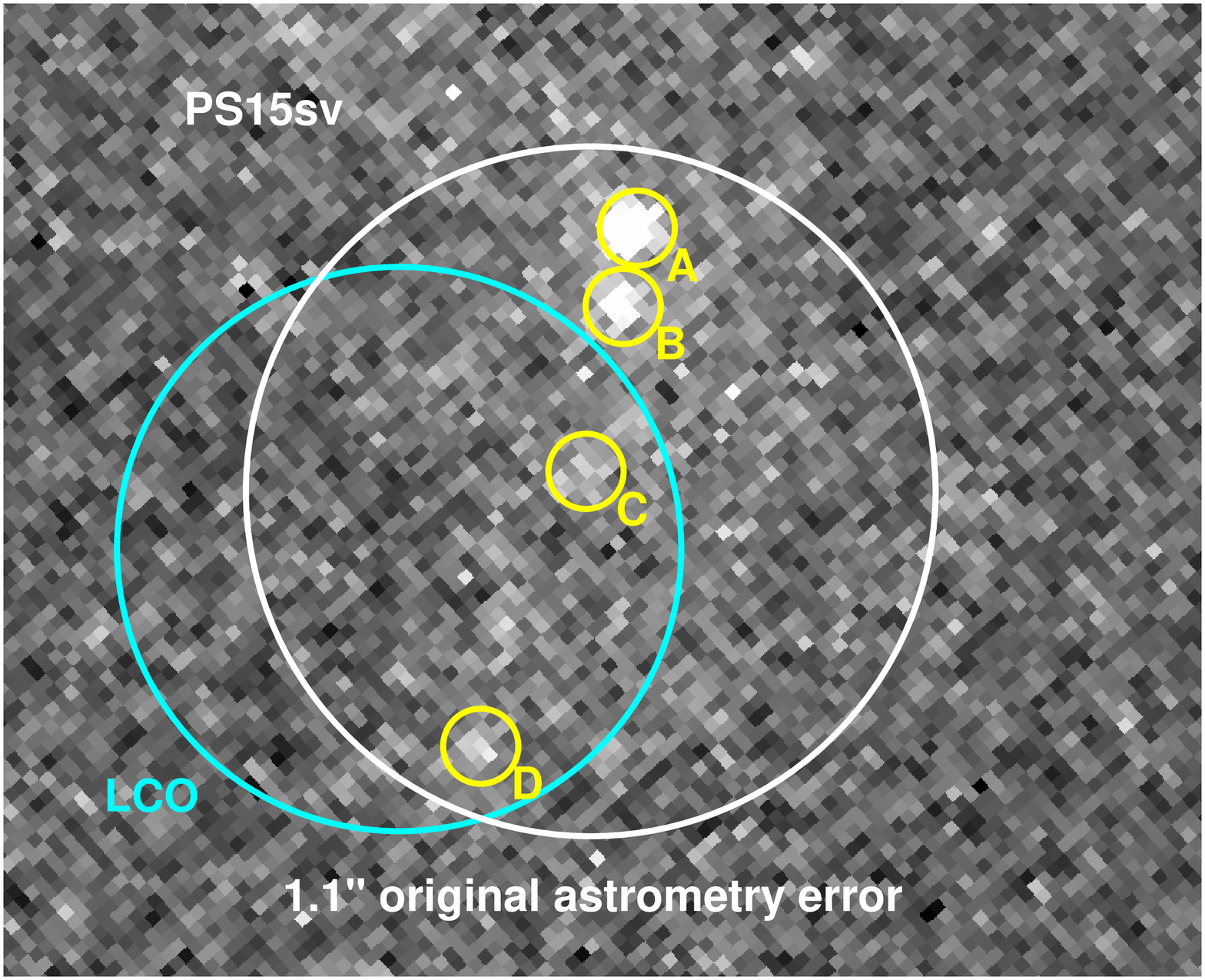}
\caption{Our {\it HST} NUV images of the fields of SN~2014ab, SN~2015bd, and PS15sv (top to bottom), all of which have relatively bright host-background emission compared to our other targets. In all images the white circle marks the initial astrometry error radius, and yellow marks detected sources. At right, for PS15sv we use images containing the SN to revise the expected location and its astrometric uncertainty (cyan circle).  \label{fig:hostbg}}
\end{center}
\end{figure}

\subsection{Image Review: Host NUV Contamination}\label{ssec:ana_snhostuv}

The NUV emission from the host galaxies of SN\,IIn 2014ab and 91T-like SNe\,Ia 2015bd and PS15sv is brighter than anticipated, as shown in Figure \ref{fig:hostbg}. For both SNe~2014ab and 2015bd, no images from the time of the SN are publicly available, so we cannot further revise the expected location or decrease the absolute astrometric error radius. SN\,Ia PS15sv was imaged by the Las Cumbres Observatory, and we use five of the best-seeing images from April 2015 in order to refine the expected location of the SN in our {\it HST} image. 

To detect sources in these images, we use mostly the same Source Extractor parameters as in Table \ref{tab:SEp}, except for more aggressive deblending parameters and a smaller background mesh size to compensate for the relatively higher NUV background. All of the detected sources that are within the astrometric error circles for these three SNe have been confirmed to be unassociated with cosmic rays. However, they all also have an extended profile (FWHM $\gtrsim 3$ pixels), indicating that they are not point sources, but associated with the underlying host-galaxy emission, and therefore not suitable candidate detections of late-time interaction. We conclude that our {\it HST} NUV imaging for these three SNe has yielded nondetections.

To estimate the limiting magnitudes of these three images, we use the same methods of injecting and recovering fake implanted PSFs as described in Section \ref{ssec:limmags}, but with tweaked Source Extractor parameters. For these images we only consider an injected source ``recovered" if it is detected as a point source (i.e., with a $1.0 \leq {\rm FWHM} \leq 2.5$), which was not a necessary condition for the other images' limiting magnitude derivations (Section \ref{ssec:limmags}). The 50\% detection limit for our SN~2014ab and SN~2015bd images is $m_{\rm F275W} = 25.5$ mag, and for PS15dv it is $m_{\rm F275W} = 25.8$ mag (reported in Table \ref{tab:obs}).

\section{CSM Interaction in SN\,Ia 2015cp}\label{sec:15cp}

We detected a NUV point source consistent with the site of SN~2015cp in our {\it HST} image obtained on 2017 Sep. 12 ($664$ days after peak brightness, $686$ days after explosion; observer frame), with an apparent magnitude of $m_{\rm F275W} = 23.0 \pm 0.1$ mag (corrected for Milky Way extinction; Section \ref{sssec:review_sn2015cp} and Figure \ref{fig:nohostbg}). 

Based on the observed NUV luminosity, $L_{\rm UV} = 7.6\times10^{25}$ $\rm erg\ s^{-1}\ Hz^{-1}$, we can immediately rule out several potential causes. First, it is too bright to be generated by interaction with low-density interstellar material (ISM), $\rho_{\rm ISM}\approx2\times10^{-24}$ $\rm g\ cm^{-3}$ (i.e., $\sim1$ hydrogen atom per $\rm cm^3$). As we discuss in Section \ref{ssec:disc_phys}, SN ejecta interactions are most luminous when $\rho_{\rm ej} \approx \rho_{\rm CSM}$ and $\rho_{\rm CSM}$ is typically $\sim10^{-17}$ $\rm g\ cm^{-3}$, and our physical models show that a significant amount of mass is required to reproduce this NUV luminosity. Second, it is too luminous to be a brightened post-impact companion star. For example, \citealt{2013ApJ...773...49P} and \citealt{2013ApJ...765..150S} show that a post-impact remnant star (PIRS) could brighten up to $10^3$--$10^4$ $\rm L_{\odot}$ in the $10$--$1000$ yr after the SN\,Ia explosion (depending on the type of companion), which corresponds to an absolute magnitude of $V\approx -4$ mag (Fig. 10 of \citealt{2013ApJ...773...49P}). At the distance of SN~2015cp this would be an apparent magnitude of $V\approx 32$ mag, well below our detection thresholds (even after accounting for the fact that the PIRS's $T\approx 10^5$ $\rm K$ blackbody emission would peak in the NUV). Third, it is too bright to be caused by the reflection of near-peak SN emission (a light echo): near-peak SNe\,Ia do not exhibit much NUV flux and it would be very faint (e.g., as seen for SN~2006X; \citealt{2008ApJ...677.1060W}) or seen at optical wavelengths \citep{2015ApJ...806..134M}. We therefore interpret this NUV point source as the result of the SN ejecta interacting with CSM in the progenitor system of SN~2015cp.

To better understand the nature of the late-time NUV emission of SN~2015cp we obtained ground-based optical and {\it HST} NUV follow-up observations in late 2017, which we present and analyze here (Sections \ref{ssec:15cp_ground} and \ref{ssec:15cp_hst}, respectively). We have also gathered all existing data for SN~2015cp from around the time of its discovery (Section \ref{ssec:15cp_main}) and obtained a spectrum of its host galaxy (Section \ref{ssec:15cp_host}), and we include these data in our analysis. Based on these data for SN~2015cp, we make a physical interpretation of its CSM's characteristics in Section \ref{ssec:15cp_phys}, and summarize our analysis of SN~2015cp in Section \ref{ssec:15cp_sum}. What we learn about SN~2015cp here we apply to our analysis of all of the nondetections from our {\it HST} NUV survey in Section \ref{sec:disc}.

\subsection{Host Galaxy}\label{ssec:15cp_host}

SN~2015cp is located 9.60\arcsec\ E and 9.96\arcsec\ N of its presumed host galaxy, a nearly face-on spiral with SDSS photometry $g=17.5$ and $r=16.7$ mag. To refine the host-galaxy redshift, on 2017 Oct. 25 we obtained a low-resolution optical spectrum of the host with the Kast Spectrograph on the 3~m Shane telescope at Lick Observatory. 

The H$\alpha$ emission line is at an observed wavelength of $\lambda = 6834 \pm 2$ $\rm \AA$, indicating a host-galaxy redshift of $z=0.0413 \pm 0.0003$, similar to that derived from the SN classification spectrum, $z\approx0.038$ \citep{2016ATel.8498....1F}. The luminosity distance to the host galaxy of SN~2015cp, assuming a flat universe with H$_{0} = 70$ $\rm km\ s^{-1}\ Mpc^{-1}$ and $\Omega_{\rm M} = 0.3$, is $167.5$ $\rm Mpc$. We adopt the values $z=0.0413$ and $d=167.5~{\rm Mpc}$ for SN~2015cp throughout this work.

\begin{figure}
\begin{center}
\includegraphics[width=8.8cm]{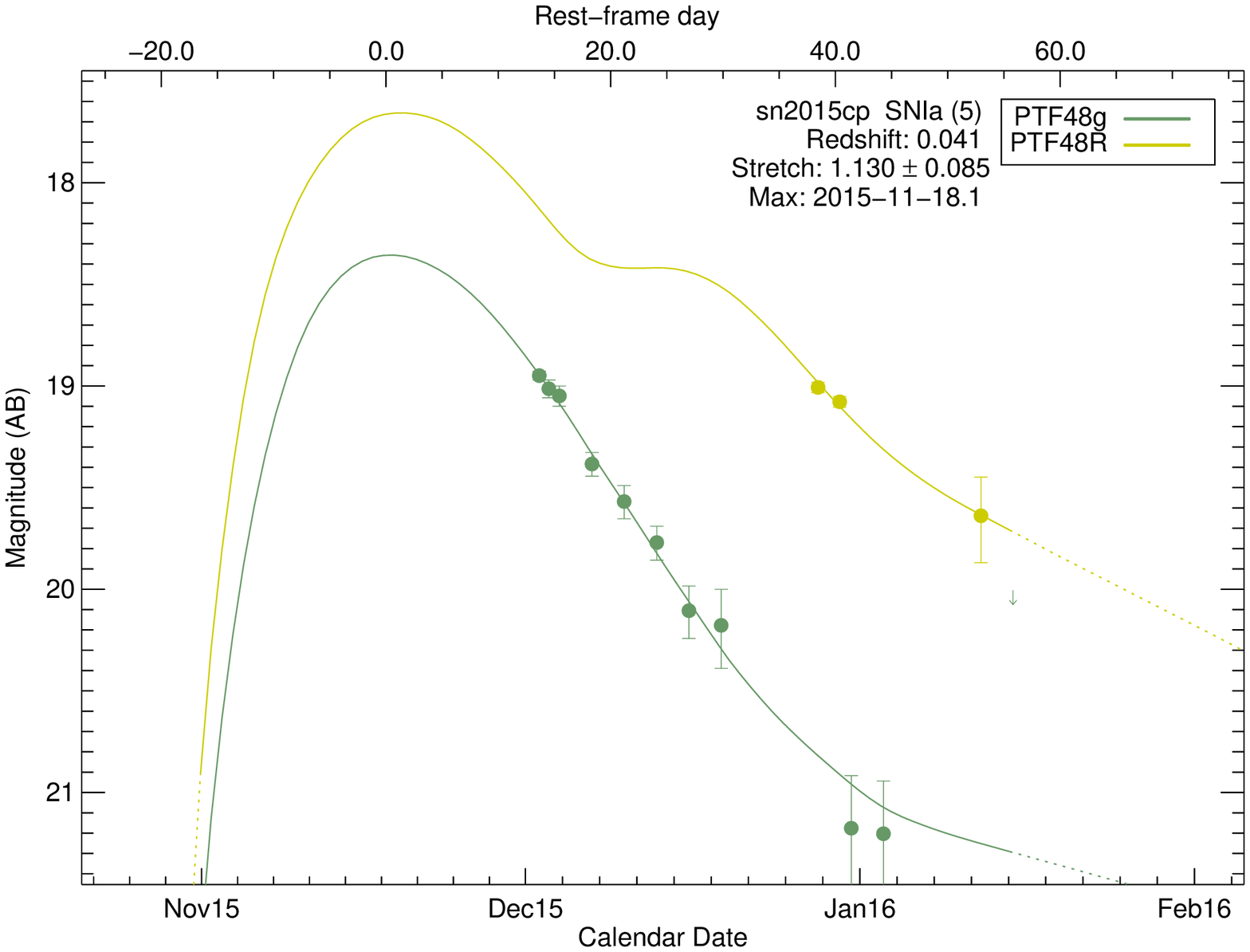}
\includegraphics[width=8.8cm]{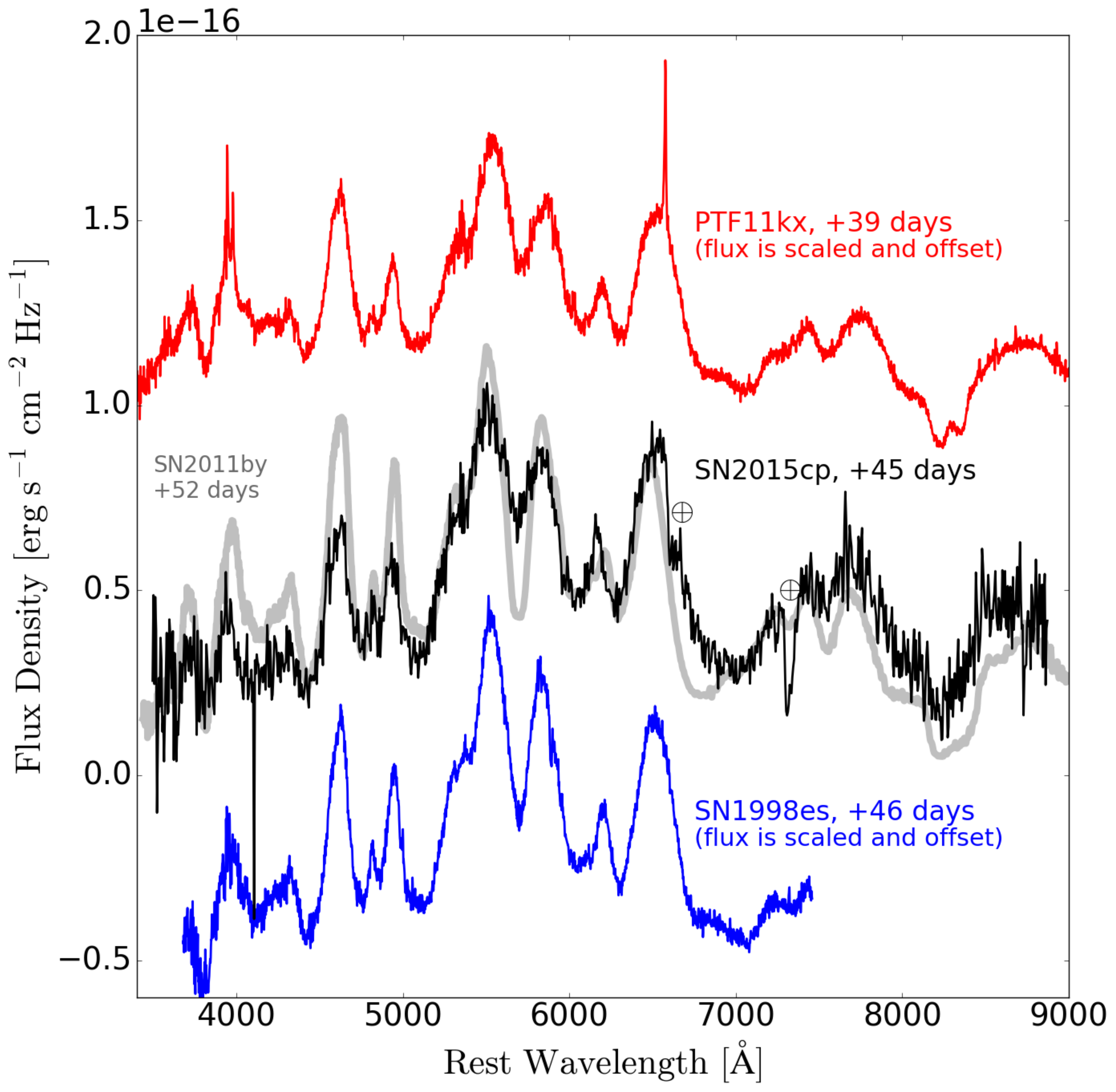}
\caption{{\it Top:} The partial iPTF light curve in $g$ and $r$, extending from mid-Dec. 2015 to mid-Jan. 2016, with the best-fit light curve from SiFTO \citep{2008ApJ...681..482C}. {\it Bottom:} The PESSTO spectrum of SN~2015cp (black line; note that the telluric absorption features were not removed) from \citealt{2016ATel.8498....1F}, with the best-fit spectral matches of SN\,Ia-CSM PTF11kx at 39~d after light-curve peak (red line; \citealt{2012Sci...337..942D}) and 91T-like SN~1998es at 46~d after light-curve peak (blue line; \citealt{2012AJ....143..126B}), flux scaled and offset to facilitate comparison. A scaled optical spectrum of the normal SN\,Ia 2011by at $+52$~d past maximum brightness is shown for comparison (gray line; \citealt{2015MNRAS.446.2073G}).}\label{fig:sn2015cp_early}
\end{center}
\end{figure}

\subsection{Light Curve and Classification Spectrum}\label{ssec:15cp_main}

The detection of SN~2015cp was first announced on 2015 Dec. 28 by the PanSTARRS survey with a discovery magnitude of $18.8$ \citep{2016ATel.8498....1F}, and a second PanSTARRS epoch of $i=20.1$ mag was obtained on 2016 Jan. 23 \citep{2017ApJ...835...64G,2015ATel.7153....1H}. The designation for the SN in this survey is PS15dpq.

We found that SN~2015cp was also covered by the iPTF survey (iPTF15fel) with a partial light curve in $g$ and $R$ that extends from mid-Dec. 2015 to mid-Jan. 2016; no other existing photometry was found for SN~2015cp. 
We fit the iPTF light curve using SiFTO \citep{2008ApJ...681..482C} as shown in Figure \ref{fig:sn2015cp_early}, holding the redshift fixed at $z=0.041$. 
Since all of the photometry is after the light curve's maximum brightness, the time and magnitude at peak are not well constrained. 
The best fit for the date of peak brightness is 2015 Nov. 18 $\pm4$ days; the fit template has a rise of 17~d (rest frame) and the estimated date of first light in the observer frame is 2015 Oct. 28 $\pm4.7$ days. 
The best fit for the peak apparent magnitudes are $m_g \approx 17.5\pm0.3$ and $m_R \approx 17.1\pm0.2$ in the AB system, in the observer frame, after correcting for Milky Way extinction. 
This light-curve fit indicates an intrinsic peak brightness of $M_B \approx -18.5 \pm 0.3$ mag, a color at peak brightness of $B-V \approx 0.38 \pm 0.07$ mag, and a light-curve stretch factor of $s=1.13\pm0.09$ (where stretch values $>1$ indicate broader/brighter light curves; \citealt{1997ApJ...483..565P}). If we include the correction for the known correlation between color and peak brightness  (i.e., the $\beta c$ correction, which accounts in part for dust extinction along the line of sight), we find $M_B \approx -19.6$ mag, which is overluminous compared to normal SNe\,Ia ($M_B \approx -19$ mag). The relatively large color correction implies that the light from SN~2015cp may have experienced significant dust extinction, which is somewhat unexpected given that SN~2015cp is well offset from its host galaxy, and may suggest extinction caused by CSM. However, since the photometry does not cover the light-curve peak, we cannot securely derive a value for dust extinction in the local environment, and do not attempt to correct for it in later analyses (Sections \ref{sec:15cp} and \ref{sec:disc}).

A spectrum of SN~2015cp was obtained on 2016 Jan. 2 by PESSTO \citep{2015A&A...579A..40S} at a phase of $45$~d after light-curve peak. The initial analysis of this spectrum classified SN~2015cp as a 91T-like SN\,Ia at $z\approx0.038$ with a spectroscopic phase of $\sim 40$~d after peak brightness \citep{2016ATel.8498....1F}; these are in agreement with both our light-curve fit and host-galaxy redshift. We used the online SN spectral classification tool GELATO\footnote{\url{gelato.tng.iac.es}} \citep{2008A&A...488..383H} to verify this classification, and found that it is 100\% identified as a SN\,Ia. The top three best-fit spectral templates are all to SN~1998es, a 91T-like SN\,Ia, at $46$--$58$~d past light-curve peak \citep{1998IAUC.7054....2J,1998IAUC.7059....1A}; no CSM interaction was reported for SN~1998es. The fourth best is to SN\,Ia-CSM PTF11kx at $39$~d past maximum light, which was near the time that CSM interaction started for PTF11kx \citep{2012Sci...337..942D}. The spectrum of SN~2015cp and examples of the best-fit GELATO templates are shown in Figure \ref{fig:sn2015cp_early}. We include for comparison a spectrum of SN\,Ia 2011by \citep{2015MNRAS.446.2073G}, a twin of the prototypically normal SN\,Ia 2011fe, which exhibits narrower features that are stronger in the blue. Although the line widths and depths of SN~2015cp do {\it appear} to be more similar to those of SN\,Ia 1991T-like PTF11kx, we do not think that a normal SN\,Ia spectral match can be ruled out for SN~2015cp based on this PESSTO spectrum. We furthermore note that the overluminous peak brightness and late-type host galaxy are also consistent with the SN~1991T-like subtype \citep[e.g.,][]{2006ApJ...648..868S,2009ApJ...691..661H}.

This spectrum does not exhibit any Na~I~D $\lambda5890$ absorption which, especially if it were blueshifted, would have indicated the potential presence of CSM. However, in this case the absence of a visible Na~I~D absorption feature is not evidence that there is no CSM, because the optical spectrum is quite noisy. Furthermore, no obvious Na~I~D absorption features were seen in a spectrum of PTF11kx at similar phase and signal-to-noise ratio, and PTF11kx exhibited CSM interaction shortly thereafter (Figure \ref{fig:sn2015cp_early}; \citealt{2012Sci...337..942D}). Given the flux noise in the $45$~d spectrum of SN~2015cp, we can rule out the presence of narrow H$\alpha$ and \ion{Ca}{2} emission lines with the same luminosity as those exhibited by PTF11kx at a similar phase (i.e., as shown in Figure \ref{fig:sn2015cp_early}) with 10$\sigma$ and 2$\sigma$ confidence, respectively. This strongly suggests that for SN~2015cp, the CSM interaction had not yet begun at $45$~d after peak brightness.

\begin{figure*}
\begin{center}
\includegraphics[width=17cm]{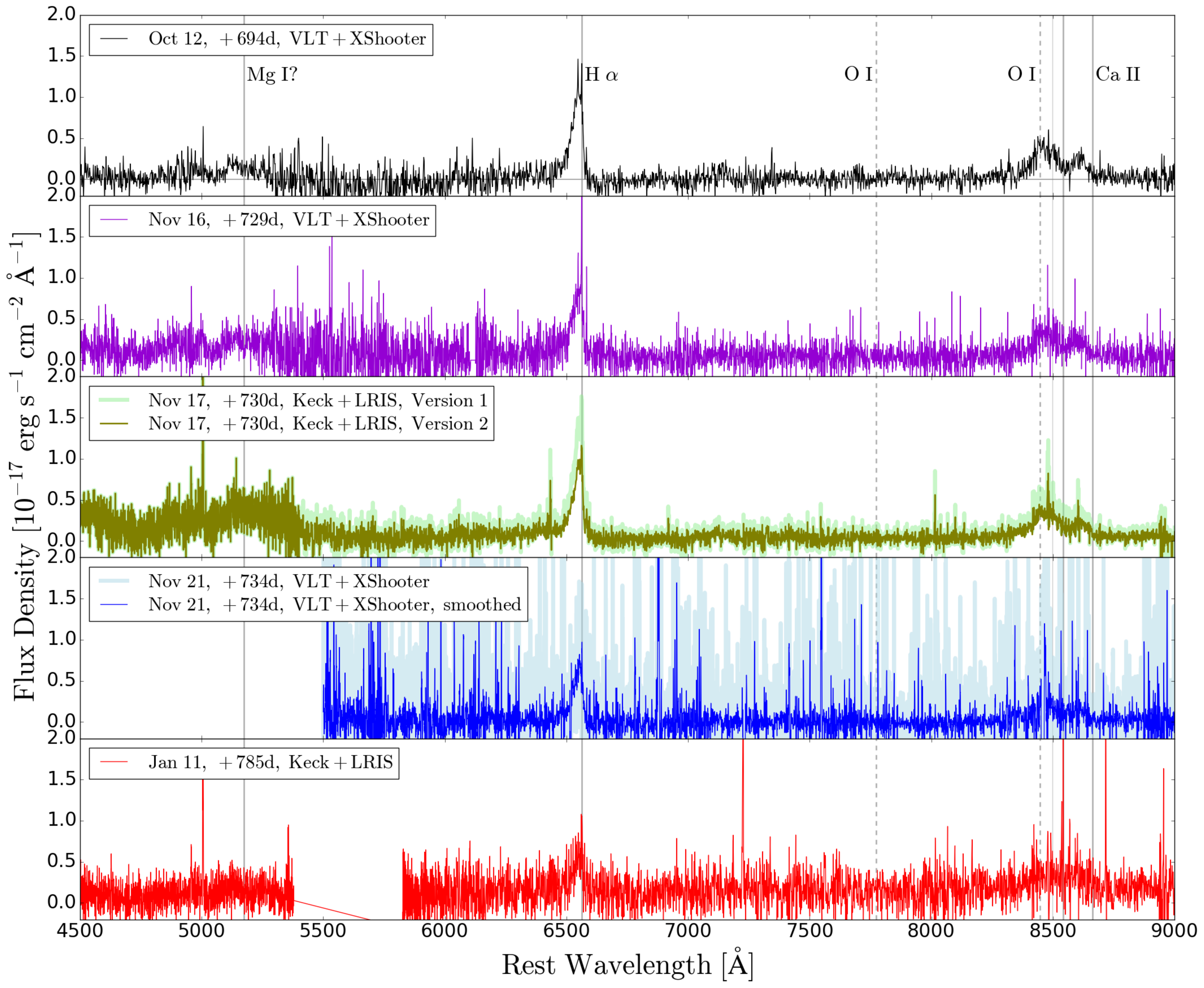}
\caption{The five epochs of spectroscopy obtained with the VLT+XShooter and Keck+LRIS. Spectra have been corrected to the host-galaxy redshift ($z=0.413$; Section \ref{ssec:15cp_host}). We have labeled our tentative identification of \ion{Mg}{1} $\lambda5174$, our detection of H$\alpha$, and all three components of the \ion{Ca}{2} near-infrared triplet with solid vertical lines. We additionally note the location of \ion{O}{1} $\lambda8446$ with a dashed vertical line, which we suggest could be contaminating the blue-edge of the \ion{Ca}{2} triplet, but cannot confirm (e.g., \ion{O}{1} $\lambda7774$, also marked with a dashed line, is not seen). In the third panel we show two reduction versions with two different standard stars used for flux calibration as an example of the systematic errors. The fourth panel shows both the original spectrum (light blue) and after smoothing with a Savitsky-Golay filter (dark blue) for clarity. \label{fig:sn2015cp_spec}}
\end{center}
\end{figure*}

\subsection{Ground-Based Optical Follow-up Observations}\label{ssec:15cp_ground}

After our {\it HST} NUV detection of SN~2015cp, we pursued ground-based optical spectroscopic observations to look for H$\alpha$ emission, which would confirm that the NUV signal was caused by the SN ejecta interacting with CSM. 

We first obtained $8\times300$~s $r$-band images on 2017 Oct. 12 with the Palomar Observatory 60~inch telescope (the seeing was $\sim1.86\arcsec$), at a phase of $694$~d since light-curve peak brightness ($30$~d after the {\it HST} observation). SN~2015cp was not detected in a deep stack of these images, which has an upper limit of $m_r = 23.6$ mag. That same night we obtained $R$-band imaging and a $2\times600$~s optical spectrum with the XShooter instrument \citep{2011A&A...536A.105V} on the Very Large Telescope (VLT) at Cerro Paranal in Chile. We detected SN~2015cp with low signal-to-noise ratio in the $R$-band image with an apparent magnitude of $m_R \gtrsim 24$ mag, and detected emission lines from a point-like trace in the spectra.

The spectrum shown in Figure \ref{fig:sn2015cp_spec} exhibits both H$\alpha$ and \ion{Ca}{2} emission. These lines are not associated with the nebular SN emission, which is dominated by forbidden lines from the nucleosynthetic products (Fe, Co, and Ni) and would be too faint to be detected by these observations. These lines of H$\alpha$ and \ion{Ca}{2} emission revealed by the VLT spectrum provided the crucial confirmation that the observed NUV luminosity was associated with SN ejecta interacting with CSM that contains H. In order to monitor the evolution in the H$\alpha$ emission, we obtained further VLT+XShooter spectra on 2017 Nov. 16 and 21, shown in Figure \ref{fig:sn2015cp_spec}.

We also obtained a Keck spectrum of SN~2015cp on 2017 Nov. 17 with the Low Resolution Imaging Spectrometer (LRIS; \citealt{1995PASP..107..375O}). We acquired $3\times1200$~s exposures with a long slit of width 1.0\arcsec, the $560$ $\rm nm$ dichroic, the $400/8500$ grating, and the $600/4000$ grism. The data were processed using standard techniques optimized for Keck+LRIS by the \texttt{CarPy} package in \texttt{Pyraf} \citep{2000ApJ...531..159K,2003PASP..115..688K}. We performed flux calibration using sensitivity functions derived from {\it each} of the two standard-star observations made that night, which led to two reduced versions with different total H$\alpha$ fluxes (Figure \ref{fig:sn2015cp_spec}). Both standards were used simply as a conservative precaution, but this has revealed a systematic error in our spectral calibrations which we cannot explain --- for example, the weather on Maunakea was stable and clear that night\footnote{See the CFHT Sky Probe plot for 2017 Nov. 17 at \url{http://www.cfht.hawaii.edu/Instruments/Elixir/skyprobe}}. Since neither the absolute H$\alpha$ flux measurements nor the shape of its flux decline curve are used in any of our analysis, this issue with the calibration of the Keck data from 2017 Nov. 17 does not affect our results. On 2018 Jan. 11 we obtained another $3\times1200$~s spectrum of SN~2015cp in a similar setup with Keck+LRIS, as shown in the bottom panel of Figure \ref{fig:sn2015cp_spec}, which provided the conclusive evidence that the H$\alpha$ flux was declining.

\begin{figure}
\begin{center}
\includegraphics[width=8.8cm]{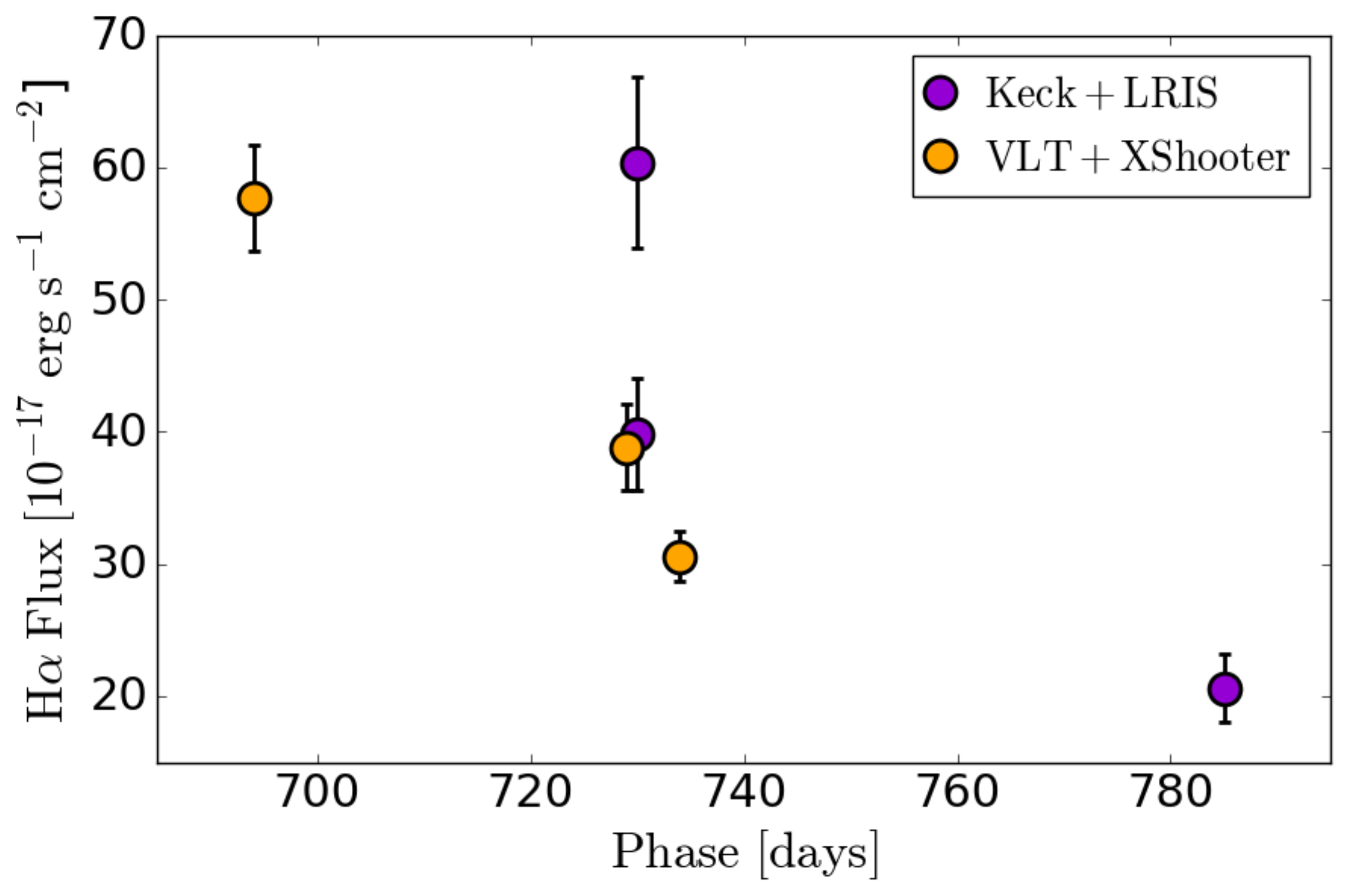}
\caption{The evolution of the integrated flux of the H$\alpha$ emission line for SN~2015cp from our spectroscopic data (circles) from VLT+XShooter (orange) and Keck+LRIS (purple). The two co-epochal Keck+LRIS data points represent the same data calibrated with two different standard stars, and demonstrate a potentially large systematic error in those data. \label{fig:sn2015cp_halpha}}
\end{center}
\end{figure}

The dominant feature in the optical spectra of SN~2015cp is the broad H$\alpha$ emission line with an asymmetric profile that exhibits a blueshifted peak. This profile is consistent with the signature of dust formation in an emitting volume that is expanding: the receding material that generates the red half of the line experiences greater extinction because of the additional dust along the line of sight. We make direct measures of the peak velocity, FWHM, and integrated flux of the H$\alpha$ emission line for all of our spectra. The latter is calculated by cutting out and interpolating over the pixels associated with host-galaxy emission (the very narrow features at $v\approx 0$ $\rm km\ s^{-1}$ in Figure \ref{fig:sn2015cp_spec}), and performing a numerical integration on the calibrated one-dimensional (1D) spectrum. Since there is no continuum flux in our spectra, we do not need to perform a continuum subtraction prior to the integration. The resulting line parameters are listed in Table \ref{tab:linepars} and the flux evolution is plotted in Figure \ref{fig:sn2015cp_halpha}, where we can see that the H$\alpha$ flux is declining by a factor of $\sim3$ in $\sim90$~d ($L_{\rm H\alpha} \propto t^{-8.6}$). Even considering the large (systematic and random) error bars on the flux measurements and the decline rate, this indicates an extremely rapid decline. For comparison, an ideal gas expanding as $V \propto t^3$ ($V$ is volume and $t$ is time) with an adiabatic equation of state $P \propto \rho^{5/3}$ ($P$ is pressure and $\rho$ is mass density) would have its optically thin free-free luminosity decreasing as $L_\textrm{ff} \propto \rho^2 T^{-1/2} V \propto t^{-2}$. 

\begin{table*} 
\begin{center} 
\caption{Hydrogen Emission Line Parameters \label{tab:linepars}} 
\begin{tabular}{llcccc}  
\hline 
\hline 
 & & & H$\alpha$~$\lambda 6563$~\AA & H$\alpha$~$\lambda 6563$~\AA & H$\alpha$~$\lambda 6563$~\AA \\ 
 Date & Instrument & Phase & Peak Velocity & FWHM & Flux \\ 
 & & [days] & [${\rm km\ s^{-1}}$] & [${\rm km\ s^{-1}}$] & [${\rm 10^{-17}\ erg\ s^{-1}\ cm^{-2}}$] \\ 
\hline 
2017-10-12 & XShooter & 694.0 & $-434\pm417$ & $2371\pm121$ & $58\pm4$ \\ 
2017-11-16 & XShooter & 729.0 & $-535\pm272$ & $2327\pm356$ & $39\pm3$ \\ 
2017-11-17 & LRIS & 730.0 & $-566\pm67$ & $1920\pm264$ & $60\pm6$ \\ 
 & & 730.0 & $-566\pm67$ & $1920\pm264$ & $40\pm4$ \\ 
2017-11-21 & XShooter & 734.0 & $-449\pm144$ & $1976\pm99$ & $31\pm2$ \\ 
2018-01-11 & LRIS & 785.0 & $-877\pm346$ & $1912\pm1526$ & $21\pm3$ \\ 
\hline 
\end{tabular} 
\end{center} 
\end{table*}

We do not detect H$\beta$ $\lambda4861$ emission, but we can estimate an upper limit on the line intensity as $3$ times the noise in that region. Using the first VLT spectrum obtained at $+694$~d, this leads to $I({\rm H\beta})<0.3 \times 10^{-17}$ $\rm erg\ s^{-1}\ cm^{-2}\ \AA^{-1}$. Assuming that there is no underlying continuum emission (or that it is the same in the regions of H$\alpha$ and H$\beta$), this puts a lower limit on the Balmer decrement of $D = I({\rm H\alpha})/I({\rm H\beta}) > 3$, which is consistent with Case B recombination in an ionized nebula and the large Balmer decrements seen in promptly interacting SNe\,Ia-CSM \citep{2013ApJS..207....3S}. However, since this is based on a noisy nondetection of H$\beta$, we do not rely on this estimate for any further analysis.

At the red end of the optical spectrum we see the \ion{Ca}{2} near-infrared triplet, appearing as a blend of Gaussian-like features. Compared to the H$\alpha$ emission line, we find that the \ion{Ca}{2} is a bit broader (FWHM $\sim 3000$ $\rm km\ s^{-1}$) and/or perhaps exhibits a slightly extended blue-side wing (at $\lambda\lesssim8440$, but it's unclear given the noise in flux), and clearly has an overall blueshift of $\sim -1400$ $\rm km\ s^{-1}$ (significantly larger than H$\alpha$; Table \ref{tab:linepars}). One explanation for a slightly extended blue-side wing might be the presence of \ion{O}{1} $\lambda 8446$, which we show as a dashed vertical line in Figure \ref{fig:sn2015cp_spec}. However, for \ion{O}{1} to contribute to the Ca triplet in this way \ion{O}{1} would have to be at rest velocity, and furthermore \ion{O}{1} is not typically seen in the late-time spectra of SN\,Ia-CSM like it is for SN\,II, as it is a common element in the ejecta of core-collapse SNe \citep{2013ApJS..207....3S,2015MNRAS.447..772F}. We do not see the \ion{O}{1} $\lambda 7774$ \AA\ line (Figure \ref{fig:sn2015cp_spec}), and unfortunately the VLT XShooter spectrum is too noisy in the region of \ion{O}{1} $\lambda 11290$ \AA\ to confirm the presence of oxygen in the spectrum. 

The additional blueshift of \ion{Ca}{2} compared to H$\alpha$ is certain, but is a phenomena we cannot yet explain. Since \ion{Ca}{2} is at redder wavelengths than H$\alpha$ the extra blueshift is unlikely to be due to more severe effects of dust extinction. Furthermore, explaining the additional \ion{Ca}{2} velocity offset with dust extinction would require significant extinction on the {\em near} side of the gas (to produce the observed zero intensity at zero velocity). A potential explanation is that perhaps the \ion{Ca}{2} line is forming in a higher-velocity region than H$\alpha$, but there is no physical motivation to expect this. An alternative interpretation is offered by \cite{2014ApJ...797..118F}, who present a well-sampled, multiwavelength temporal series of observations for SN\,IIn 2010jl, which also exhibits blueshifted spectral features. Instead of dust extinction, \cite{2014ApJ...797..118F} find it more likely that the emitting CSM has been accelerated by the SN's radiation. We ultimately conclude that the SN~2015cp spectral data are insufficient to distinguish between these competing explanations. As a final note in support of dust extinction as the root cause of the observed blueshift for H$\alpha$, we note that other SNe\,Ia-CSM such as PTF11kx have exhibited increasing infrared emission after the ejecta pass through the CSM, indicating dust formation in the post-shock nebula \citep[e.g.,][]{2017ApJ...843..102G}.

The decline rate of the \ion{Ca}{2} near-infrared triplet's flux is not as securely measured as that of H$\alpha$'s flux owing to the noise at the red end of the spectrum, which prevents a detailed analysis. Although the \ion{Ca}{2} appears to be declining more slowly than H$\alpha$, because of the large uncertainties on the flux measurements we cannot rule out that they are co-evolving and that the two features are both connected to the shock evolution. This is discussed further by H18 (their Figure 2 provides a direct comparison of the \ion{Ca}{2} and H$\alpha$ flux decline rates).

These spectra also potentially exhibit a \ion{Mg}{1} $\lambda 5174$ feature with a broad or blueshifted profile, but this has a very low signal-to-noise ratio that prohibits further analysis. 

\begin{figure}
\begin{center}
\includegraphics[width=8.8cm]{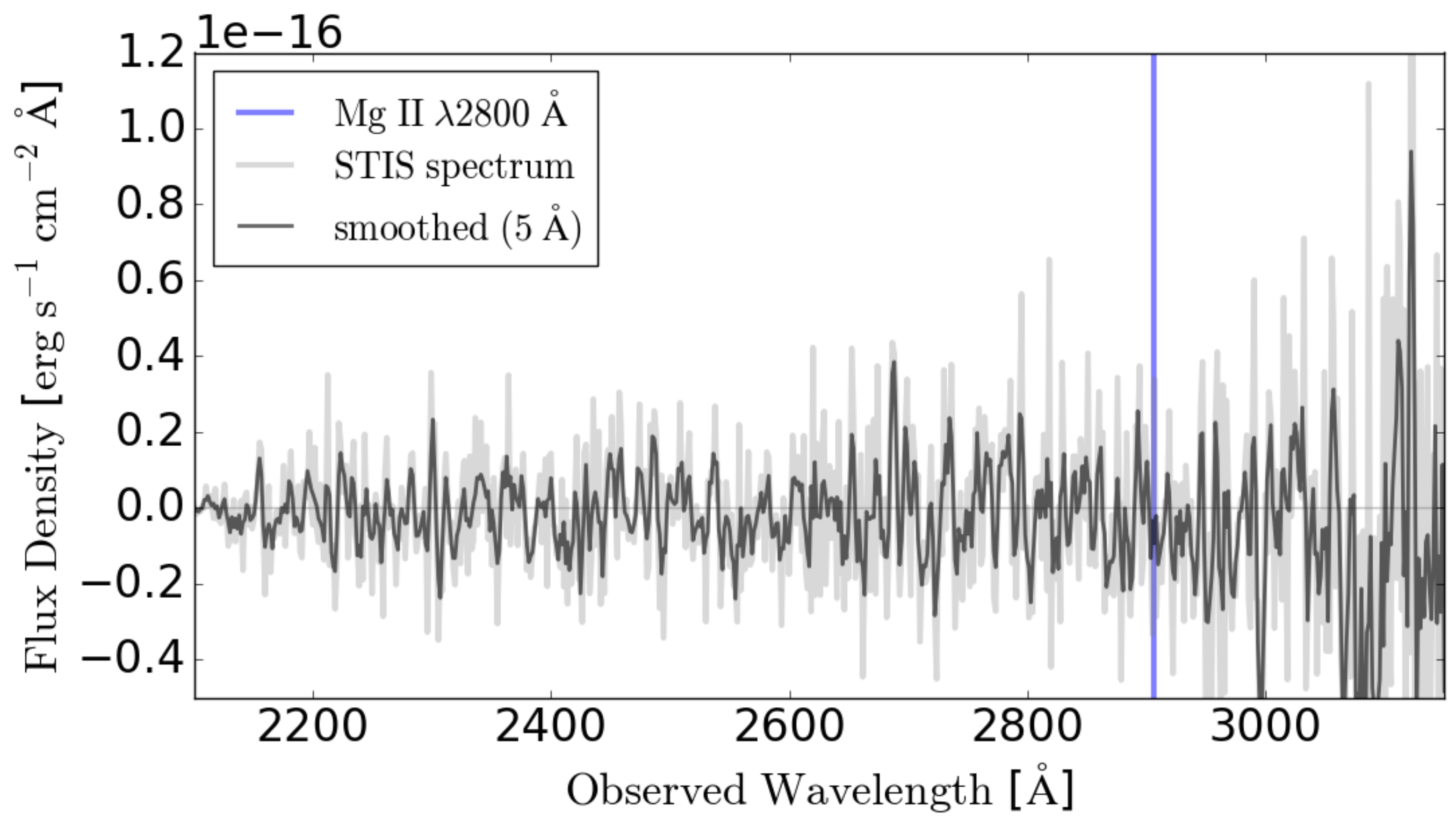}
\includegraphics[width=8.8cm]{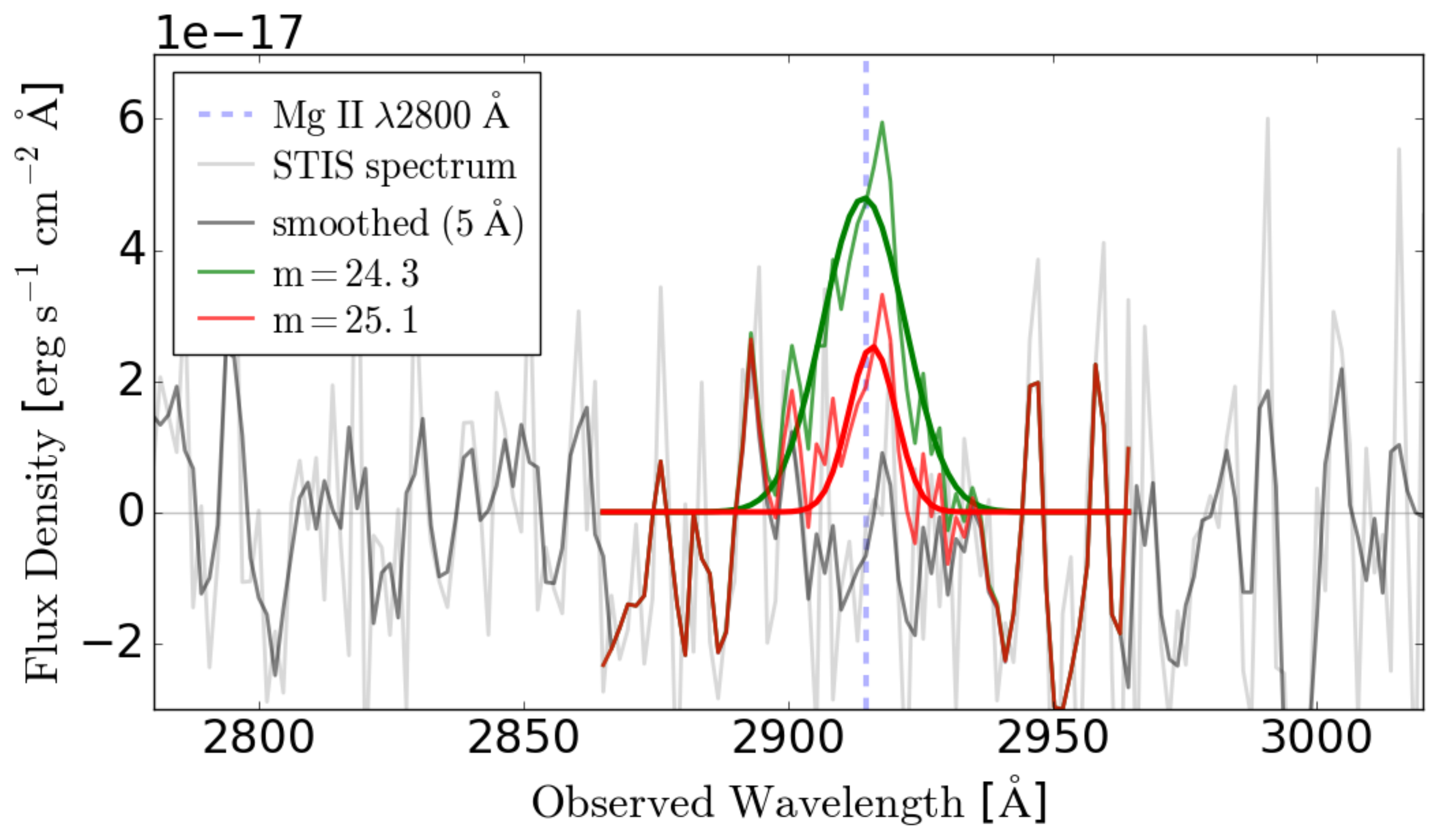}
\caption{{\it Top:} An {\it HST} STIS NUV spectrum obtained on 2017 Nov. 25 (thick gray line), smoothed with a Savitsky-Golay filter (black line), in the region of \ion{Mg}{2} $\lambda2800$ (blue line). {\it Bottom:} Simulated emission lines that would result in $m_{\rm F275W} = 24.3$ mag (i.e., equivalent to our photometric detection; green) or $m_{\rm F275W} = 25.1$ mag (i.e., equivalent to our derived Gaussian-line detection limit; red) are planted in the original spectrum, smoothed, and fit with a Gaussian to demonstrate our detection-limit measurement that is described in the text.   \label{fig:sn2015cp_stis}}
\end{center}
\end{figure}

\subsection{{\it HST} NUV Follow-up Observations}\label{ssec:15cp_hst}

After confirming the interaction origin of our NUV detection, we submitted an {\it HST} Director's Discretionary Time (DDT) proposal requesting one NUV spectrum with the STIS MAMA G230L grating and a time series of NUV photometric monitoring with the WFC3 F275W filter. We requested the spectrum to determine the nature of the NUV emission, and we suspected that the NUV emission would be line dominated for two reasons:
(1) our initial analysis of the NUV detection indicated that bremsstrahlung (free-free) and synchrotron emission could not account for the observed luminosity (see also \ref{ssec:disc_phys}), and (2) the \ion{Mg}{2} $\lambda 2800$ resonance line was covered by {\it HST}'s F275W filter for this SN. We also proposed for several closely spaced F275W imaging epochs followed by a slower cadence in order to determine the decline rate of the emission, because this would give some indication of the timescale of the CSM interaction and the radial extent of the material. Our request for a STIS spectrum was granted, with the photometric time series conditional on the confirmed presence of \ion{Mg}{2} (GO-15407; PI Graham).

On 2017 Nov. 25, at a phase of $738$~d since the light-curve peak, {\it HST} executed our program and obtained a $12,991$~s spectrum with the STIS NUV MAMA G230L grating and the $52\times0.5\arcsec$ slit. Our choice of the $0.5\arcsec$ width was to ensure that SN~2015cp was within the slit at the time of the exposure; as described in the STIS Data Handbook, {\it ``Once the telescope acquires its guide stars, your target will be within $0.2$--$0.3\arcsec$ of the aperture center"} \citep{STISData_Handbook}. Unfortunately, absolutely no source is seen at the expected location in the processed two-dimensional and 1D spectra. In Figure \ref{fig:sn2015cp_stis} we show the extracted 1D spectrum (from the STIS pipeline's {\tt x1d} task), and it is clearly dominated by noise with no emission. To estimate our detection limit for a Gaussian-shaped emission line of \ion{Mg}{2} $\lambda2800$, we inject simulated Gaussians into the original STIS data. We use a range of FWHM and peak velocities that are similar to our observations of H$\alpha$, and a range of fluxes that would yield an apparent magnitude similar to and fainter than our initial NUV detection of $m_{\rm F275W} = 24.3$ mag. We then smooth the spectrum with a Savitsky-Golay filter of $5$~\AA\ (i.e., less than our expected FWHM), and attempt to fit a Gaussian function using randomized initial parameters and {\tt scipy.optimize.curve\_fit}. We consider the planted emission line successfully recovered if the fit has a peak wavelength $\lambda_{\rm peak} < \sigma/2$, and also has a similar FWHM and peak intensity; this process is illustrated by the planted-and-recovered Gaussians in Figure \ref{fig:sn2015cp_stis}.

From this procedure we report an upper limit on the emission-line brightness at $+738$~d of $m_{\rm F275W} \approx 25.1$ mag, the magnitude at which our emission-line recovery fraction drops to 50\% (and we checked that this result is the same if we do not smooth the spectrum). As reported in Section \ref{sssec:review_sn2015cp}, our NUV detection at $+664$~d had $m_{\rm F275W} \approx 24.3$ mag. 
If 100\% of that flux was from an emission line of \ion{Mg}{2} $\lambda2800$ declining at the same rate as H$\alpha$ and \ion{Ca}{2}, then the expected integrated flux at $+738$~d is $m_{\rm F275W} \approx 25.3$ mag. This is slightly fainter than our upper limit from the STIS spectrum, and so our nondetection remains consistent with our hypothesis that \ion{Mg}{2} line emission is the underlying cause of our NUV detection.

If the NUV luminosity at $+664$~d was the result of continuum emission, then we would not have detected it in our STIS spectrum at $+738$~d unless the brightness had \textit{increased} to $m_{\rm F275W} \leq 22.3$ mag (i.e., the magnitude which would provide an average signal-to-noise ratio of $\sim1$ per resolution element in the spectrum). Additionally, we note that if the NUV luminosity at $+664$~d was due to high-temperature blackbody emission with $f_{\rm NUV} \approx 3 \times 10^{-18}$ $\rm erg\ s\ cm^2\ \AA$, then its optical component would be undetectable in our optical spectra (Figure \ref{fig:sn2015cp_spec}). Therefore, we conclude that our observations cannot rule out continuum emission as the underlying cause of our NUV detection, and furthermore note that it is implausible given that the optical spectrum is line dominated.

Unfortunately, the photometric monitoring aspect of our {\it HST} DD proposal was conditional on detecting the \ion{Mg}{2} line, so we were unable to proceed with our program to obtain multiple epochs of photometry to further understand the NUV luminosity evolution. 

\subsection{Physical Interpretation}\label{ssec:15cp_phys}

In this section we use our NUV and optical follow-up observations to deduce the most likely characteristics of the SN~2015cp's H$\alpha$ and NUV emission and their evolution, and to make a preliminary physical interpretation of the amount and location of CSM in its progenitor system. Detailed physical modeling of the CSM properties of SN~2015cp can be found in H18, who incorporate radio and X-ray follow-up observations of SN~2015cp to produce quantitative constraints.

As described in Section \ref{ssec:15cp_main}, the PESSTO spectrum of SN~2015cp is best fit by templates from the 91T-like subclass of SNe\,Ia, such as SN\,Ia 1998es and SN\,Ia-CSM PTF11kx. The SN ejecta of PTF11kx were observed to interact with CSM starting $\sim 60$~d after its explosion, marked by the appearance of H and Ca emission features \citep{2012Sci...337..942D}. The evolution of PTF11kx's spectral features was monitored for hundreds of days afterward \citep{2013ApJ...772..125S}, generating a significantly better temporal series of data for PTF11kx than we have for SN~2015cp. For this reason we base most of our preliminary physical interpretation of SN~2015cp's CSM on a comparison with PTF11kx. 

In Section \ref{ssec:15cp_main} we presented the optical light curve of SN~2015cp, from which we determined the date of explosion to be 2015 Oct. 28 $\pm4.7$ days. The fact that the 2016 Jan. 2 spectrum does not exhibit any signature of CSM interaction (e.g., a broad, ${\rm FWHM} \approx 2000~{\rm km~s^{-1}}$ H$\alpha$ emission line, as seen for PTF11kx at a similar phase in Figure \ref{fig:sn2015cp_early}) indicates that the ejecta of SN~2015cp did not impact the CSM earlier than $\sim 60$~d after explosion. Assuming a maximum ejecta speed of $v_{\rm ej,max} \approx 30,000$~$\rm km\ s^{-1}$, this indicates that the CSM was at a distance of at least $R_{\rm CSM}>10^{16}~{\rm cm}$ from the WD in the SN~2015cp progenitor system. Since the ejecta of PTF11kx were observed to begin interacting with CSM at $\sim 60$~d after explosion, we conclude that the CSM in the system of SN~2015cp was more distant than that of PTF11kx, though perhaps only marginally.

\cite{2013ApJ...772..125S} show that the luminosity of the broad H$\alpha$ line of PTF11kx increased for $\gtrsim 60$~d after the onset of CSM interaction, and then exhibited a plateau of $200$--$400$~d before declining. 
The start of the H$\alpha$ luminosity decline has been interpreted as indicating the end of interaction \citep{2013ApJ...772..125S,2017ApJ...843..102G}; in other words, the time at which the ejecta had fully passed through the CSM. However, the spectral time series of PTF11kx is too sparse to accurately measure the H$\alpha$ luminosity decline rate or the true duration of the plateau. 
We have shown in Figure \ref{fig:sn2015cp_halpha} that SN~2015cp has a declining $L_{\Hal}$ throughout our observations, which indicates that our discovery occurred well after the ejecta-CSM interaction occurred, and perhaps very soon after the shock had swept through the CSM given the rapidity of the decline (as discussed in Section \ref{ssec:15cp_ground}). The H$\alpha$ light curve of PTF11kx may also have declined more rapidly in the $\sim100$~d after its plateau than at much later times (e.g., Figure 4 of \citealt{2017ApJ...843..102G}), but the H$\alpha$ light curve is too sparsely sampled to be sure. A direct comparison of the $L_{\Hal}$ light curves of SN~2015cp and PTF11kx is presented by H18 (their Figure 2).

We find that the $\Hal$ line profiles of SN~2015cp and PTF11kx are similar in width and shape between the two events, but that the integrated luminosity is different --- for SN~2015cp, $L_{\Hal}$ is about an order of magnitude lower than that of PTF11kx at a similar phase. Whether this is due to a less dense CSM for SN~2015cp, or a difference in the CSM radial extents for the two events, is not possible to distinguish from our data. Unfortunately, no NUV observations of PTF11kx were obtained, so we cannot extend our comparison of these two SNe\,Ia-CSM any further. 

As a final aspect of our physical interpretation for SN~2015cp, we consider the possible mechanisms that could have produced the observed NUV emission, and what they may reveal about the nature of the CSM in the SN~2015cp system. The main open question is whether the  NUV signal was from a \ion{Mg}{2} $\lambda 2800$ emission line or from a continuum process like optically thin bremsstrahlung. As described in Section \ref{ssec:15cp_hst}, our STIS nondetection does not allow us to rule out either. However, one argument in favor of line emission is that an improbably large CSM mass, $\sim 10$~$\rm M_\odot$, would be required to produce enough continuum emission (see \S\ref{ssec:disc_phys}). 
Another argument in favor is that the models of \cite{1994ApJ...420..268C} show that the \ion{Mg}{2} $\lambda 2800$ line can be luminous enough to account for the NUV emission --- specifically, up to $8$ times more luminous than H$\alpha$ when the ejecta of a SN\,II interact with a slow, dense wind. This is not an identical scenario to SNe\,Ia, but similar physical mechanisms are at play. These models are consistent with our observation that $L_{\rm Mg~II}/L_{\rm H\alpha} < 6$ (based on our late-time {\it HST} imaging and ground-based optical spectra for SN~2015cp, presented above).
Unfortunately, the requisite modeling to evaluate the density and extent of CSM associated with such NUV line emission remains to be done, and is beyond the scope of this work. 

\subsection{SN~2015cp Summary}\label{ssec:15cp_sum}

We have demonstrated that the ensemble NUV through optical observations of SN~2015cp are consistent with an overluminous, possibly SN~1991T-like SN\,Ia, for which the ejecta interacted with CSM containing H no earlier than $\sim 60$~d after explosion. This sets the inner CSM radius of SN~2015cp to be $R_{\rm CSM}>10^{16}$~$\rm cm$, farther than that of SN\,Ia-CSM PTF11kx (but perhaps not by much). Our {\it HST} NUV detection at $686$~d after explosion sets the maximum inner radius of the CSM to be $R_{\rm CSM}<10^{17}$~$\rm cm$. During the $\sim 120$~d after our NUV detection, which itself was nearly 2~yr after explosion, our optical and NUV follow-up observations reveal rapidly declining H$\alpha$ and \ion{Ca}{2} emission, which is typically associated with the interaction of CSM and SN ejecta. The rapid decline suggests that we caught SN~2015cp soon after that emission peaked in brightness, which in turn implies that the interaction-driven peak in H$\alpha$ luminosity was probably fainter than that of PTF11kx, and that the CSM of SN~2015cp has either a lower density or a smaller radial extent than in PTF11kx. Finally, we argue that our NUV photometric detection was likely produced by line emission rather than thermal continuum emission (although the latter could not be ruled out by our observations), but that without more detailed modeling of line-emission mechanisms our NUV detection alone cannot place any independent constraints on the CSM properties in the SN~2015cp progenitor system.

\section{Discussion}\label{sec:disc}

The majority of our {\it HST} observations (70 out of 72) yielded nondetections, which indicates that NUV emission from late-onset CSM interaction in SNe\,Ia is short-lived, faint, and/or rare. In Section \ref{ssec:disc_nuvlims} we convert our detections and upper limits into intrinsic spectral luminosities, and compare with the observed emission from different types of SNe. In Section \ref{ssec:disc_rate} we use a fiducial NUV light curve and our nondetections to place limits on the occurrence rate of CSM interaction events like SN~2015cp. In Section \ref{ssec:disc_phys} we present a toy model for the NUV emission and use our nondetections to constrain the amount of CSM, and its radial distribution, in SN\,Ia progenitor systems. 

\subsection{Intrinsic NUV Luminosity Limits}\label{ssec:disc_nuvlims}

\begin{figure*}
\begin{center}
\includegraphics[width=17cm]{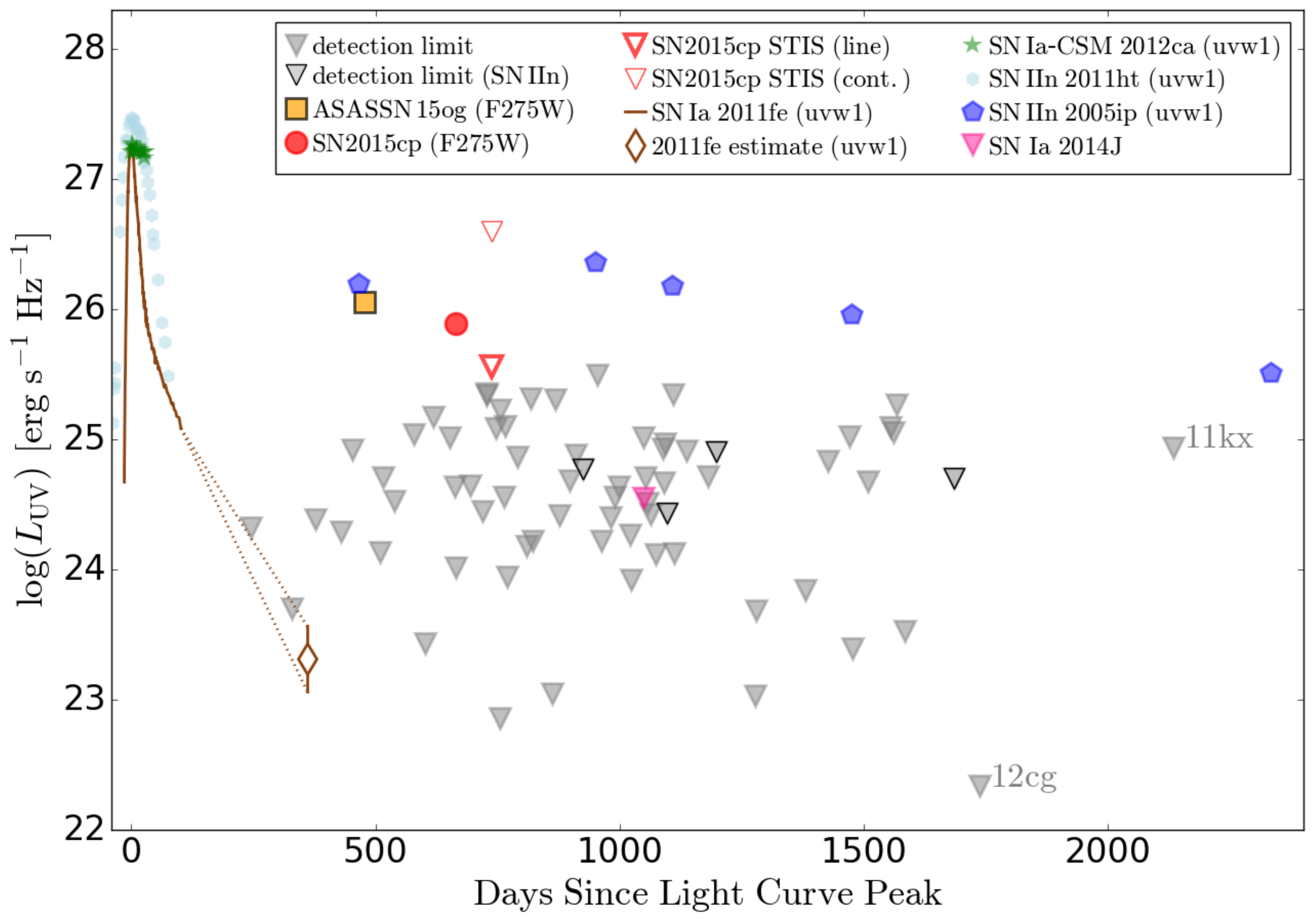}
\caption{NUV luminosities and/or limits as a function of phase in days after SN peak brightness (see Table \ref{tab:absuv}). Inverted gray triangles mark the 50\% limiting magnitudes from our WFC3 F275W imaging survey (Table \ref{tab:obs}); markers edged in black denote targets from our survey with a SN\,IIn classification, which means CSM interaction was observed at or around the time of discovery and initial monitoring. Our WFC3 F275W detections of ASASSN-15og and SN~2015cp are marked by an orange square and a red circle, respectively. Our {\it HST} STIS nondetection of SN~2015cp is indicated by two open red triangles, which represent the $L_{\rm UV}$ limits if the NUV spectrum was dominated by line or continuum emission (heavier and lighter weight symbols, respectively). All inverted triangles have been shifted so that their top side is aligned with the limit, and given a width of $\pm 15$~d for clarity. This width is larger than the uncertainty in most of our estimates of the optical light-curve peak, but equivalent to the largest error in peak date (for SNe\,Ia that were discovered late). Our luminosity limits for SNe\,Ia SN~2014J (pink symbol) and 2012cg (labeled) have been corrected for their considerable host-galaxy extinction; PTF11kx is also labeled because it is discussed in Section \ref{ssec:disc_nuvlims}. The brown line represents the near-peak {\it Swift} UVOT {\it uvw1} observations of normal SN\,Ia 2011fe \citep{2012ApJ...753...22B}, and is extended with dotted brown lines to our late-time estimate of the \textit{uvw1} luminosity of SN~2011fe, represented with an open brown diamond. Green stars represent several epochs of near-peak {\it Swift} UVOT {\it uvw1} photometry of SN~2012ca (i.e., not a full light curve as shown for SN~2011fe and SN~2011ht) from \cite[][who classify SN~2012ca as a core-collapse SN instead of a SN\,Ia]{2016MNRAS.459.2721I}. The blue pentagons and light-blue hexagons represent {\it Swift} UVOT observations of SNe\,IIn 2005ip and 2011ht in the \textit{uvw1} filter \citep{2012ApJ...756..173S,2012ApJ...751...92R}. All points represent spectral luminosities derived from AB magnitudes, and have been corrected for Milky Way extinction. \label{fig:absuv}}
\end{center}
\end{figure*}

In order to analyze our full sample of late-time NUV observations for this diverse set of SNe, we must convert our NUV apparent magnitudes or magnitude limits to intrinsic spectral luminosities, and the dates of our {\it HST} observations to phase in days after SN peak brightness. For each of our 70 undetected targets, we convert the limiting magnitudes that represent the 50\% detection probability, as listed in Table \ref{tab:obs}, into an upper limit on the intrinsic NUV luminosity by correcting for Milky Way extinction \citep{2011ApJ...737..103S} and using the distances to each object from Table \ref{tab:targets}. For our two detected SNe, ASASSN-15og and SN~2015cp, we convert the apparent NUV magnitudes into an intrinsic NUV luminosity in the same way. For all objects, we use the {\it HST} observation dates from Table \ref{tab:obs} and the dates of peak brightness from Table \ref{tab:absuv}, which come from a combination of private (e.g., PTF) and public (e.g., The Astronomer's Telegrams) resources (column 3 of Table \ref{tab:absuv}), to estimate the SN's phase at the time of every {\it HST} observation. The phases and derived spectral luminosity limits are also listed in Table \ref{tab:absuv}. In Figure \ref{fig:absuv} we plot our NUV luminosity limits vs. phase as inverted gray triangles. 

\begin{table*} 
\begin{center} 
\caption{Data Associated with Figure \ref{fig:absuv}} 
\label{tab:absuv} 
\begin{tiny} 
\begin{tabular}{lclcc} 
\hline 
\hline 
Target Name & Peak Date & Peak Date Reference(s) & Phase & $\log(L_{\rm UV})$ \\ 
            & [MJD]     &                  & [d]    & [${\rm erg\ s^{-1}\ Hz^{-1}}$] \\ 
\hline 
ASASSN-14co & $     56816 \pm    7 $ & \protect{iPTF} & 1180 & <24.79 \\ 
ASASSN-14dc & $     56828 \pm    7 $ & \protect{\cite{2014ATel.6284....1C}} & 1198 & <24.98 \\ 
ASASSN-14eu & $     56880 \pm    2 $ & \protect{Las Cumbres Observatory} & 990 & <24.64 \\ 
ASASSN-14ew & $     56884 \pm    4 $ & \protect{Las Cumbres Observatory} & 962 & <24.3 \\ 
ASASSN-14lo & $     57000 \pm    5 $ & \protect{\cite{2014ATel.6811....1A}} & 808 & <24.26 \\ 
ASASSN-14lq & $     56996 \pm    7 $ & \protect{\cite{2014ATel.6814....1Z}} & 878 & <24.49 \\ 
ASASSN-14lw & $     57012 \pm    5 $ & \protect{\cite{2014ATel.6832....1C,2014Ap&SS.354...89B}} & 768 & <24.02 \\ 
ASASSN-15de & $     57076 \pm    5 $ & \protect{\cite{2015ATel.7103....1C}} & 652 & <25.09 \\ 
ASASSN-15hy & $  57153.20 \pm 0.40 $ & \protect{\cite{2018MNRAS.475..193F}} & 695 & <24.72 \\ 
ASASSN-15jo & $     57160 \pm    5 $ & \protect{\cite{2015ATel.7553....1H}} & 666 & <24.09 \\ 
ASASSN-15nr & $     57244 \pm    5 $ & \protect{\cite{2015ATel.7882....1B}} & 540 & <24.61 \\ 
ASASSN-15og & $     57234 \pm   14 $ & \protect{\cite{2015ATel.7932....1M}} & 477 & 26.06 \\ 
ASASSN-15sh & $     57320 \pm    7 $ & \protect{\cite{2015ATel.8268....1D}} & 516 & <24.78 \\ 
ASASSN-15ut & $     57392 \pm    4 $ & \protect{\cite{2014Ap&SS.354...89B}} & 330 & <23.77 \\ 
LSQ14fmg & $     56926 \pm    5 $ & \protect{\cite{2014ATel.6495....1T}} & 330 & <23.77 \\ 
LSQ15aae & $     57120 \pm    5 $ & \protect{Las Cumbres Observatory} & 746 & <25.17 \\ 
LSQ15adm & $     57106 \pm   10 $ & \protect{Las Cumbres Observatory} & 728 & <25.43 \\ 
LSQ15bxe & $     57364 \pm   10 $ & \protect{\cite{2016ATel.8518....1F}} & 452 & <24.99 \\ 
MasterOT0442 & $     56890 \pm   30 $ & \protect{\cite{2014ATel.6487....1S,2014ATel.6488....1O}} & 926 & <24.85 \\ 
OGLE-2014-SN-107 & $     56956 \pm    5 $ & \protect{\cite{2014ATel.6612....1T}} & 756 & <25.3 \\ 
OGLE-2014-SN-141 & $     56982 \pm    5 $ & \protect{\cite{2014ATel.6706....1D}} & 726 & <25.41 \\ 
PS1-13dsg & $     56544 \pm   10 $ & \protect{\cite{2013ATel.5456....1D}} & 1472 & <25.1 \\ 
PS1-14oo & $     56720 \pm    5 $ & \protect{\cite{2014ATel.5937....1C}} & 1094 & <25.04 \\ 
PS15cwx & $  57348.50 \pm 0.30 $ & \protect{\cite{2018MNRAS.475..193F}} & 618 & <25.25 \\ 
PS15sv & $     57114 \pm    5 $ & \protect{\cite{2015ATel.7308....1W}} & 910 & <24.96 \\ 
PSNJ02+42 & $     56864 \pm   10 $ & \protect{\cite{2014ATel.6378....1S}} & 898 & <24.76 \\ 
PSNJ08+48 & $     57026 \pm   10 $ & \protect{\cite{2014ATel.6852....1Z}} & 790 & <24.94 \\ 
PSNJ23-15 & $  57259.40 \pm 0.02 $ & \protect{\cite{2018MNRAS.475..193F}} & 790 & <24.94 \\ 
PTF11kx & $     55589 \pm    1 $ & \protect{PTF} & 2134 & <25.01 \\ 
iPTF13asv & $     56430 \pm    1 $ & \protect{iPTF} & 1509 & <24.75 \\ 
iPTF13daw & $     56546 \pm    3 $ & \protect{iPTF} & 1509 & <24.75 \\ 
iPTF13dud & $     56603 \pm    1 $ & \protect{iPTF} & 1509 & <24.75 \\ 
iPTF13ebh & $     56624 \pm    3 $ & \protect{iPTF} & 1380 & <23.92 \\ 
iPTF13s & $     56339 \pm    1 $ & \protect{iPTF} & 1561 & <25.13 \\ 
iPTF14abk & $     56750 \pm    1 $ & \protect{iPTF} & 1561 & <25.13 \\ 
iPTF14aqs & $     56786 \pm    3 $ & \protect{iPTF} & 1108 & <25.42 \\ 
iPTF14atg & $     56800 \pm    1 $ & \protect{iPTF} & 1111 & <24.2 \\ 
iPTF14bdn & $     56826 \pm    3 $ & \protect{iPTF} & 1022 & <24.0 \\ 
iPTF14fpg & $     56932 \pm    2 $ & \protect{iPTF} & 1074 & <24.2 \\ 
iPTF14fyq & $     56930 \pm    3 $ & \protect{iPTF} & 818 & <25.39 \\ 
iPTF14gmo & $     56948 \pm    3 $ & \protect{iPTF} & 868 & <25.38 \\ 
iPTF14gnl & $     56956 \pm    3 $ & \protect{iPTF} & 1050 & <25.09 \\ 
iPTF14ibo & $     56984 \pm    3 $ & \protect{iPTF} & 954 & <25.57 \\ 
iPTF14sz & $     56703 \pm    1 $ & \protect{iPTF} & 1091 & <24.74 \\ 
iPTF15agv & $     57138 \pm    3 $ & \protect{iPTF} & 662 & <24.71 \\ 
iPTF15akf & $     57138 \pm    5 $ & \protect{iPTF} & 580 & <25.12 \\ 
iPTF15clp & $     57130 \pm   14 $ & \protect{\cite{2015ATel.7410....1G}} & 822 & <24.3 \\ 
iPTF15eod & $     57346 \pm    7 $ & \protect{\cite{2015ATel.8263....1T}} & 378 & <24.47 \\ 
iPTF15go & $     57053 \pm    1 $ & \protect{iPTF} & 763 & <24.64 \\ 
iPTF15wd & $     57112 \pm    2 $ & \protect{iPTF} & 764 & <25.18 \\ 
iPTF16abc & $     57499 \pm    1 $ & \protect{iPTF} & 245 & <24.4 \\ 
SN2012cg & $  56079.50 \pm 0.75 $ & \protect{\cite{2012ApJ...756L...7S}} & 1738 & <22.41 \\ 
SN2012gl & $     56228 \pm   14 $ & \protect{\cite{2012CBET.3302....2S}} & 1584 & <23.6 \\ 
SN2013I & $     56302 \pm   20 $ & \protect{\cite{2013CBET.3386....1T}} & 1686 & <24.77 \\ 
SN2013bh & $     56386 \pm    5 $ & \protect{iPTF} & 1566 & <25.35 \\ 
SN2013dn & $     56447 \pm   10 $ & \protect{\cite{2015MNRAS.447..772F}} & 1555 & <25.17 \\ 
SN2013dy & $  56501.10 \pm 0.50 $ & \protect{\cite{2015MNRAS.452.4307P}} & 1278 & <23.11 \\ 
SN2013gh & $     56528 \pm    3 $ & \protect{\cite{2016yCat..35920040F}} & 1478 & <23.47 \\ 
SN2013gv & $     56635 \pm    1 $ & \protect{iPTF} & 1136 & <24.99 \\ 
SN2013hh & $     56636 \pm    4 $ & \protect{\cite{2013CBET.3754....1K}} & 1278 & <23.76 \\ 
SN2014E & $     56664 \pm    1 $ & \protect{iPTF} & 1278 & <23.76 \\ 
SN2014I & $     56676 \pm    4 $ & \protect{\cite{2016PASA...33...55C}} & 1056 & <24.59 \\ 
SN2014J & $     56688 \pm    1 $ & \protect{\cite{2014ApJ...788L..21A}} & 1048 & <24.63 \\ 
SN2014R & $     56708 \pm    5 $ & \protect{\cite{2014Ap&SS.354...89B}} & 1064 & <24.5 \\ 
SN2014ab & $     56724 \pm   20 $ & \protect{\cite{2014CBET.3826....1H}} & 1098 & <24.51 \\ 
SN2014ai & $     56747 \pm    1 $ & \protect{iPTF} & 1021 & <24.35 \\ 
SN2014aj & $     56752 \pm   20 $ & \protect{\cite{2014CBET.3844....2E}} & 980 & <24.48 \\ 
SN2014ap & $     56738 \pm    5 $ & \protect{\cite{2016PASA...33...55C}} & 980 & <24.48 \\ 
SN2014aw & $     56780 \pm    7 $ & \protect{iPTF} & 1054 & <24.78 \\ 
SN2014bn & $     56822 \pm    5 $ & \protect{iPTF} & 1090 & <25.01 \\ 
SN2014ch & $     56814 \pm    5 $ & \protect{\cite{2014ATel.6205....1W}} & 1426 & <24.91 \\ 
SN2014dl & $     56934 \pm    4 $ & \protect{Las Cumbres Observatory} & 998 & <24.72 \\ 
SN2014dt & $     56950 \pm    7 $ & \protect{\cite{2018MNRAS.474.2551S}} & 862 & <23.13 \\ 
SN2014eg & $     56990 \pm    2 $ & \protect{Las Cumbres Observatory} & 862 & <23.13 \\ 
SN2015F & $  57106.50 \pm 0.02 $ & \protect{\cite{2017MNRAS.464.4476C}} & 602 & <23.51 \\ 
SN2015aw & $     57228 \pm    5 $ & \protect{\cite{2015ATel.7815....1M}} & 510 & <24.21 \\ 
SN2015bd & $     57344 \pm   10 $ & \protect{\cite{2015ATel.8393....1S}} & 430 & <24.37 \\ 
SN2015bp & $     57112 \pm    2 $ & \protect{\cite{2014Ap&SS.354...89B}} & 754 & <22.94 \\ 
SN2015bq & $     57088 \pm    3 $ & \protect{iPTF} & 718 & <24.53 \\ 
SN2015cp & $     57344 \pm    4 $ & \protect{iPTF} & 664 & 25.88 \\ 
\hline 
\end{tabular} 
\end{tiny} 
\end{center} 
\end{table*}

\subsubsection{Host-Galaxy Extinction Corrections}\label{sssec:disc_nuvlims_hgex}

Although we have applied a correction for Milky Way extinction, doing the same for host-galaxy dust is much more difficult because estimates of host extinction are only reliably derived from well-sampled multiband light curves or narrow absorption features in high-resolution spectra around the time of peak brightness. The two most nearby objects in our sample, SNe~2014J and 2012cg (both labeled in Figure \ref{fig:absuv}), were very well observed and also well known to suffer from significant host-galaxy extinction. 
For SN~2014J we have applied a correction of $A_{\rm F275W,host}\approx 7.5$ mag \citep{2014ApJ...788L..21A}.
For SN~2012cg we have applied a correction of $A_{\rm F275W,host}\approx1.1$ mag (from $E(B-V)=0.18$ mag; \citealt{2016ApJ...820...92M}). 

For an estimate of the average amount that host-galaxy extinction might affect our results, we consider the work of \cite{2015MNRAS.446.3768H}. They collect the observed values of host extinction for a large sample of SNe\,Ia, and find that $A_{V,{\rm host}} < 0.6$ mag in $\sim80\%$ of the cases, and that the distribution of $A_V$ values is approximately flat between $0$ and $0.6$ mag. For an extinction law similar to that of the Milky Way (which is not necessarily appropriate for all SNe\,Ia; e.g.,  \citealt{2015MNRAS.453.3300A}), this implies that $A_{\rm F275W,host} \lesssim 1.1$ mag for most of our objects. For the few cases to which this might apply, this correction would shift a point in the positive-ordinate direction by about the height of three symbols in Figure \ref{fig:absuv}, but the majority would be a smaller shift --- a small overall impact on our total survey results.

\subsubsection{ASASSN-15og and SN~2015cp}

In Figure \ref{fig:absuv} we plot our NUV detections of ASASSN-15og and SN~2015cp as an orange square and red circle, respectively (with no estimated corrections for host-galaxy extinction). In the case of ASASSN-15og, the NUV detection and its luminosity is not surprising; it was classified as a Type\,IIn/SN\,Ia-CSM because it exhibited the signatures of CSM interaction around the time of discovery, as discussed in Section \ref{sssec:review_asassn15og}. In the case of SN~2015cp, the late-time NUV detection and its luminosity are more surprising, and clearly indicate late-onset interaction between the SN ejecta and CSM, as discussed in Section \ref{sec:15cp} (see also the physical analysis of SN~2015cp in H18). We also plot two NUV upper limits for SN~2015cp based on the {\it HST} STIS spectrum at $+738$~d: a heavy- and light-weight inverted open red triangle represents the limits if the NUV luminosity was dominated by continuum or line emission, respectively (as described in Section \ref{sec:15cp}). 

\subsubsection{Comparing NUV Emission for SNe}\label{sssec:disc_compNUV}

Here we compare our observations with examples of NUV detections for other types of SNe, with and without CSM interaction: SN\,Ia 2011fe, which appeared to explode into a clean environment; SNe\,Ia that exhibit a strong signature of CSM emission (SNe\,Ia-CSM); and SNe\,IIn, in which the emission is dominated by CSM interaction starting from the time of explosion. In several cases, we have chosen to compare our F275W observations with the {\it Swift} UVOT's \textit{uvw1} filter because it has a similar effective wavelength, $\lambda_{uvw1} = 2600$~\AA.

{\bf SN\,Ia 2011fe ---} For a comparison with the late-time NUV luminosity of normal SNe\,Ia, we include the {\it Swift} UVOT {\it uvw1} light curve of prototypical SN\,Ia 2011fe from \cite{2012ApJ...753...22B}. As with the other {\it Swift} data, we convert from Vega to AB magnitudes and apply the small correction for Milky Way dust along the line of sight. SN~2011fe has been shown to have exploded in a relatively clean environment \citep{2012ApJ...750..164C,2012ApJ...746...21H,2013AA...549A..62P,2013AA...554A..27P}, and it exhibited a normal peak NUV \textit{uvw1} brightness and evolution (i.e., normal by comparison to the NUV brightness for a sample of SNe\,Ia in Table 9 of \citealt{2010ApJ...721.1608B}). The NUV photometric coverage from \cite{2012ApJ...753...22B} continues to phase of $\sim100$~d after peak. For an estimate of the intrinsic NUV flux at later times we turn to \cite{2017MNRAS.467.2392F}, who show with a NUV-optical spectrum at $+360$~d that the flux at $\sim2700$~\AA\ is about $30$--$100$ times lower than the flux in the $B$ band ($\lambda \approx 4450$~\AA). Since the apparent $B$-band magnitude of SN~2011fe was $\sim18$ mag at $+360$~d past peak \citep{2015MNRAS.446.2073G}, we use the flux ratio to estimate the intrinsic NUV emission from SN~2011fe at $+360$~d. We can see from Figure \ref{fig:absuv} that there is no danger of the emission from the cooling SN\,Ia ejecta contaminating our survey.

{\bf SN\,Ia-CSM ---} In Figure \ref{fig:absuv} we plot the {\it Swift} UVOT {\it uvw1} brightness of SN\,Ia-CSM 2012ca around the time of its light-curve peak as green stars in Figure \ref{fig:absuv} (from \citealt{2016MNRAS.459.2721I}, after converting from Vega to AB magnitudes and correcting for Milky Way extinction). This NUV emission was well in excess of normal SNe\,Ia. The NUV photometric coverage ends shortly after peak brightness, but \cite{2016MNRAS.459.2721I} show that the $U$-band brightness of SN~2012ca continues with a slow decline. Owing to the overall high-energy output of SN~2012ca, \cite{2016MNRAS.459.2721I} argue that it was a core-collapse SN and not a SN\,Ia. On the other hand, \cite{2015MNRAS.447..772F} argue that the late-time spectral observations are more consistent with a thermonuclear explosion. What is certain is that the ejecta of SN~2012ca impacted nearby CSM with a high-density component, as further evidenced by the recent identification of X-rays at a phase of $500$--$800$~d past optical maximum \citep{2018MNRAS.473..336B}. This is an extreme example of a SN\,Ia-CSM event, and not the type of phenomenon our {\it HST} survey was designed to discover. 

In Section \ref{sec:15cp} we discussed how the late-time H$\alpha$ emission from SN~2015cp resembles that of PTF11kx, and how they were likely similar SN\,Ia progenitor systems in terms of their CSM content. We also obtained an image of PTF11kx during our {\it HST} NUV survey, but did not detect it, and we have labeled the upper limit for that image in Figure \ref{fig:absuv}. This nondetection was previously published by \cite{2017ApJ...843..102G} and, as they point out, completely expected unless there had been new or renewed CSM interaction. We include it here mainly for completeness, and to demonstrate that very late-time NUV snapshots would miss a PTF11kx-like CSM interaction event. Unfortunately, to our knowledge there were no other NUV observations of PTF11kx that could be incorporated into Figure \ref{fig:absuv}.

{\bf SNe\,IIn ---} For a comparison with the late-time NUV luminosity of SNe in which the emission is dominated by CSM interaction, we plot the {\it Swift} UVOT {\it uvw1}-filter photometry for SNe\,IIn 2011ht \citep{2012ApJ...751...92R} and 2005ip \citep{2012ApJ...756..173S}, after converting from Vega to AB magnitudes \citep{2011AIPC.1358..373B} and correcting for Milky Way extinction \citep{2011ApJ...737..103S}. For SN\,IIn 2011ht we have also applied the host-galaxy extinction correction as derived by \cite{2012ApJ...751...92R}. We chose to compare to these two SNe\,IIn because they represent the extrema of NUV behavior in this heterogeneous class. SN\,IIn 2011ht rose and declined relatively quickly, reached a peak absolute magnitude of only $M_V\approx-17$, and was suggested to be SN ejecta interacting with a nearby shell of material released by the progenitor star in the year before explosion \citep[i.e., short-lived CSM interaction][]{2012ApJ...751...92R}. SN\,IIn 2011ht also exhibited qualities similar to those of ``SN impostors" \citep{2012ApJ...760...93H}, but was ultimately confirmed as a core-collapse SN based on its nebular-phase spectra \citep{2013MNRAS.431.2599M}. We note that SN~2011ht is not the only SN\,IIn seen to transition to a typical SN\,IIP after peak brightness (e.g., SN~2007pk; \citealt{2013A&A...555A.142I}). By comparison, SN\,IIn 2005ip also reached a peak absolute magnitude of $M_B\approx-17$, but continued to exhibit the signatures of ongoing CSM interaction to late times: e.g., a variable late-time light curve that does not monotonically decline, and spectra with blackbody continuum emission \citep{2012ApJ...756..173S} and narrow line emission such as \ion{Mg}{2} $\lambda2800$.
The diversity in the SN\,IIn class is driven by the radial extent and density of the CSM in these systems, but generally the CSM density is higher for SNe\,IIn compared to SNe\,Ia-CSM \citep{1997ARA&A..35..309F}, so the fact that the detection limits for our SNe\,Ia rule out any late-time SN\,IIn-like emission is not surprising.
Given this diversity, both our detection of ASASSN-15og and our nondetections of the other SNe\,IIn included in our sample (marked with black outlines in Figure \ref{fig:absuv}) are consistent with the Type IIn class. We included these SNe\,IIn because they were more {\it likely} to show NUV emission from CSM interaction at late times, but did not expect them all to yield a detection owing to the inherent diversity in the SN\,IIn class. For this reason we cannot use our SN\,IIn nondetections to (for example) further constrain our survey's detection efficiency. 

\subsection{Observational Constraints on the Occurrence Rate of CSM Interaction for our Targeted SNe\,Ia}\label{ssec:disc_rate}

We use our {\it HST} images' detection limits (Section \ref{ssec:limmags}), and our detection of SN~2015cp, to constrain the fraction of SN\,Ia systems that have CSM ($f_{\rm CSM}$), the occurrence rate of CSM interaction, and observational attributes of the NUV emission's peak, duration, and decline rate. Since our targeted sample of SNe\,Ia is overrepresentative of the subtypes that are more likely to be associated with CSM interaction (Section \ref{ssec:targs}), our results will serve as an upper limit. In the following paragraphs we introduce the key components to this analysis --- the probabilistic expectation value, control time, detection efficiency, and NUV light-curve parameters. For simplicity, in this section we assume a single shell of CSM.

\smallskip
{\bf Expected Number of Detections ---} The number of late-time NUV detections expected in our survey, $N_{\rm exp}$, is represented by

\begin{equation}\label{eq:nexp}
N_{\rm exp} = f_{\rm CSM} \sum_{i=1}^{i=N_{\rm SNIa}} P({\rm occ} \leq t_{\rm obs})  \times P({\rm det} | {\rm occ}),
\end{equation}

\noindent where $f_{\rm CSM}$ is the primary unknown we constrain with this analysis: the fraction of SNe\,Ia in our sample that have CSM within $R_{\rm CSM} \approx 3\times10^{17}$ $\rm cm$. \cite{2012ApJ...761..182M} show that $\sim10^{17}$ $\rm cm$ is about the maximum distance of a CSM shell that is formed by recurrent nova systems (see their Figure 4). By design, $3\times10^{17}$ $\rm cm$ is the distance traveled by SN ejecta with $v\approx20,000$ $\rm km\ s^{-1}$ in $+1700$ days, which is the maximum phase probed by our survey. 

The sum term in Equation \ref{eq:nexp} calculates the total number of detections that we would expect if {\it all} of our targeted SNe had CSM within $R_{\rm CSM} \approx 3\times10^{17}$ $\rm cm$, and it has two components. The first, $P({\rm occ} \leq t_{\rm obs})$, is the probability that the interaction occurred before the date of our observation, $t_{\rm obs}$, given that there was CSM within $R_{\rm CSM} \approx 3\times10^{17}$ $\rm cm$. We assume that the probability density distribution for CSM distance is flat, although this depends on the properties of the recurrent nova system and thus the distribution of (for example) mass-transfer rates in those systems \citep{2012ApJ...761..182M}. The second, $P({\rm det} | {\rm occ})$, is the probability that the NUV emission would be detected in our {\it HST} image, given that it occurred:

\begin{equation}\label{eq:pdet}
P({\rm det} | {\rm occ}) = \Delta t / t_{\rm obs}, 
\end{equation}

\noindent where $\Delta t$ is the control time as described below. The value of $P({\rm det} | {\rm occ})$ is always $\leq 1$. 

We exclude from Equation \ref{eq:nexp} the 5 SNe with classifications that include SN~IIn, which means they exhibited clear and strong signatures of CSM interaction at $<100$~d after optical peak, and the 3 classified at early times as SNe\,Ia-CSM (Table \ref{tab:targets}), because our goal is to constrain the fraction of SNe\,Ia in our sample with {\it late}-onset interaction.

\smallskip
{\bf Control Time ---} The concept of control time, $\Delta t$, also referred to as the visibility window, is the length of time over which a variable or transient object is detectable, in days:

\begin{equation}\label{eq:ct}
\Delta t = \int_{0}^{\infty} \eta(m(t))dt,
\end{equation}

\noindent where $\eta(m)$ is the survey's detection efficiency (defined below) and $m(t)$ is a fiducial light curve of the transient or variable object (defined below). For use in Equation \ref{eq:pdet} we impose that the maximum value for the control time is the SN's phase at the time of our observation: $\Delta t \leq t_{\rm obs}$.

\smallskip
{\bf Detection Efficiency ---} The detection efficiency of an imaging survey, $\eta(m)$, is the probability of detecting a source of apparent magnitude $m$ with a given threshold of confidence. In Section \ref{ssec:limmags} we determined the detection efficiency of our survey by injecting point sources with a variety of fluxes into our {\it HST} images and calculating the fraction that were detected at the 2$\sigma$ level (Table \ref{tab:SEp}). The functional form of $\eta(m)$ has two components: a bright-end flat regime in which $\eta(m) = \eta_{\rm max}$ for $m < m_{\eta_{\rm max}}$, and a faint-end decline in which the probability of detection drops with an approximately linear slope down to $\eta=0$ at $m = m_{\rm 0}$. This functional form holds for all of our {\it HST} images, and the average parameter values are $\eta_{\rm max}\approx 0.954$, $m_{\eta_{\rm max}} \approx 25.6$ mag, a faint-end slope of $-0.15$ per $0.1$ mag, and $m_{\rm 0} \approx 26.3$ mag.

\smallskip
{\bf The NUV Light Curve --- } The full evolution of NUV emission from late-onset CSM interaction for SNe\,Ia might be heterogeneous, as it is for SNe\,IIn, but this diversity has not yet been observed. Instead, we must estimate a probable functional form for a NUV light curve; to do this, we base our fiducial NUV light curve of CSM interaction, $m(t)$ in Equation \ref{eq:ct}, on the analytical models of \cite{2016ApJ...823..100H}. They show that the radio light curve from CSM interaction exhibits a sharp rise when the ejecta impact the CSM, flattens out while the ejecta pass through the CSM, and then rapidly declines after shock breakout. Based on this we use a simple functional form for the spectral luminosity $L_{\rm NUV}$:

\begin{equation}
  L_{\rm NUV}(t) =
    \begin{cases}
      L_{\rm max} & \text{if $t \leq \mathcal{W}$}\\
      L_{\rm max} \times \Phi^{ (t-\mathcal{W})/100} & \text{if $t > \mathcal{W}$}.\\
    \end{cases}       
\end{equation}

\noindent The time $t$ is in days after the ejecta impact the CSM, and $L_{\rm NUV}$ is in units of $\rm erg\ s^{-1}\ Hz^{-1}$. In this model, the luminosity exhibits a plateau of width $\mathcal{W}$ days followed by a fractional flux decline of $\Phi$ every $100$~d. This parameterization is not a direct derivation from the radio light curves, but adopted as a simple way to represent the two main light-curve features: the width of the plateau ($\mathcal{W}$) and the rate of the decline ($\Phi$). To use this light-curve model in Equation \ref{eq:ct}, we convert the spectral luminosity to an apparent magnitude in the {\it HST} F275W filter.

\begin{figure}
\begin{center}
\includegraphics[width=8.8cm]{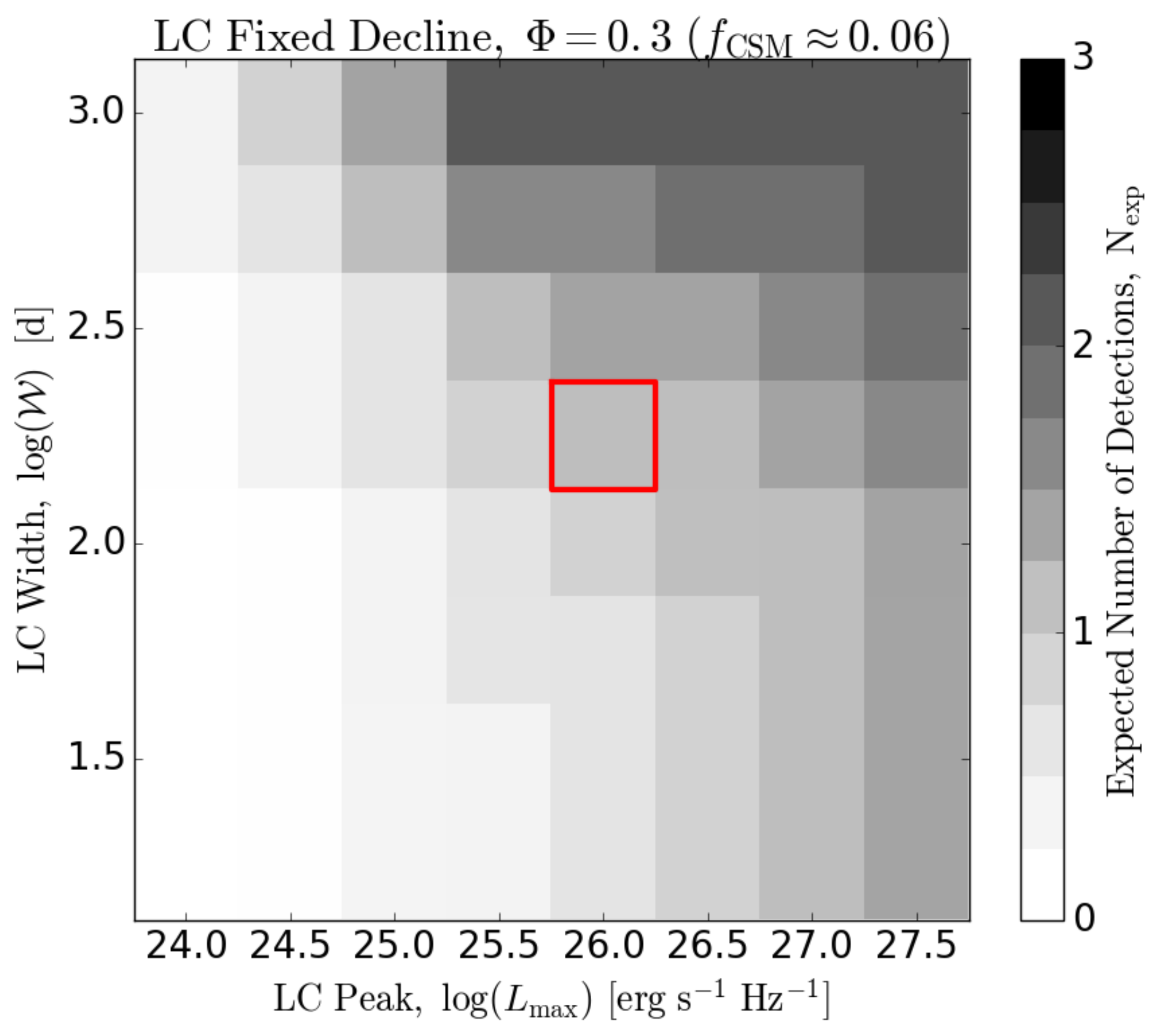}
\includegraphics[width=8.8cm]{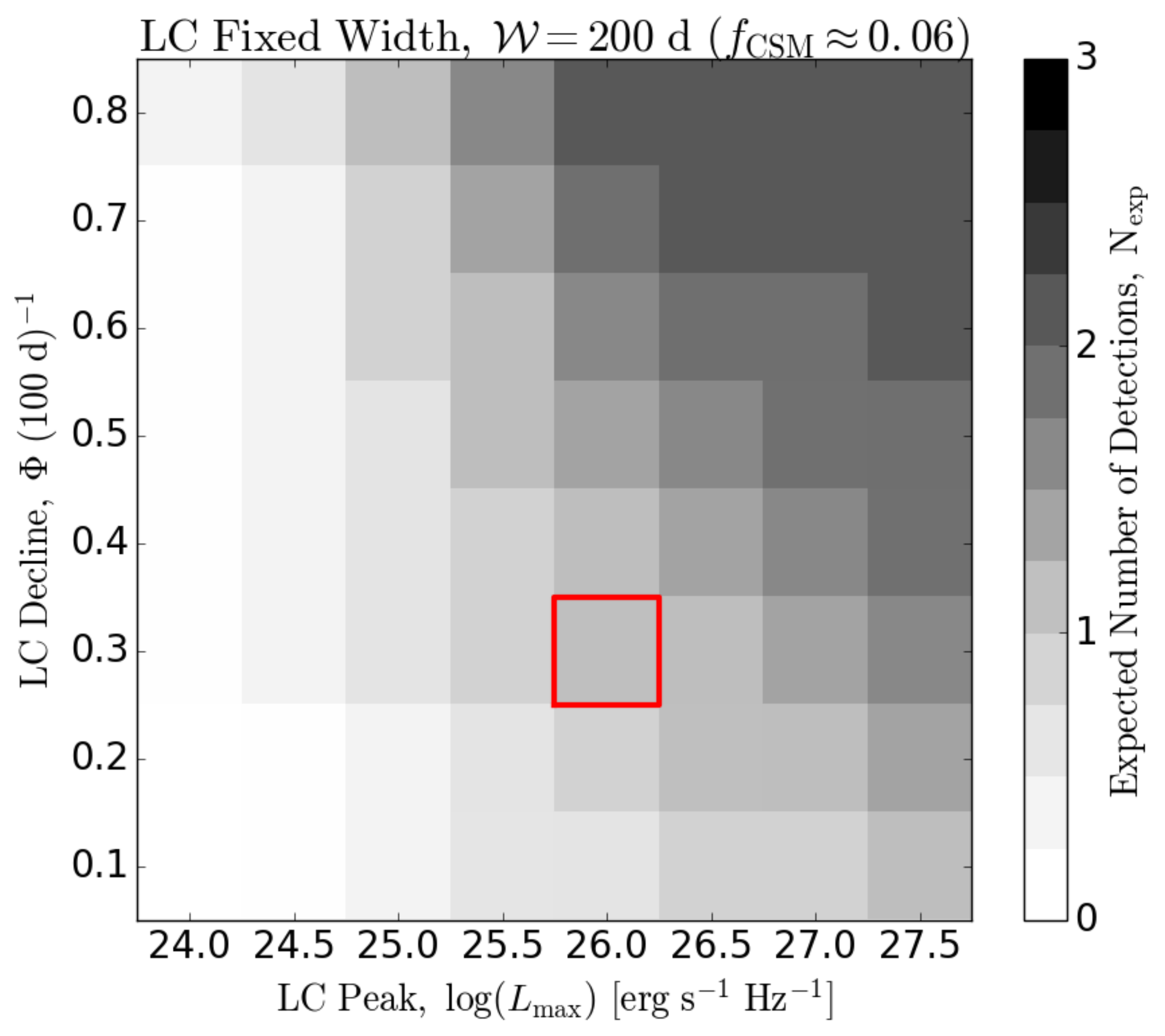}
\caption{Grids of the NUV light-curve (LC) parameters for CSM interaction: width ($\mathcal{W}$) or decline rate ($\Phi$)  vs. peak luminosity ($L_{\rm max}$), in the top and bottom panels (respectively). Boxes are shaded by the expected number of detections {\it if} the fraction of our observed SNe\,Ia that have CSM within $R_{\rm CSM} \approx 3\times10^{17}$ $\rm cm$ is $f_{\rm CSM} = 0.06$ (color bar). Red outlines the grid box of the fiducial SN~2015cp-like light curve; the value of $f_{\rm CSM}=0.06$ has been set to yield $N_{\rm exp}=1$ in this box to match our observations. \label{fig:nexp}}
\end{center}
\end{figure}

\smallskip
{\bf Results ---} To demonstrate the methodology we calculate $N_{\rm exp}$ for a hypothetical scenario in which all of our targeted SNe\,Ia had CSM within $R_{\rm CSM} \approx 3\times10^{17}$ $\rm cm$ ($f_{\rm CSM}=1$), and all exhibited emission like that of SN~2015cp: a peak intrinsic spectral luminosity of $L_{\rm NUV} = 7.6 \times 10^{25}$ $\rm erg\ s^{-1}\ Hz^{-1}$ that lasts for $\mathcal{W} = 200$~d, followed by a fractional flux decline of $\Phi = 0.3$ every $\sim100$~d. These values come from our analysis of SN~2015cp in Section \ref{sec:15cp}: the peak-luminosity estimate is a lower limit based on our single detection, the width estimate is based on our comparison of SN~2015cp to PTF11kx, and we assume that the NUV and H$\alpha$ fluxes would have a similar decline rate. 

The control time for this fiducial, SN~2015cp-like light curve in our {\it HST} survey is $\Delta t \approx 330$~d, and if all of our targets exhibited emission like SN~2015cp we would have expected $N_{\rm exp}=17.4$ detections in our survey. The fact that we only detected one suggests that the fraction of our targets with CSM is closer to $f_{\rm CSM} = 0.06$. Of our assumptions for the three light-curve parameters --- the peak, width, and decline rate --- the width is the one that is most uncertain: it is a time span based on a single epoch and essentially unconstrained. To illustrate the effect of our width assumption on our results, $\mathcal{W}=50$~d (a minimum based on the light-travel time across a shell of $R\sim10^{17}$ $\rm cm$) results in $f_{\rm CSM}=0.09$, and $\mathcal{W}=600$~d (a maximum considering the nondetections at $<50$~d after peak brightness, and $t_{\rm obs}=664$~d results in $f_{\rm CSM}=0.03$.

In reality, the amount and distribution of CSM in SN progenitor systems will have some intrinsic diversity, and \cite{2016ApJ...823..100H} show that, for example, thicker CSM shells result in a wider light curve and denser shells in a brighter maximum luminosity.  We therefore extend our analysis to incorporate a suite of light curves with peak luminosities in the range of $24.0 \leq \log{L_{\rm NUV}} \leq 27.5$ $\rm erg\ s^{-1}\ Hz^{-1}$, widths in the range of $20 \lesssim \mathcal{W} \leq 1000$~d, and fractional decline rates in the range of $0.1 \leq \Phi \leq 0.8$. These ranges represent the extent of what might be both physically possible and detectable by our survey; in particular, the shallower decline rates represent the slower NUV light-curve evolution seen for SNe\,IIn (e.g., Figure \ref{fig:absuv}).

In Figure \ref{fig:nexp} we show the number of expected detections in this light-curve parameters space, normalized by the fiducial results of $f_{\rm CSM}=0.06$. In the top panel we hold the fractional flux decline rate fixed at $\Phi=0.3$ and show $N_{\rm exp}$ in the width--peak plane, and in the bottom panel we hold the light-curve width fixed at $\mathcal{W}=200$~d and show $N_{\rm exp}$ in the decline-peak plane. Red boxes mark the location of the fiducial SN~2015cp-like light curve in this parameter space. We find that even when the brightest, longest, slowest light curves of CSM interaction are considered, we would expect $N_{\rm exp}<3$ detections with our {\it HST} survey (for $f_{\rm CSM}=0.06$). \cite{1986ApJ...303..336G} show that the Poisson limits at $2\sigma$ confidence for a single detection are $0.173$--$3.3$, which indicates that this entire range of light-curve parameters is statistically consistent with our observations. 

Our single-epoch survey is unable to put any stronger constraints on the fraction of distant CSM in SN\,Ia progenitor systems, or the occurrence rates or light-curve parameters of its associated NUV emission. As a final caveat, we note that since our SNe\,Ia were targeted for being more likely to be associated with CSM interaction (Section \ref{ssec:targs}), even the constraints that we can provide should be considered an upper limit.

\subsection{Physical Constraints on the CSM Properties for our Targeted SNe\,Ia}\label{ssec:disc_phys}

We can use the models of \citet[][hereafter H16]{2016ApJ...823..100H} to convert our upper limits on the NUV luminosity into upper limits on the CSM mass.
As shown by H16, the interaction shock heats up the CSM and the ejecta, increasing the volume of hot gas until the ejecta have swept through all of the CSM, at which point the shock-accelerated CSM rapidly expands and cools.
These models assume that the CSM has low density ($\rho < 10^{-17}~\mathrm{g~cm^{-3}}$) to avoid significant photon trapping. 
A low density is also appropriate for our systems: a dense CSM at a distance of $R_{\rm CSM} > 10^{16}$ $\rm cm$ would imply a larger mass than a red giant companion star could be expected to generate.
Additionally, such high densities and large masses of CSM would cause significant dust extinction and narrow, blueshifted absorption features in the spectra obtained around the time of peak light, which are not generally seen for SNe\,Ia.

In order to derive upper limits on the CSM mass, we assume that the time of our observations, $t_{\rm obs}$, corresponds to the time of the peak of the CSM interaction-driven NUV light curve, $t_p$, and use the H16 models to constrain the bremsstrahlung luminosity from the CSM at the moment the shock reaches the outer edge of the CSM. This assumption does not lead to the most conservative estimates on the amount of CSM, but was chosen as an intuitive and practical option for interpreting our single-epoch nondetections. If we instead assume that $t_p$ was any time in the hundreds of days before $t_{\rm obs}$, the upper limits on the peak luminosity (and CSM mass) remain essentially unbounded; to avoid this contrived assumption, future surveys for late-onset CSM interaction need to include multiple epochs of imaging. Owing to the assumption of $t_p = t_{\rm obs}$, we omit PTF11kx from this analysis because we know it was well past the peak of its interaction at the time of our observation ($t_{\rm obs} = 2134$~d), and we also omit any targeted SN with a classification of Type\,IIn because this spectral type indicates that it exhibited clear signatures of strong CSM interaction around the time of peak brightness (when it was classified).
Any emission from the reverse shock would also contribute to our observed luminosity -- although it would be a small contribution, including it would lower the CSM mass required to generate a given luminosity.

First, we will summarize some of the main results of H16 that are required for this analysis. We start by setting up the physical parameters for the CSM and ejecta density, location, mass, and volume before moving on to modeling the NUV emission from the shocked CSM. H16 describe a CSM shell of constant density $\rho_{\rm CSM}$ with an inner radius of $R_{\rm in}$, an outer radius of $R_{\rm out}$, and a fractional extent of

\begin{equation}\label{eq:F_R}
  F_R \equiv R_{\rm out}/R_{\rm in} ~. 
\end{equation}

\noindent We consider shells with $1.1 \leq F_R \leq 10$, where the lower value is like a nova shell \citep{2012ApJ...761..182M} and the upper represents a very thick shell.
For PTF11kx, observations indicated $F_R \approx 5$ \citep[e.g.,][]{2017ApJ...843..102G}. 
A strong shock forms in the CSM when ejecta of approximately $\rhocsm$ reaches the inner edge of the CSM.
As in H16, this sets the impact time $\timp$, because the outer ejecta density profile is described by
\begin{equation}\label{eq:rhoej}
  \rho_{\rm ej}(r,t) = \frac{0.124\,M_{\rm ej}}{(v_t\,t)^3} \left( \frac{r}{v_t\,t} \right)^{-10} ~,
\end{equation}
where $r$ is the radial distance from the SN center, $t$ is time since explosion, $M_{\rm ej}$ is the total mass of the ejecta, and $v_t$ is the transition velocity between the outer ejecta and inner ejecta. For a typical SN\,Ia, the transition velocity is $v_t = 1.104 \times 10^{9}$ $\rm cm\ s^{-1}$, and the mass $M_{\rm ej} = 1.38$ $\rm M_{\odot}$ $= 2.74 \times 10^{33}$ $\rm g$. 
For the family of models presented by H16, which have $\rhoej(R_{\rm in}, \timp) = \rhocsm/0.33$, this profile leads to 
the relation
\begin{equation}\label{eq:Rin}
  R_{\rm in} = 3.43 \times 10^9\,\timp^{7/10}\,\rhocsm^{-1/10} ~.
\end{equation}
H16 found that the time at which the shock overtakes the edge of the CSM shell ($t_p$) is related to the time of impact and 
the fractional extent by
\begin{equation}\label{eq:t_p}
  t_p = 0.983\,\timp\,F_R^{9/7} ~. 
\end{equation}
Because these models are appropriate for low-density shells, the evolution assumes a gamma-law equation of state with adiabatic index $\gamma_{\rm ad}=5/3$. Thus, the strong shock increases the CSM (and ejecta) density by a factor of four: shocked CSM has $\rho_s = 4\rhocsm$, and the initial volume of the shell is simply expressed by
\begin{equation}\label{eq:Vcsm}
  V_\csm = (4\pi/3)\,R_{\rm in}^3\, (F_R^3 - 1) ~. 
\end{equation}
At the time we are considering, $t_p$, the CSM volume is $V_{s,p} = V_\csm/4$ owing to shock compression.

Now we turn to H16's models for the bremsstrahlung emission from the shocked CSM at $t_p$, the peak time of the interaction-driven luminosity.
For the densities and shell extents that we are considering, the bremsstrahlung process is very securely in the optically thin regime --- and recall that we are calculating at the time that all of the CSM has been shocked.
In the optically thin regime, the spectral luminosity ($\mathrm{erg~s^{-1}~Hz^{-1}}$) can be estimated from the emissivity ($\varepsilon_\nu$) and emitting volume $V_s$ to be $L_\nu = \varepsilon_\nu V_s$.

The emissivity in cgs units ($\rm erg\ s^{-1}\ cm^{-3}\ Hz^{-1}$), from Equation 5.14b of \cite{RL1979}, is
\begin{equation}\label{eq:E_ff}
  \mathcal{E}_{\nu}^{\mathit{ff}} \approx 6.8 \times 10^{-38} \, Z^2 \, n_e \, n_i \, T^{-1/2} \, e^{-h \nu / kT}\, \bar{g}_{\mathit{ff}},
\end{equation}
\noindent where $Z$ is the charge on the ion, $n_e$ and $n_i$ are the particle densities of electrons and ions (respectively), $T$ is the temperature, and the frequency is $\nu \approx 1.11 \times 10^{16}$ $\rm s^{-1}$ (from the pivot wavelength of the {\it HST} WFC3 F275W filter, $\lambda = 2704$ \AA). We set the thermally averaged Gaunt factor $\bar{g}_{\mathit{ff}}=1$, appropriate for an order-of-magnitude estimate at near-optical wavelengths.
The temperature of the shocked CSM depends on its density, but it will be high ($10^8$--$10^9$ K) owing to its low density and the high shock speed.
Given that the energy of a NUV photon is $h \, \nu \approx 6.6 \times 10^{-11}$ $\rm erg$, the exponential term in the emissivity is approximately unity.
We furthermore assume the simple case of complete ionization of H-dominated material in the emitting region, such that $Z=1$ and $n_e = n_i = \rho_s/m_p = 4\rhocsm/m_p$, where $m_{\rm p}$ is the proton mass. 

Using $\rho_s = 4\rhocsm$, $M_\csm = \rhocsm V_\csm$, and Equations \ref{eq:F_R}--\ref{eq:Vcsm}, we find that the NUV
bremsstrahlung spectral luminosity of the fully shocked CSM shell at the moment the shock reaches its outer edge is 
\begin{equation}\label{eq:mass_to_lum}
  L_{\nu} \approx 1.63 \times 10^{-31} \, T_s^{-1/2} \, t_{\rm obs}^{-3} \, M_\csm^{17/7} \, y(F_R),
\end{equation}
where $y(F_R) = F_R^{-3/7} \, (1-F_R^{-3})^{-10/7}$. The $y(F_R)$ term is dominated by the first factor for $F_R^3 \approx 1$ and by the second factor for $F_R^3 \gg 1$.
As in Equation \ref{eq:E_ff}, all units are cgs.
From this equation it is clear that a maximum allowed CSM mass can be found from a luminosity upper limit, for a known $t_{\rm obs}$, once values for $T_s$ and $F_R$ are assumed. For example, an observation of a NUV luminosity limit $L_{\nu, {\rm NUV}} = 10^{25}~{\rm erg~s^{-1}~Hz^{-1}}$ rules out the presence of a nova-like shell ($F_R = 1.1$) more massive than $5$~$\rm M_\odot$ (assuming that the post-shock temperature is $T_s = 10^8$~K).

\begin{figure*}
\begin{center}
\includegraphics[width=18cm]{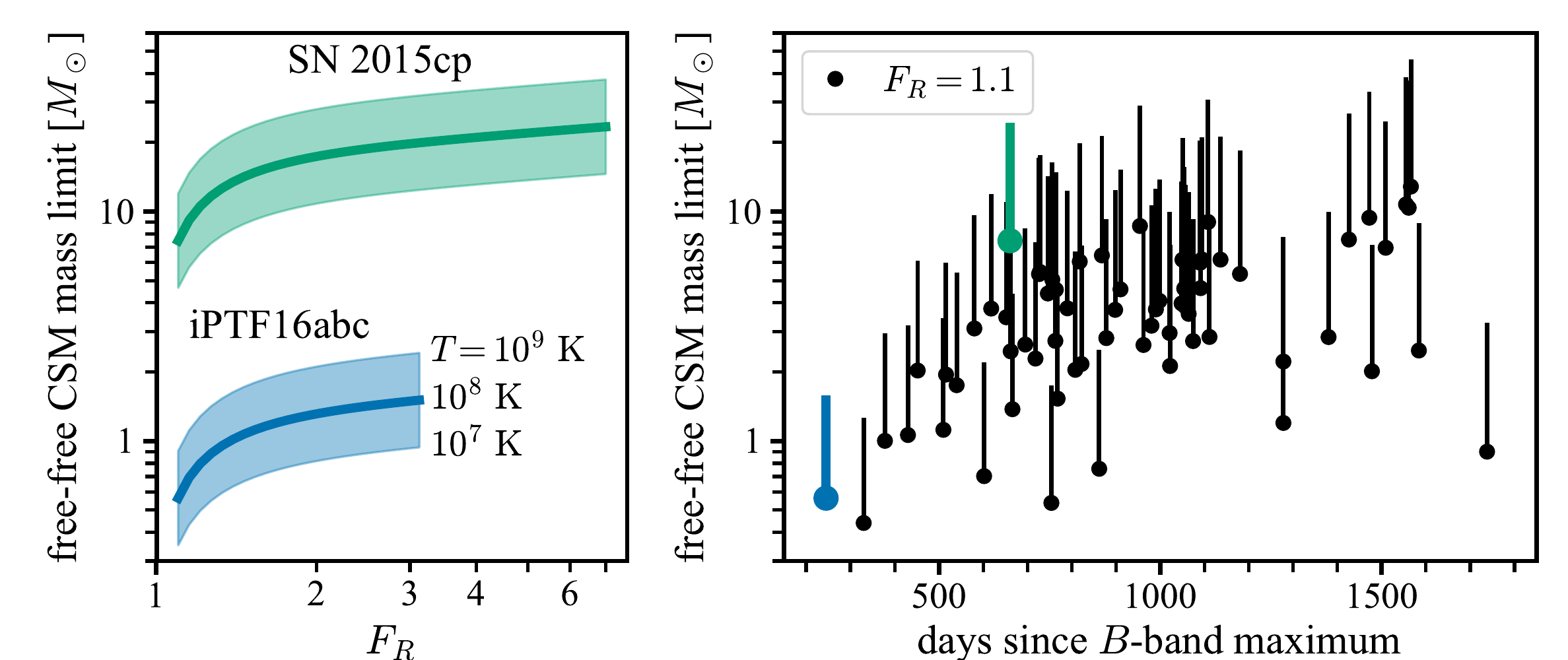}
\caption{
Maximum CSM mass allowed by our NUV nondetections (or detection, in the case of SN 2015cp)
given that shocked CSM must radiate by the free-free (bremsstrahlung) process. 
{\it Left:} The dependence of Equation \ref{eq:mass_to_lum} on $F_R$ and $T$ for two
 of the SNe in our sample. Because iPTF16abc was observed at an earlier phase than SN~2015cp, its domain of possible $F_R$ values is smaller. Very high CSM masses would be required to explain the NUV detection
 of SN~2015cp with free-free emission, favoring the interpretation of that detection as line emission.
 {\it Right:} The CSM mass limits for our total sample --- a literal conversion of the luminosity limits in Figure \ref{fig:absuv} into mass limits --- assuming $T = 10^8~{\rm K}$. For each SN, we show the limit for $F_R = 1.1$ (circles) through its maximum possible $F_R$ --- i.e., the minimum and maximum values of the curves from the left panel. The tightest constraints ($<1$ $\rm M_{\odot}$) are obtained from observations at earlier phases (denser CSM) and for more nearby SNe. 
\label{fig:physics}}
\end{center}
\end{figure*}

The behavior of Equation \ref{eq:mass_to_lum} is shown in the left panel of Figure \ref{fig:physics} for two of the SNe in our sample: the positive detection of SN~2015cp and our earliest nondetection, iPTF16abc --- which was well studied for its peculiarity at early times \citep{2017A&A...606A.111F,2018ApJ...852..100M,2018MNRAS.480.1445D}. The tighter mass constraint for iPTF16abc is mostly because the luminosity is limited to be fainter than the detection of SN~2015cp (see Figure \ref{fig:absuv}), but also because the observation time of iPTF16abc is earlier and thus requires a closer, denser shell that would be more luminous for a given mass than a more-distant shell. The two curves in the left panel of Figure \ref{fig:physics} represent the extrema of the full range of constraints provided by all of our NUV {\it HST} observations.

The final results of this analysis --- the upper limits on the CSM mass for all of our observed SNe\,Ia, based on our models of the bremsstrahlung (free-free) luminosity from the hot shocked CSM --- are shown in the right panel of Figure \ref{fig:physics}. We illustrate the full range of allowed $F_R$ for a given SN, after imposing $\timp \geq 60$~d since explosion, which puts a lower limit on $R_{\rm in}$ (to incorporate the fact that none of these SNe\,Ia exhibited the signatures of interaction during their initial observations).
In the right-hand plot of Figure \ref{fig:physics}, circles mark the mass limit for an $F_R=1.1$ shell, and the vertical lines extend up to the mass limit at the maximum possible $F_R$ for each SN (i.e., the horizontal axis of the left-hand plot). We note that the CSM densities that these limits imply are consistent with the optically thin assumption (limiting how high the CSM density can be) and the assumption that SNe\,Ia surely do not have ejecta faster than 45,000~${\rm km~s^{-1}}$ (limiting how low the CSM density can be).

Figure \ref{fig:physics} demonstrates that, when we only consider bremsstrahlung emission from the shocked CSM, our survey was sensitive to CSM shells with masses as low as a few $\rm M_{\odot}$ for our closest targets. Based on this analysis alone, it would appear that our survey was insensitive to small amounts of CSM: several tenths of a solar mass in some cases, but $>1$ $\rm M_{\odot}$ in most cases, could have remained undetected. However, as we explain below, we find that this insensitivity is more related to our assumption of bremsstrahlung emission for the  physical modeling than to any fault of this survey's observational depth or design.

For example, in Figure \ref{fig:physics} we see that the lower limit on the free-free CSM mass of SN~2015cp is $\sim 7$ $\rm M_{\odot}$, which is an unreasonably high CSM mass. For comparison, the CSM mass limit for SN~2015cp based on radio follow-up observations is $M_\csm \lesssim 0.5~M_\odot$ (H18; but we note that obtaining that limit depended heavily on the NUV discovery and optical follow-up data). This discrepancy illustrates why we suspect the NUV emission to originate from atomic line emission (\ion{Mg}{2} at 2800~\AA) rather than continuum emission. Unfortunately, line emission is much more complicated than bremsstrahlung to approach theoretically --- it requires many more assumptions about the state of the gas. Such a tool for translating our upper limits into CSM masses based on line emission is not yet in hand, and we consider its development beyond the scope of this work.

One possible alternative approach assumes that the $\Hal$ line is formed through radiative recombination during the cooling of shocked gas \citep[][e.g.,]{Ofek+2013}, but as this results in CSM masses (or magnesium abundances) that are too high to be physically feasible, we do not attempt to apply it to our inferred NUV lines in this work. An appropriate and accurate model of the line formation for the detached CSM shell scenario, like that presented by \citet{1994ApJ...420..268C} for SNe~II interaction with a wind, will (in the future) allow us to utilize the results of our survey to deduce more sophisticated and stringent mass limits. We cannot apply the models of \cite{1994ApJ...420..268C} directly to our observations because the hydrodynamics of SNe\,Ia are very different from the SNe\,II that they have modeled: SNe\,Ia have one tenth the ejecta mass of SNe\,II, the mass and distribution of their CSM are quite different, and SNe\,Ia have steeper density gradients in their outer layers and their fastest-moving ejecta travels at about twice the speed compared to SNe\,II. Although \cite{1994ApJ...420..268C} find a general trend that SNe with a steeper gradient for the ejecta's radial density distribution leads to more luminous emission from the reverse shock (i.e., their figure 3), this cannot be directly extended to our discussion about a potential contribution to the NUV emission from the reverse shock in the ejecta of SN~2015cp. Additionally, timescales for the decline of NUV emission dominated by contributions from the reverse-shock are expected to be much longer, perhaps years compared to the few-week decline we infer from our {\it HST} follow-up nondetection for SN\,2015cp. In lieu of such models for SNe\,Ia, we again stress the advantage of multiwavelength observations that include X-ray and radio bands (as done by H18) for constraining the CSM properties.

\section{Conclusions}\label{sec:conc}

In this work we presented the results of the first systematic survey for late-onset CSM interaction in SNe\,Ia. Our {\it HST} NUV snapshot survey obtained images with the F274W filter for the sites of 72 SNe at $1$--$3$~yr after their explosion. The majority of these observations yielded nondetections with a limit of $m_{\rm F275W} < 25.8$ mag (AB), but two SNe were detected: ASASSN-15og, a SN\,IIn that exhibited strong signatures of CSM interaction at epochs near peak brightness, and SN~2015cp, a SN\,Ia that did not. Our single NUV detection of SN\,Ia 2015cp at $664$~d demonstrated that late-onset CSM interaction does occur, and that it has likely been missed in typical SN\,Ia observations that focus on the first month after peak brightness.

We provided an in-depth analysis of SN~2015cp that includes photometry and spectroscopy obtained $<40$~d after its light-curve peak, combined with our NUV through optical follow-up observations at $>664$~d. We find that SN~2015cp is similar to PTF11kx, a SN\,Ia whose ejecta were observed to impact CSM in its progenitor system $\sim40$~d after light-curve peak. Ground-based optical spectroscopy revealed intermediate-width H$\alpha$ and \ion{Ca}{2} emission that are classic signatures of CSM interaction. The observed decline in spectral flux over $\sim100$~d  indicated that we caught SN~2015cp after the peak of its interaction-driven emission, and that the ejecta had fully passed through the CSM. We demonstrated how the emission mechanism for the observed NUV flux is difficult to interpret, especially as we had only a single photometric detection. Although we find that the NUV emission is more likely to be line- than continuum-dominated, we cannot rule out the latter. 

We converted our detections (and the limiting magnitudes of our nondetections) into upper limits on the intrinsic NUV spectral luminosity, and compared with NUV observations of SNe\,IIn, SNe\,Ia-CSM, and normal SNe. Our nondetections rule out any late-time CSM interaction at the level exhibited by SNe\,IIn, as expected. We performed a statistical analysis of our survey's detection efficiency for events like SN~2015cp, and posit that such CSM interaction events are rare for SN\,Ia progenitor systems ($f_{\rm CSM} < 0.06$). Since our targeted sample is not representative of all SNe\,Ia, but composed of SNe\,Ia that are more likely to exhibit CSM interaction, this should be considered a maximum upper limit. The true rate of observable CSM interaction events like SN~2015cp and PTF11kx could be significantly lower. 

We applied theoretical models of free-free emission for optically thin shells of CSM and showed that our {\it HST} survey is sensitive to cases in which the SN\,Ia ejecta interacted with several solar masses of material. However, most of our NUV observations --- and in particular our NUV detection of SN~2015cp --- are consistent with tens of solar masses of CSM (or more), which is more than a recurrent nova system is expected to produce (e.g., \citealt{2012ApJ...761..182M}). We discussed how the most likely explanation is that the NUV emission is dominated by lines and not continuum, but that the modeling required to derive CSM mass estimates does not yet exist and is beyond the scope of this paper.

We recommend that in the future, such late-time surveys for CSM interaction have follow-up observations for detections built into their original proposal. The time to prepare and execute a separate proposal (in this case, we used {\it HST} Director's Discretionary Time) is too much of a risk given that a sparsely sampled late-time survey is most likely to detect interaction-driven emission when it is at its peak, after which the decline can be very rapid. We furthermore recommend that radio and/or X-ray observations be included, as H18 demonstrate the constraining power of these wavelengths in their analysis of SN~2015cp. In the LSST era, when $\sim18,000$ square degrees of the southern sky is surveyed $\sim3$ times weekly to $r\approx24.7$ mag, photometric detections of a late-time rise in H$\alpha$ emission at the sites of SNe\,Ia will be done automatically. This would have been sufficient to monitor the rise and decline of the emission from CSM interaction from SN~2015cp. Such future observations will allow us to make more detailed model fits for CSM characteristics, and better constrain the frequency of CSM interaction in SNe\,Ia progenitor systems.

\section*{Acknowledgements}

This work is based in part on observations made with the NASA/ESA {\it Hubble Space Telescope}, obtained (from the Data Archive) at the Space Telescope Science Institute (STScI), which is operated by the Association of Universities for Research in Astronomy (AURA), Inc., under National Aeronautics and Space Administration (NASA) contract NAS 5-26555. These observations are associated with program GO-14779 (PI Graham), for which support was provided by NASA through a grant from STScI, and with program GO-15407 (PI Graham).

This work makes use of data from Las Cumbres Observatory, the Supernova Key Project, the BANZAI pipeline \citep{banzai}, and the Supernova Exchange, all of which are funded in part by National Science Foundation (NSF) grant AST-1313484.
This work is based in part on observations from the Low Resolution Imaging Spectrometer at the Keck-1 telescope.
We are grateful to the staff at Keck Observatory for their assistance. The W.~M. Keck Observatory is operated as a scientific partnership among the California Institute of Technology, the University of California, and NASA; it was made possible by the generous financial support of the W.~M.\ Keck Foundation. 
The authors wish to recognize and acknowledge the very significant cultural role and reverence that the summit of Maunakea has always had within the indigenous Hawaiian community.  We are most fortunate to have the opportunity to conduct observations from this mountain.
We thank Daniel Perley and Brad Cenko for the use of, and assistance with, their Keck LRIS imaging and spectroscopy reduction pipeline\footnote{Dan Perley's pipeline can be found at \url{http://www.astro.caltech.edu/$\sim$dperley/programs/lpipe.html}}.

This work is based in part on observations made with the Kast Spectrograph on the Shane 3~m telescope at Lick Observatory. A major upgrade to the Kast spectrograph was recently made possible by a generous gift from the Heising-Simons Foundation as well as William and Marina Kast.  
Research at Lick Observatory is partially supported by a generous gift from Google.
This work is based in part on observations made with ESO telescopes at the La Silla Paranal Observatory under programmes ID 099.D-0683(A) (PI Maguire), and ID 198.A-0915 (PI Sullivan).
This work has made use of publicly available PanSTARRS data. Operation of the Pan-STARRS1 telescope is supported by NASA under grants NNX12AR65G and NNX14AM74G issued through the NEO Observation Program.
This work is based (in part) on observations collected at the European Organisation for Astronomical Research in the Southern Hemisphere, Chile as part of PESSTO (the Public ESO Spectroscopic Survey for Transient Objects) ESO program 191.D-0935.

This paper uses data from SDSS-IV, published in the public domain. Funding for the Sloan Digital Sky Survey IV has been provided by the Alfred P. Sloan Foundation, the U.S. Department of Energy Office of Science, and the Participating Institutions. SDSS acknowledges support and resources from the Center for High-Performance Computing at the University of Utah. 
This paper made use of data from the Dark Energy Camera Legacy Survey (DECaLS), NOAO Proposal ID \#2014B-0404 by PIs David Schlegel and Arjun Dey, from the Cerro Tololo Inter-American Observatory. NOAO is operated by AURA, Inc., under a cooperative agreement with the NSF.

We would like to thank LIGO, and the teams doing optical gravitational wave (GW) follow-up observations, for the detection of SN~2015cp. This event was originally discovered by PanSTARRS in imaging data obtained as part of a search for the optical counterpart of a GW trigger in 2015. Our Cycle 24 {\it HST} visits for SN~2015cp and 12 other targets were conducted in late 2017 when additional {\it HST} Snapshots were needed to fill in the gaps in {\it HST}'s schedule left owing to the impact of follow-up triggers for GW170817.

This research has made use of a variety of online tools, including the VizieR catalogue access tool, CDS, Strasbourg, France (the original description of the VizieR service was published by \citealt{2000A&AS..143...23O};
NASA's Astrophysics Data System Bibliographic Services;
the NASA/IPAC Extragalactic Database (NED), which is operated by the Jet Propulsion Laboratory, California Institute of Technology, under contract with NASA;
the Open Supernova Catalog \citep{2017ApJ...835...64G};
the Weizmann interactive supernova data repository, WISERep (\url{http://wiserep.weizmann.ac.il}; \citealt{2012PASP..124..668Y});
the GEneric cLAssification TOol (GELATO; \url{gelato.tng.iac.es}; \citealt{2008A&A...488..383H});
and the AstroBetter blog and Wiki.
This research made use of {\tt astropy}, a community-developed core {\tt Python} package for astronomy \citep{2013A&A...558A..33A}, as well as {\tt scipy} \citep{citescipy}, {\tt matplotlib} \citep{4160265}, and {\tt numpy} \citep{5725236}.

A.V.F.'s group at U.C. Berkeley has been supported by Gary \& Cynthia Bengier, the Richard \& Rhoda Goldman Fund, the Christopher R. Redlich Fund, the TABASGO Foundation, and NSF grant AST--1211916. 
M.S. acknowledges support from EU/FP7-ERC grant \#615929.
K.J.S. is supported by NASA through the Astrophysics Theory Program (NNX17AG28G).
K.M. acknowledges support from the UK STFC through an Ernest Rutherford Fellowship and from Horizon 2020 ERC Starting Grant (grant \#758638).





\bibliographystyle{apj}
\bibliography{apj-jour,ms}

\label{lastpage}


\end{document}